\documentclass{article}
\pdfoutput=1
\usepackage[dvipsnames]{xcolor}
\usepackage{jheppub}
\usepackage{lineno}
\usepackage{style}
\usepackage{braket}	
\usepackage{tikz}
\usepackage{pgfplots}
\usepackage{bm}
\usetikzlibrary{decorations.markings}
\usetikzlibrary{decorations.pathmorphing}
\usetikzlibrary{arrows.meta}
\usepackage{graphicx}
\usepackage{mathrsfs}

\title{\boldmath Mixed signals in the IR: \\
Positivity bounds with indefinite species}

\author[a,b,c]{Claudia de Rham,}
\author[a,d]{Sumer Jaitly,}
\author[a,b,e,f]{and Greg Kaplanek}

\affiliation[a]{Abdus Salam Centre for Theoretical Physics, Imperial College, London, SW7 2AZ, UK}
\affiliation[b]{Perimeter Institute for Theoretical Physics, 31 Caroline St N, Waterloo, ON, N2L 2Y5, Canada}
\affiliation[c]{CERCA, Department of Physics, Case Western Reserve University, 10900 Euclid Ave, Cleveland, OH 44106, USA}
\affiliation[d]{Scuola Normale Superiore and INFN, Piazza dei Cavalieri 7, 56126, Pisa, Italy}
\affiliation[e]{Department of Electrical Engineering and Computer Science, Syracuse University, NY 13210, USA}
\affiliation[f]{Institute for Quantum \& Information Sciences, Syracuse University, NY 13210, USA}

\emailAdd{c.de-rham@imperial.ac.uk, sumer.jaitly@sns.it, gkaplane@syr.edu}

    \abstract{In theories with multiple particle species standard fixed-$t$ positivity bounds do not directly apply to 2-to-2 definite species scattering amplitudes when the initial and final state are not the same (inelastic processes). These inelastic amplitudes are nevertheless constrained by positivity bounds indirectly, by considering scattering states which are arbitrary superpositions of definite species two-particle states. While these ‘superposition bounds' have been studied and utilised extensively in the past, earlier analyses typically consider cases insensitive to relative particle masses and IR branch cuts. Here we derive new families of bounds that take account and depend explicitly on mass differences between species making \textit{no assumption of weak-coupling}. We emphasise unusual non-analyticities induced by the IR mass difference within the superposition amplitude and use fixed (backwards) angle dispersion relations to prove our bounds. We then discuss extensions of our results to ‘improved bounds', with implications worth exploring for pions and other EFTs of the Standard Model and Beyond, particularly where IR branch cuts are non-negligible.}

\begin{document}
\maketitle
\flushbottom

\section{Introduction}
\label{sec:intro}

First investigated in the 1960s \cite{Hearn:1962zz,Martin:1965jj,Goldberg:1968zza,Martin:1969ina,PhysRev.166.1768,Ananthanarayan:1994hf,PhysRevD.31.3027}, recent years have witnessed a remarkable modern revival of the $S$-matrix bootstrap programme \cite{Paulos:2016fap, Paulos:2016but, Paulos:2017fhb, Guerrieri:2018uew, Doroud:2018szp, He:2018uxa, Cordova:2018uop, Homrich:2019cbt, EliasMiro:2019kyf, Correia:2020xtr, Bose:2020shm, Bose:2020cod, Hebbar:2020ukp,Guerrieri:2021tak, Chen:2021pgx, EliasMiro:2022xaa, Tourkine:2023xtu, He:2023lyy,Haring:2023zwu,Guerrieri:2023qbg, Eckner:2024ggx, Guerrieri:2024ckc, Bhat:2024agd, Copetti:2024dcz, Guerrieri:2024jkn, Gumus:2024lmj, He:2025gws,Correia:2025uvc,deRham:2025mjh,deRham:2025vaq}. 
From a fundamental perspective, constraining the space of low-energy effective field theories (EFTs) that admit a standard (Wilsonian) UV completion relies on basic principles such as unitarity, locality, Lorentz invariance and causality implemented through analyticity  of the $S$-matrix \cite{Martin:1969ina}. These principles lead to {\it positivity bounds}, which impose rigorous constraints on EFT couplings and have proven invaluable in charting the landscape of consistent theories \cite{Adams:2006sv,Jenkins:2006ia,Luo:2006yg,Nicolis:2009qm,Bellazzini:2014waa,Keltner:2015xda,Bellazzini:2016xrt,Bonifacio:2016wcb,Cheung:2016wjt,Bai:2016pae,deRham:2017avq,deRham:2017imi, deRham:2017zjm,Herrero-Valea:2019hde,Melville:2019gux, Bellazzini:2020cot,Arkani-Hamed:2021ajd,Noumi:2021uuv,Noumi:2022zht,Chandrasekaran:2018qmx,Cheung:2025nhw,
Kennedy:2020ehn,Wang:2020jxr,Bern:2021ppb, Caron-Huot:2021rmr, deRham:2022hpx, Baumgart:2022yty,deRham:2021bll, Li:2022tcz,Albert:2022oes,Haring:2022sdp,Henriksson:2022oeu,Herrero-Valea:2022lfd,CarrilloGonzalez:2023cbf, Berman:2023jys, Berman:2024wyt, Albert:2024yap, Caron-Huot:2024lbf,Ye:2024rzr,Chang:2025cxc, Beadle:2025cdx, Cheung:2025krg,
Tolley:2020gtv,Caron-Huot:2020cmc,Sinha:2020win,Du:2021byy,Chowdhury:2021ynh,Arkani-Hamed:2020blm,Chiang:2021ziz,Xu:2023lpq,
Cheung:2014ega,Cheung:2016yqr,deRham:2018qqo,Hamada:2018dde,Alberte:2019xfh,Alberte:2019zhd,deRham:2022sdl,deRham:2022gfe,Alberte:2021dnj,
Zhang:2018shp,
Zhang:2020jyn,Li:2021lpe,Remmen:2019cyz,Remmen:2020vts,Bonnefoy:2020yee,Yamashita:2020gtt,Remmen:2022orj,Chen:2023bhu,Wan:2024eto}. These bounds are particularly useful in guiding EFT constructions and excluding regions of parameter space that would otherwise appear viable from a purely low-energy consistency perspective.

In EFTs involving multiple fields, as is the case in the Standard Model, the proliferation of independent couplings can rapidly become overwhelming, threatening the efficiency of the systematic EFT approach. For instance, considering operators of dimension-8 alone, the Standard Model EFT (SMEFT) already contains 44,807 couplings \cite{Murphy:2020rsh,Li:2021tsq}. Previous studies have shown, however, that while the number of couplings grows significantly with the number of fields, so too does the number of independent positivity bounds \cite{Zhang:2020jyn, Li:2021lpe,Remmen:2019cyz,Remmen:2020vts,Bonnefoy:2020yee,Yamashita:2020gtt,Remmen:2022orj,Chen:2023bhu,Wan:2024eto}. These bounds have proven remarkably effective at severely restricting the allowed parameter space, and in some cases even forbidding specific interactions altogether.

A key advance in this program has been the use of indefinite helicity scattering states when deriving positivity bounds \cite{Cheung:2014ega,Cheung:2016yqr,deRham:2018qqo,Hamada:2018dde,Alberte:2019xfh,Alberte:2019zhd,deRham:2022sdl,deRham:2022gfe,Alberte:2021dnj}. This approach enables the derivation of compact bounds on the EFT parameter space without resorting to non-linear constraints as derived in \cite{Tolley:2020gtv,Sinha:2020win,Caron-Huot:2020cmc,Arkani-Hamed:2020blm,Chiang:2021ziz,Du:2021byy,Chowdhury:2021ynh,Xu:2023lpq}. In this work, we extend the indefinite-state methodology to more exotic superpositions of external states, where the species have different masses and the superposed states also may no longer be center-of-mass energy eigenstates. Our motivation is to rigorously derive analytic bounds on the superposition amplitude without making any simplifying assumption on the IR analytic structure of the amplitudes, thereby providing further justification for their validity.

Deriving positivity bounds in general can be reduced to two steps: first, expressing the amplitude as an integral over its discontinuity across the real axis, and second, imposing unitarity to obtain inequalities on this integral. An essential ingredient in establishing these bounds is a careful treatment of the non-analytic structure of scattering amplitudes at low energies. When the $S$-matrix element between superposition states is considered, it can be decomposed into a linear combination of definite species amplitudes which notably contains fixed angle backwards limit amplitudes in addition to the usual forward limit ones. We make use of dispersive representations of these amplitudes in the backward limit ($\theta = \pi$), in particular the form presented by Goldberg \cite{Goldberg:1968zza}, to prove positivity bounds for unequal mass superposition scattering. As our bounds are derived with no weak coupling assumption to remove IR branch cuts we require that in a superposition of two species, the ratio of the heavier and the lighter mass should not exceed\footnote{Coincidentally this condition ensures the absence of anomalous thresholds \cite{Correia:2022dcu}.} $\sqrt2$, to ensure that the superposition amplitude has a region of analyticity on the real $s$ axis. This restriction can be considered severe, however in the final section of the work we explain how one can relax this condition by way of an improved amplitude if loop corrections can be computed accurately in the EFT.

Our analysis builds on  previous work such as \cite{Remmen:2020vts,Bonnefoy:2020yee,Yamashita:2020gtt,Remmen:2022orj,Chen:2023bhu,Wan:2024eto}, which considered superpositions of helicities in the SMEFT context, primarily in the massless limit. By allowing for species with different masses, we aim to understand how positivity bounds constrain interactions in more realistic settings, including applications to glueball EFTs and chiral perturbation theory. For instance, in the latter case, the octet of pseudoscalar mesons $(\pi, K, \eta)$ provides a natural laboratory, where the $K$ and $\eta$ masses are close enough that our bounds may offer non-trivial insights into the allowed couplings.\\

The paper is organized as follows: In Section~\ref{sec:setup}, we review the general framework for positivity bounds and indefinite states in the equal mass case. Section~\ref{sec:Unequal_1} then dives into the core of this work by introducing our extension to unequal mass (indefinite species) superpositions and discusses the non-analytic features that emerge and their dispersive representation. We see that introducing the mass difference and treating it carefully generically weakens the bounds that one would obtain by approximating the masses to be equal. Then we consider a generalised family of superposition amplitudes in Section~\ref {sec:gen} which involve a sum of two-particle states with independent centre-of-mass energies, leading to a family of generalised unequal mass bounds. These bounds are complementary to those obtained in the previous section and in both cases we discuss the percentage by which the new bounds are weaker when applied to a tree-level EFT. We then in Section~\ref {sec:Improved_Bounds} briefly discuss how by computing in the EFT beyond tree level it is possible to to relax the upper bound on the heavier species mass in the superposition.

Appendix~\ref{App:kinematics} presents some standard results for $2\to 2$ scattering and the relevant Mandelstam variables that enter the inelastic $t$-channel amplitude in the forward and backward limits. Details related to the non-trivial derivation of the backward limit of the dispersion relation in the non-equal mass superposed states amplitude is given in Appendix~\ref {app:backwards_dr}. The proof of the generalised superposition bounds are given in Appendices~\ref {app:gen_bound_proof} and \ref {App:Alt_Bound1}. Finally we apply our generalized bounds to a specific two-field scalar EFT in Appendix~\ref {App:ToyEFT} and compared with causality bounds and an example of a (partial) UV completion.

\paragraph{Conventions, notation and terminology:} We shall deal with scattering amplitudes between two scalar fields $\phi$ and $\chi$. For elastic scattering processes, where the initial two particles scatter to themselves e.g. $\phi\phi\xr\phi\phi$ or $\phi\chi\xr\phi\chi$, we refer to the process with a shorthand omitting repetition of the particle names, e.g. $\phi\phi\xr\phi\phi$ is `$\phi\phi$' and $\phi\chi\xr\phi\chi$ is `$\phi\chi$' scattering. The configuration $\phi\phi\xr\chi\chi$ is often referred to as the `$t$-channel amplitude' and any amplitude between particles of definite species is often referred to as a `sub-amplitude'.

We work with Mandelstam variables, in the all in-going convention
\begin{eqnarray} \label{Mandelstams}
s = - (k_1 + k_2)^2 \ , \qquad t = - (k_1 + k_3)^2 \qquad \mathrm{and} \qquad u = -  (k_1 + k_4)^2 \ ,
\end{eqnarray}
subject to the usual relation $s+t+u = \sum_i m_i^2$. For the arguments of amplitudes $\mathcal{A}$, we frequently switch between Mandelstam variables $s,t$ (with $u$-dependence suppressed) and the cosine of the scattering angle in the centre-of-mass frame (see Appendix \ref{App:kinematics}) such that
\begin{equation}
\mathcal{A}(s,t) = \mathcal{A}(s; \cos \theta) \ .
\end{equation}
When the second argument of an amplitude appears after a semicolon, it denotes the cosine of the scattering angle, $\cos \theta$. For example, the backwards $\theta = \pi$ limit of $\phi\chi \to \phi\chi$ is written as $\mathcal{A}_{\phi\chi}(s;-1)$.

%%%%%%%%%%%%%%%%%%%%%%%%%%%%%%

\section{Superposition positivity bounds}
\label{sec:setup}

In this section we review the standard derivation of positivity bounds on the forward limit scattering amplitude and how this is used to constrain effective field theories. Superposition scattering states are then introduced as a means of obtaining more restrictive positivity bounds as can be seen in many multiple-field or non-scalar examples of EFTs \cite{Cheung:2014ega,Cheung:2016yqr,deRham:2018qqo,Hamada:2018dde,Alberte:2019xfh,Alberte:2019zhd,deRham:2022sdl,deRham:2022gfe,Alberte:2021dnj}.

We consider throughout this paper a theory of two real scalar fields $\phi$ and $\chi$, which have masses $m_\phi$ and $m_\chi$ respectively. Without loss of generality, we will assume that $m_\phi \leq m_\chi$. As a further simplifying assumption we impose a $\mathbb{Z}_2\times \mathbb{Z}_2$ symmetry which forbids any cubic couplings. 
This symmetry requirement can be relaxed to a $\mathbb{Z}_2$ subgroup which allows for interaction terms with odd numbers of $\phi$ \textbf{or} $\chi$ fields, but not both. This allows pole terms to arise in the amplitude which we may freely subtract and proceed with the derivation of positivity bounds as below. Another potential consequence of such cubic interactions is the appearance of \textit{anomalous thresholds} in the amplitude, however to prove superposition bounds for unequal masses: $m_\chi>m_\phi$, we will require $m_\chi<\sqrt{2}m_\phi$ to ensure an analytic region on the real $s$ axis. Coincidentally, this same upper bound on the heavier mass precisely ensures that anomalous thresholds \textit{do not} arise.

%%%%%%%%%%%%%%%%%%%%%%%%%%%%%
\subsection{Dispersion relations and positivity bounds}
\label{sec:superpos}

The physical principles of locality, causality, unitarity and Lorentz invariance, combined with the framework of local quantum field theory lead to the dispersion relation -- an integral formula for the exact non-perturbative scattering amplitude in terms of its imaginary part (or more generally, its discontinuities). From the point of view of analytic $S$-matrix theory, the physical scattering amplitude for some process is more fundamentally defined as a particular limit of an \textit{analytic amplitude} that is a function of complex Mandelstam variables from Eq.~(\ref{Mandelstams}). In particular, this complex function has specific branch cuts and the limiting values of the function as one approaches these cuts gives the physical scattering amplitude for not only the original scattering process, but other scattering processes that are related to it by crossing symmetry.

To illustrate these points, consider a theory of a single real scalar particle $\phi$ of mass $m$. The fixed-$t$ scattering amplitude for the $\phi\phi\xr\phi\phi$ process -- which we call the $s$-channel\footnote{Not to be confused with the $s$-channel exchange diagram in the Feynman diagram expansion. The word `channel' here specifies what initial and final particles are involved in the scattering process and refers to a non-perturbative scattering amplitude rather than any individual terms in a series expansion.} -- is an analytic function in $s$ away from isolated poles and branch cuts along the real axis \cite{Martin:1965jj} so we can express it via Cauchy's integral formula, $\cA(s,t)=\frac{1}{2\pi\ri}\oint_\mathcal C \rd\mu \,\cA(\mu,t)/(\mu-s)$ where the contour encloses $\mu=s$ in a counter-clockwise sense. For the remainder of this work we shall only consider \textit{pole-subtracted} amplitudes (which we continue to denote with $\cA$) which have had their simple poles explicitly subtracted off\footnote{The derivation of the dispersion relation and subsequent positivity bounds are not affected by this modification provided that the residue of the pole terms 
% themselves 
do not grow as fast or faster than $|s|^2$ 
or in other words, provided that we are dealing with poles with associated spin $<2$, otherwise
% at large $|s|$, as if they did then 
the resulting pole-subtracted function would not itself obey the Froissart/Jin-Martin bound even if the exact amplitude does.
}. By analyticity we can deform this contour to $|s|\xr\infty$, wrapping around the branch cuts to obtain
\begin{equation}\label{eq_ftdr}
    \mathcal{A}_s(s, t)=\int_{4 m^2}^{\infty} \frac{\mathrm{d} \mu}{\pi} \frac{\operatorname{Disc}_\mu \mathcal{A}_s(\mu, t)}{(\mu-s)}+\int_{4 m^2}^{\infty} \frac{\mathrm{d} \mu}{\pi} \frac{\operatorname{Disc}_\mu \mathcal{A}_u(\mu, t)}{(\mu-u)}+\frac{1}{2\pi\ri}\int_{\infty}\rd\mu\,\frac{\cA_{s}(\mu,t)}{(\mu-s)}\,,
\end{equation}
where the discontinuity in $s$ across the branch cut of scattering channel $c$ has been defined as
\begin{equation}
    \operatorname{Disc}_s \mathcal{A}_c(s, t) \equiv \lim _{\epsilon \rightarrow 0^{+}} \frac{1}{2 \mathrm{i}}\left(\mathcal{A}_c(s+\mathrm{i} \epsilon, t)-\mathcal{A}_c(s-\mathrm{i} \epsilon, t)\right)\,.
\end{equation}
We include the subscripts $s,u$ and to highlight the fact that due to crossing symmetry, the `left-hand' branch cut at negative real $s$ of $\cA_{\rm S}(s,t)$ is the physical `right-hand' branch cut of the $u$-channel amplitude $\cA_u(s,t)$. That is, to arrive at the result above one must change variables in the integrals at negative $\mu$ and then use the crossing symmetry, $\cA_{\rm S}(s,t,u)=\cA_u(u,t,s)$. The analytic continuation of the amplitude satisfies the Schwarz reflection principle $\cA_{s}(s,t)=\cA_{s}(s^*,t)^*$ for real $t$ and so the above discontinuity reduces to the imaginary part of the amplitude: $\operatorname{Disc}_s\mathcal{A}_c(s,t)=\operatorname{Im}\mathcal{A}_c(s,t)$. 

The final term in the expression is the closure of the contour integral in the asymptotic high-energy region $|\mu|\xr\infty$, which cannot be explicitly computed without a precise UV theory at hand and may diverge. Nevertheless, the  Froissart/Jin-Martin bound (a consequence of locality through polynomial boundedness and unitarity) implies that the growth of the amplitude at large $|s|$ is sufficiently soft that $\lim_{|s|\xr\infty}\cA(s,t)/s^3\xr0$ \cite{Froissart:1961ux,Jin:1964zza,Azimov:2011nk}, implying that taking two $s$ derivatives of \eqref{eq_ftdr} gives a vanishing integral at infinity. We shall consider the pole subtracted amplitude obtained by literally subtracting the simple pole terms from both sides of the above equation. The resulting expression can then be integrated in $s$ twice, generating two constants of integration and giving
\begin{equation}\label{eq_ftdr2}
    \mathcal{A}_s(s, t)=a_0(t)+a_1(t)s+\int_{4 m^2}^{\infty} \frac{\mathrm{d} \mu}{\pi} \frac{\operatorname{Im}\mathcal{A}_s(\mu, t)}{(\mu-s)}+\int_{4 m^2}^{\infty} \frac{\mathrm{d} \mu}{\pi} \frac{\operatorname{Im} \mathcal{A}_u(\mu, t)}{(\mu-u)}\,.
\end{equation}
In our example the $s$ and $u$-channel processes are the same as all external particles are identical, hence $\cA_{s}(s,t,u)=\cA_u(s,t,u)$, which leads to further simplification of the above formula. In particular, the two imaginary parts are now identical, the subtraction coefficient $a_1(t)$ must vanish.

\paragraph{Positivity bounds:} The simplest positivity bound follows directly from the above by taking two $s$ derivatives and the forward limit (vanishing scattering angle, or $t\xr 0^-$ in this case) to obtain
\begin{equation}\label{eq_pb}
    \del_s^2\mathcal{A}_s(s, t\xr0^-)=2\int_{4 m^2}^{\infty} \frac{\mathrm{d} \mu}{\pi} \frac{\operatorname{Im}\mathcal{A}_s(\mu, 0)}{(\mu-s)^3}+2\int_{4 m^2}^{\infty} \frac{\mathrm{d} \mu}{\pi} \frac{\operatorname{Im} \mathcal{A}_u(\mu, 0)}{(\mu+s-4m^2)^3}\,.
\end{equation}
Unitarity of the $S$-matrix in the form of the optical theorem demands that the imaginary part of an elastic (of the form $\ket{i}\xr\ket{i}$) scattering amplitude appearing in the integrands are positive for physical centre-of-mass energies. Since the denominators are both positive for $s$ in the interval $0<s<4m^2$ we immediately have the positivity bound: $\del_s^2\cA(s,t\xr0^-)>0$ for $0<s<4m^2$. So far, this is a bound on the exact non-perturbative scattering amplitude derived from physical consistency principles, and has no dependence on low energy effective field theory. Since the left hand side of the bound can be evaluated at low energies in an effective field theory, these consistency conditions can be used to constrain the space of effective field theory parameters (a.k.a. Wilson coefficients). 

The above bound is the most basic of its kind that can be derived from the various consistency conditions outlined, and indeed in recent years there has been an explosion in techniques and tools developed 
to leverage the technology of dispersion relations to strongly constrain effective field theory parameter spaces. Making justice to all of them here is beyond the scope of this work, but we refer the reader to bounds beyond the forward limit \cite{Bellazzini:2014waa,deRham:2017avq,deRham:2017zjm}, improved bounds \cite{Bellazzini:2014waa,Bellazzini:2017fep,deRham:2017xox,Alberte:2020bdz}, bounds from positive moments \cite{Bellazzini:2020cot,Arkani-Hamed:2020blm,Chiang:2021ziz,Wan:2024eto}, non-linear bounds that either make use to the null constraints from full crossing symmetry or non-linear relations inferred from the full positive properties of Legendre/Gegenbauer polynomials   \cite{Tolley:2020gtv,Sinha:2020win,Caron-Huot:2020cmc,Arkani-Hamed:2020blm,Chiang:2021ziz,Du:2021byy,Chowdhury:2021ynh,Xu:2023lpq} and to multi-positivity bounds for higher order amplitudes in \cite{Chandrasekaran:2018qmx,Cheung:2025nhw}, without mentioning new numerical and bootstrap techniques.
One such technique, which we shall now explore, is that of using superposition scattering states as opposed to using states with fixed quantum numbers.

\subsection{Superposition positivity bounds}
Unitarity of the $S$-matrix in principle is a far more constraining requirement than simply positivity of the imaginary part of the scattering amplitude. Recall that if we split the $S$-matrix as $S=1+\ri T$, unitarity implies
\begin{eqnarray}\label{eq:unitarity}
S^{\dagger} S = 1 \qquad \qquad \iff \qquad \qquad \frac{1}{2\ri} \big[ T - T^{\dagger} \big] = \frac12 T^{\dagger} T =\frac12\sum_n T^{\dagger}\ket{n}\bra{n}T\,,
\end{eqnarray}
and as this is an operator equation we can sandwich it between \textit{any} state $\ket{\psi}$ to obtain $\operatorname{Im}\braket{\psi|T|\psi}=\frac12|\hspace{0.5mm} T\ket{\psi}|^2=\frac12\sum_n |\bra{n}T\ket{\psi}|^2>0$. The statement of unitarity is therefore evidently a non-linear constraint on the scattering amplitude and so is much stronger than the basic inequality on the right hand side (which is sometimes simply referred to as \textit{positivity}). Inserting a resolution of the identity on the Hilbert space gives the last equality above, which teaches us that the imaginary part of the amplitude is generated by transitions from the initial state to arbitrary virtual (intermediate) states denoted `$\ket n$'. We shall make use of the fact that positivity of the imaginary part holds for arbitrary scattering states and not just eigenstates of quantum numbers like helicity or flavour, and by considering this more general class of states it is possible to obtain stronger bounds on the amplitude. There have been numerous examples of this technique being used to constrain effective field theories in the literature \cite{Sinha:2020win,Cheung:2016yqr,deRham:2018qqo,Hamada:2018dde,Alberte:2019xfh,Alberte:2019zhd,Chowdhury:2021ynh,Du:2021byy,deRham:2022sdl,deRham:2022gfe,Alberte:2021dnj,
Zhang:2018shp,Zhang:2020jyn,Li:2021lpe,Remmen:2019cyz,Remmen:2020vts,
Bonnefoy:2020yee,Yamashita:2020gtt,Remmen:2022orj,Chen:2023bhu,Wan:2024eto}.

It is worth noting that there are other works which consider positivity bounds derived from superpositions of states \cite{Andriolo:2018lvp,Andriolo:2020lul}. See for example \cite{Alberte:2021dnj} which considers photon states which have indefinite helicities, and \cite{Remmen:2019cyz,Yamashita:2020gtt} apply similar techniques in the context of the SMEFT. Also \cite{Bellazzini:2023nqj}\footnote{In reference this work, note that loops break the analytic assumption hence affecting the applicability of these results to theories where no tree-level completion is expected to hold as is in the case of massive gravity \cite{Keltner:2015xda,deRham:2017xox}.} which uses similar technology to show that inelastic amplitudes are always bounded by elastic amplitudes. Finally \cite{Li:2021lpe} showed that the problem of deriving optimal bounds from superpositions can be cast into a geometric one involving the \textit{spectrahedron}.

\subsection{Equal mass scalar superposition bounds}
\label{sec:eqmass_proof}

Consider a theory of two scalars $\phi$ and $\chi$ with equal masses $m_\phi=m_\chi=m$ and take $\ket{\psi}$ to be an initial state consisting of a superposition of the four distinct two-particle configurations
\begin{equation}\label{eq:2pstate}
    \ket{\psi(s)} = \alpha_{\phi\phi}\ket{\phi\phi;s}+\alpha_{\phi\chi}\ket{\phi\chi;s}+\alpha_{\chi\phi}\ket{\chi\phi;s}+\alpha_{\chi\chi}\ket{\chi\chi;s}
\end{equation}
where all states are in the centre-of-mass frame (where the total 3-momentum is zero --- see Appendix \ref{App:kinematics}) and have a centre-of-mass energy squared equal to $s$. Taking the matrix element of $T$ in this state and stripping off the four-momentum conserving delta function leaves a superposition amplitude
\begin{equation}\label{eq:eqmass_superamp}
\begin{aligned}
    \cA_{\rm S}(s) \equiv \frac{\braket{\psi|T|\psi}}{(2\pi)^4\delta^4(0)} &= |\alpha_{\phi\phi}|^2\cA_{\phi\phi}(s,0)+(|\alpha_{\phi\chi}|^2+|\alpha_{\chi\phi}|^2)\cA_{\phi\chi}(s,0)+|\alpha_{\chi\chi}|^2\cA_{\chi\chi}(s,0)\\
    & \quad +2\operatorname{Re}(\alpha_{\phi\chi}\alpha^*_{\chi\phi})\cA_{\phi\chi\xr\chi\phi}(s,0)+2\operatorname{Re}(\alpha_{\phi\phi}\alpha^*_{\chi\chi})\cA_{\phi\phi\xr\chi\chi}(s,0) \ .
\end{aligned}
\end{equation}
Recall, we make a simplifying assumption of $\mathbb{Z}_2\times\mathbb{Z}_2$ symmetry so that transitions such as $\phi\phi\xr\phi\chi$ are forbidden. In addition, we have used crossing symmetry and time reversal invariance of the matrix elements to relate all scattering processes to the five terms above. By unitarity \eqref{eq:unitarity} this amplitude has a positive imaginary part within the region $s>4m^2$ as it is of the form $\sim\braket{\psi|T|\psi}$. The novel constituents of this superposition amplitude are the two terms on the second line, which are not forward limit elastic amplitudes and hence would not typically appear in positivity bounds. It is the appearance of these terms which gives the superposition amplitude technique its advantage. 

By crossing the out-going $\phi$ and $\chi$ we can view the first term on the second line as the backwards scattering angle limit $\theta\xr\pi$ of the $\phi\chi\xr\phi\chi$ amplitude, which means that it is \textit{not} a fixed-$t$ amplitude, since $t=4m^2-s$. Motivated by unitarity, we make the assumption that $\cA_{\rm S}(s)$ only has branch cuts when $s$ is sufficiently large to produce multi-particle intermediate states, i.e. $s>4m^2$. Due to crossing symmetry of the individual terms on the RHS of \eqref{eq:eqmass_superamp} we can infer the existence of a left hand cut along $s<0$. More precisely, if we take the limit of $\cA_{\rm S}(s)$ approaching $s<0$ \textit{from below} in the complex $s$ plane we may use crossing symmetry to write,
\begin{equation}
    \lim_{\epsilon\xr0^+}\cA_{\rm S}(-|s|-\ri\epsilon) \equiv \lim_{\epsilon\xr0^+}\cA^{\times}_{S}(s'+\ri\epsilon)\,,\qquad s'\equiv |s|+4m^2>0,
\end{equation}
where the crossed superposition amplitude is given by,
\begin{equation}\label{eq:eqmass_superamp_crossed}
\begin{aligned}
    \cA_{\rm S}^\times(s) \equiv &|\alpha_{\phi\phi}|^2\cA_{\phi\phi}(s,0)+(|\alpha_{\phi\chi}|^2+|\alpha_{\chi\phi}|^2)\cA_{\phi\chi}(s,0)+|\alpha_{\chi\chi}|^2\cA_{\chi\chi}(s,0)\\&+2\operatorname{Re}(\alpha_{\phi\chi}\alpha^*_{\chi\phi})\cA_{\phi\phi\xr\chi\chi}(s,0)+2\operatorname{Re}(\alpha_{\phi\phi}\alpha^*_{\chi\chi})\cA_{\phi\chi\xr\chi\phi}(s,0)\,.
\end{aligned}
\end{equation}
This crossed amplitude is not in general elastic, i.e. of the form $\braket{\psi|T|\psi}$, and so the discontinuity across its branch cut will not necessarily be positive by unitarity. We must make the additional restriction that the coefficients $\alpha_{ij}$ are factorisable in the sense that $\alpha_{ij}=A_i B_j$, which ensures that there exists a state $\ket{\psi^\times}$ of the form \eqref{eq:2pstate} for which $\cA_{\rm S}^\times(s) = \braket{\psi^\times|T|\psi^\times}/((2\pi)^4\delta^4(0))$. To this end we choose the normalised parametrisation:
\begin{equation}
    A_i = (\cos\theta_A,\sin\theta_A\re^{\ri\varphi_A})\,,\quad B_i = (\cos\theta_B,\sin\theta_B\re^{\ri\varphi_B})\,,
\end{equation}
for arbitrary real angles $\bm\Theta\equiv(\theta_{A,B},\varphi_{A,B})$, in which the superposition state and its crossed partner are given by,
\begin{equation}
\begin{aligned} \label{in_mu_state}
    \ket{\rm \psi } = & \cos\theta_A \cos\theta_B \ket{\phi\phi}+\cos\theta_A \sin\theta_B \re^{\ri\varphi_B}\ket{\phi\chi}\\
    &+\cos\theta_B \sin\theta_A \re^{\ri\varphi_A}\ket{\chi\phi}+\sin\theta_A\sin\theta_B\re^{\ri(\varphi_A+\varphi_B)}\ket{\chi\chi}\\
    \ket{\rm \psi ^\times} = & \cos\theta_A \cos\theta_B \ket{\phi\phi}+\cos\theta_A \sin\theta_B \re^{-\ri\varphi_B}\ket{\phi\chi}\\
    &+\cos\theta_B \sin\theta_A \re^{\ri\varphi_A}\ket{\chi\phi}+\sin\theta_A\sin\theta_B\re^{\ri(\varphi_A-\varphi_B)}\ket{\chi\chi}\,.
\end{aligned}
\end{equation}
With these restrictions we can be sure that the discontinuity across the left hand cut is positive by unitarity and we can immediately write down the superposition positivity bound:
\begin{equation}\label{eq:eqmass_superbound}
    \del_s^2\cA_{\rm S}(s) = \frac{2}{\pi}\int_{4m^2}^{\infty}\rd\mu\,\frac{\operatorname{Im}\cA_{\rm S}(\mu,\bm{\Theta})}{(\mu-s)^3}+\frac{2}{\pi}\int_{4m^2}^{\infty}\rd\mu\,\frac{\operatorname{Im}\cA^\times_{\rm S}(\mu,\bm\Theta)}{(\mu+s-4m^2)^3}>0\,,\quad 0<s<4m^2\,.
\end{equation}
Due to the freedom in the angles $\bm\Theta$ one obtains a continuum of linear positivity bounds which must be considered in unison to obtain the strongest bound on the definite species amplitudes within the superposition. For future convenience we define the following superposition dependent coefficients:
\begin{equation} \label{eq:anglecoeff}
  \begin{split}
\sfc_1 & = \cos^2 \theta_A  \cos^2 \theta_B  \\
\sfc_2 & = \tfrac{1}{2} \big[ 1 - \cos(2\theta_A) \cos(2\theta_B) \big] \\
\sfc_3 & = \sin^2  \theta_A  \sin^2 \theta_B
  \end{split}
\qquad \qquad
  \begin{split}
\sfc_b & = \tfrac{1}{2} \sin 2 \theta_A  \sin 2 \theta_B \cos( \varphi_A - \varphi_B) \\
\sfc_t & = \tfrac{1}{2} \sin 2 \theta_A  \sin 2 \theta_B \cos( \varphi_A + \varphi_B)\,.
  \end{split}
\end{equation}

\subsection{Toy EFT Application}
\label{sec:toyEFT}

Consider a toy EFT of the aforementioned scalars given by \cite{Andriolo:2020lul,Ye:2024rzr}
\begin{eqnarray} 
\mathcal{L}_{\mathrm{EFT}} & = & - \frac{1}{2} (\partial \phi)^2 - \frac{1}{2} m_\phi^2 \phi^2 - \frac{1}{2} (\partial \chi)^2  - \frac{1}{2} m_{\chi} \chi^2  \label{EFTaction} \\
&\ & \qquad \qquad + \lambda_\phi (\partial \phi )^4 + \lambda_1 ( \partial \phi \cdot \partial \chi )^2 + \lambda_2 (\partial \phi)^2 (\partial \chi)^2 + \lambda_\chi (\partial \chi )^4+\ldots  \ \notag
\end{eqnarray}
Computing tree-level scattering amplitudes (see Appendix \ref{App:ToyEFT}) and inserting into the superposition amplitude \eqref{eq:eqmass_superamp}, one uses the positivity bound (\ref{eq:eqmass_superbound}) for equal masses $m_{\phi} = m_\chi$ to find
\begin{equation}
\begin{aligned} \label{eqMassEFT-b}
\partial_s^2 A_{\mathrm{S}}(s) |_{\mathrm{EFT}} = 8 \sfc_1 \lambda_\phi +  (\sfc_b+\sfc_t) \left( \lambda_{1} + 2 \lambda_{2} \right) + 2\sfc_2 \lambda_{1} + 8 \sfc_3  \lambda_\chi >0
\ . \end{aligned}
\end{equation}
Had one considered only the three definite species scattering configurations corresponding to the angle choices $(\theta_A,\theta_B)\in\{(0,0),(\frac{\pi}{2},\frac{\pi}{2}),(0,\frac{\pi}{2})\}$ for elastic $\phi\phi\xr\phi\phi$, $\chi\chi\xr\chi\chi$ and $\phi\chi\xr\phi\chi$ respectively, the bounds read
\begin{equation} \label{eq:stdbound}
\lambda_\phi \geq 0\ , \quad \lambda_\chi \geq 0 \ , \quad \lambda_1 \geq 0 \ .
\end{equation}
One can instead fully vary over all real values of the angles $\theta_{A,B}$ and $\varphi_{A,B}$ in an elementary calculation shown in Appendix \ref{App:EqMass} to find a much stronger constraint on the EFT parameters ({\it e.g.}~see also Refs.~\cite{Andriolo:2018lvp,Andriolo:2020lul,Li:2021lpe}),
\begin{equation} \label{tightestsec3}
- \lambda_1 - 2 \sqrt{\lambda_\phi \lambda_\chi} \leq \lambda_2 \leq 2 \sqrt{\lambda_\phi \lambda_\chi} \ , 
\end{equation}
which not only bounds $\lambda_2$ (otherwise unconstrained from definite species bounds). Interestingly this constraint now involves a \textit{non-linear} combination of EFT coefficients, and leads to a compact bound on $\lambda_2$. This is remarkable because the positivity bound is linear in the amplitude, which itself at tree level is linear in the quartic Wilson coefficients, and so the appearance of the non-linearity is entirely owing to the superposition technique, as of course is well-known and already pointed out in the literature. Our goal now is to explore in detail the same technique in the case where the masses of the two particles are \textit{not} the same. 

It is worth noting that a partial UV completion which reduces to the EFT of Eq.~(\ref{EFTaction}) (upon integrating out the heavy fields) and populates the full region of parameter space satisfying \eqref{eq:stdbound} and \eqref{tightestsec3} is presented in Appendix~\ref{App:partialUV}. 
Interestingly, regardless of whether the particle masses are equal or different, this partial UV completion reproduces exactly the bound in Eq.~(\ref{tightestsec3}). Furthermore, by requiring the absence of superluminal signal propagation (following the approach of~\cite{Adams:2006sv}) one again arrives at the same bound, as demonstrated in Appendix~\ref{App:causality}. 

\vspace{3mm}

% Interestingly one returns exactly the expected bounds from Eq.~(\ref{causality_result}).

%%%%%%%%%%%%%%%%%%%%%%%%%
%%%%%%%%%%%%%%%%%%%%%%%%%

\section{Unequal mass superposition bound}
\label{sec:Unequal_1}

The above derivation relies on the fact that the masses of both low energy particles $\phi$ and $\chi$ are the same and so the normal thresholds (branch cuts from multi-particle production) all start at $4m^2$ and lie within the region of physical kinematics (i.e. real momenta). When the particles do not have the same mass however, the situation is not this simple. Assuming now that $m_\phi\neq m_\chi$ and without loss of generality letting $m_\phi<m_\chi$ one immediate observation is that the state we defined in \eqref{eq:2pstate} becomes un-physical when the centre of mass energy is not large enough for an on-shell $\ket{\chi\chi}$ state to exist. Due to this, within the region $4m_\phi^2<s<4m_\chi^2$ unitarity does not apply automatically and must be proven to hold with some additional assumptions (this is often referred to the `extended unitarity' region) \cite{Mandelstam:1960zz}; we shall assume that the amplitude is unitary for this range of $s$.

A further complication is that the Mandelstam variables for fixed angle scattering have non-trivial relationships to one another when the masses are not equal. For example the momentum transfer of the backwards angle $\phi\chi\xr\phi\chi$ scattering amplitude goes from being linearly related to the Mandelstam variable $s$ to being a hyperbola of $s$. Assuming branch cuts exist in this amplitude at the usual normal (and extended unitarity) thresholds (in the sense of maximal analyticity) therefore leads to a non-trivial analytic structure in the complex $s$ plane which modifies the derivation of the dispersion relation. In this section we shall elaborate on the analytic structure of the superposition amplitude and derive a dispersion relation for it. Following this we shall use the dispersion relation to derive an analogous positivity bound to \eqref{eq:eqmass_superbound}.\\

As we are now working with particles of two distinct non-zero masses, we make repeated use of the following parameters
\begin{eqnarray} \label{DeltaSigmadef}
\Delta := m_\chi^2 - m_\phi^2>0 \qquad \mathrm{and} \qquad \Sigma := 2 m_\phi^2 + 2 m_\chi^2\,.
\end{eqnarray}
For compactness we define new notation for the different scattering amplitudes in the superposition,
\begin{equation} \label{Ashorthands}
(\cA_{1},\cA_{2},\cA_{3}) \equiv (\cA_{\phi\phi},\cA_{\phi\chi},\cA_{\chi\chi}) \qquad \mathrm{and} \qquad \cA_{\phi\phi\xr\chi\chi}\equiv\cA_t
\end{equation}
so that,
\begin{equation}\label{eq:eqenergysuper}
    \cA_\mathrm{S}(s)=\sum_{i=1}^{3}\sfc_i \cA_{i}(s;1)+\sfc_b \cA_{\phi\chi}(s;-1)+\sfc_t \cA_{t}(s;1)
\end{equation}
Again, the above assumes time reversal symmetry (Hermitian analyticity), $\mathbb{Z}_2\times\mathbb{Z}_2$ symmetry (to eliminate matrix elements with odd numbers of $\chi$ and $\phi$ particles), and the second argument of the functions on the RHS is the cosine of the scattering angle. Additionally, we have used the fact that the scattering process $\phi\chi\xr\chi\phi$ in the forward limit (i.e. zero scattering angle between the initial $\phi$ and final $\chi$ momenta) is physically identical to the backwards limit of the process $\phi\chi\xr\phi\chi$. Note, from their definitions we see that $\sfc_i \geq0$ and the sign of $\sfc_b$ and $\sfc_t$ can be chosen freely by choosing the $\varphi_{A,B}$ angles. We will refer to each of the definite species scattering amplitudes on the RHS of \eqref{eq:eqenergysuper} as `sub-amplitudes'.\\

All positivity bounds rest on a dispersion relation for the amplitude. The most straightforward route to obtaining a dispersion relation for the superposition amplitude is to write dispersion relations for each sub-amplitude (i.e.~each term on the right hand side of the sum \eqref{eq:eqenergysuper}) individually and then take the sum. For the three sub-amplitudes that are elastic and in the forward limit we can use the standard result to easily write down their dispersion relations, however the backwards and $t$-channel sub-amplitudes are not as straightforward as they do not correspond to fixed-$t$ amplitudes. In the following section we describe the analytic structure of the $\phi\phi\xr\chi\chi$ and backwards limit $\phi\chi$ amplitudes and provide dispersion relations for each.

\begin{figure}
    \centering
    \begin{tikzpicture}[scale=1.2]
  % Draw horizontal axis
  \draw[->] (-6,0) -- (6,0) node[anchor=west] {$s$};

  % Phiphi cut
  \draw[blue!50, line width = 1.2mm] (-6,0.2) -- (-4,0.2);
  \draw[blue!50, line width = 1.2mm] (2,0.2) -- (6,0.2);

  % Phichi cut
  \draw[purple!50, line width = 1.2mm] (-6,0.4) -- (-2,0.4);
  \draw[purple!50, line width = 1.2mm] (4,0.4) -- (6,0.4);

  % chichi cut
  \draw[orange!50, line width = 1.2mm] (-6,0.6) -- (0,0.6);
  \draw[orange!50, line width = 1.2mm] (2,0.6) -- (6,0.6);

  % Dashed lines
  \draw[dotted, thick, black] (4,0.4) -- (4,-0);
  \draw[dotted, thick, black] (2,0.6) -- (2,-0);
  \draw[dotted, thick, black] (-4,0.2) -- (-4,-0);
  \draw[dotted, thick, black] (-2,0.4) -- (-2,-0);
  \draw[dotted, thick, black] (0,0.6) -- (0,-0);

  % Labels
  \draw[fill=black] (0,0) circle (1pt) node[below] {$ 4\Delta$};
  \draw[fill=black] (-2,0) circle (1pt) node[below] {$(m_\chi-m_\phi)^2$};
  \draw[fill=black] (-4,0) circle (1pt) node[below] {$0$};
  \draw[fill=black] (2,0) circle (1pt) node[below] {$4m_\phi^2$};
  \draw[fill=black] (4,0) circle (1pt) node[below] {$(m_\chi+m_\phi)^2$};
  
\end{tikzpicture}
    \caption{Real $s$ branch cut structure of the three elastic forward limit scattering amplitudes, assuming maximal analyticity. All cuts lie on the real $s$ axis and the vertical separation is purely for visual aid. The orange line denotes branch cuts of forward limit $\chi\chi\xr\chi\chi$, the purple of $\phi\chi\xr\phi\chi$ and the blue of $\phi\phi\xr\phi\phi$ scattering respectively. We see that a sum of these three functions is analytic in the gap between $4\Delta$ and $4m_\phi^2$.}
    \label{fig:elastic_cuts}
\end{figure}
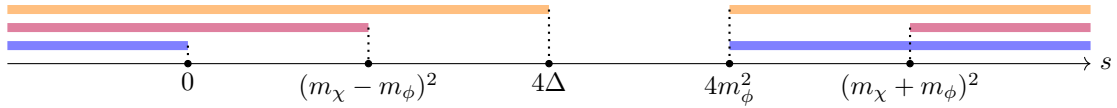

\paragraph{Elastic forward limit amplitudes:}For the elastic configurations in the forward limit we have branch cuts along the left and right hand side of the real $s$ axis which start at multi-particle thresholds in the $s$ and $u$ channels. This structure is summarised in figure \ref{fig:elastic_cuts}. As the centre-of-mass energy of each definite species amplitude is the same, the position of the $u$-channel branch points are shifted due to the different definitions of the variable `$u$' depending on the particle content of each amplitude. In particular, the start of the left hand cut for $\cA_{\phi\phi}(s,0)$ is $s=0$, whereas for $\cA_{\chi\chi}(s,0)$ it starts at the point $s=4\Delta$ and for $\cA_{\phi\chi}(s,0)$ it starts at $s=(m_\phi-m_\chi)^2$. We assume that $\cA_{\chi\chi}(s,0)$ has a cut from $s>4\mph^2$ due to $\phi$ loops/two-particle states. 

If we are strict about taking into account these branch cuts, the superposition amplitude will only have a cut-free region on the real axis if $4m_\phi^2 > 4\Delta$, or phrased differently,
\begin{equation}
m_\phi\leq m_\chi<\sqrt{2}m_\phi \ .
\end{equation}
In practice, it is often possible to make simplifying assumptions regarding the low-energy loop corrections that allow one to ignore these branch cuts and so have a larger region of analyticity for the EFT amplitude. We shall proceed assuming this upper bound on the heavier mass to ensure that we have a region on the real $s$ axis where $\cA_{\rm S}$ and its $s$-derivatives can safely be evaluated without making any further assumptions about whether or not the cuts can be ignored.

%%%%%%%%%%%%%%%%%%%%%%%%%%%%%%%

\subsection{Dispersion relation for \texorpdfstring{$\cA_t$}{}}
\label{sec:tchan}
We begin by defining a hyperbolic function $h(s)$ that appears as the momentum transfer of the $\phi\chi\xr\phi\chi$ amplitude at fixed backwards scattering angle $\theta=\pi$:
\begin{equation} \label{hs_def}
h(s) \equiv \Sigma - s - \frac{\Delta^2}{s}\,.
\end{equation}
The forward limit $\phi\phi\xr\chi\chi$ amplitude, which recall we refer to as the ``$t$-channel amplitude'' $\cA_t$, has a momentum transfer given by the function (see Eq.~(\ref{t_tcurve})), 
\begin{equation}
    t_{\text{forward limit }\phi\phi\xr\chi\chi} =\frac12\left(2m_\phi^2+2m_\chi^2-s+\sqrt{s-4m_\phi^2}\sqrt{s-4m_\chi^2}\right)
\end{equation}
which can be related to the inverse of the hyperbola $h(s)$ defined above in Eq.~(\ref{hs_def}):
\begin{equation}\label{eq:h_inverse}
    h_{\pm}^{-1}(s)=\frac12\left(\Sigma-s\pm\sqrt{(s-\Sigma)^2-4\Delta^2}\right)\equiv \frac{1}{2}\left(\tau(s)\pm\sigma(s)\right)\,,
\end{equation}
with the definitions
\begin{equation} \label{sigmatau_def}
\tau \equiv \Sigma-s \qquad \mathrm{and} \qquad \sigma\equiv+\sqrt{(s-\Sigma)^2-4\Delta^2}
\end{equation}
where the latter is defined to be the \textit{positive} root. Note that due to the square roots, $h_\pm^{-1}(s)$ is complex for $4m_\phi^2<s<4m_\chi^2$. Additionally, as we have merged the arguments of the two square roots in equation \eqref{eq:h_inverse} we emphasise that the relationship between $t_{\text{forward limit }\phi\phi\xr\chi\chi}$ and $h_\pm^{-1}$ is for real $s$:
\begin{equation}
    t_{\text{forward limit }\phi\phi\xr\chi\chi}(s) =\begin{cases}
        h^{-1}_{+}(s)\,,\quad\text{if}\quad s>4m_\phi^2 \\
        h^{-1}_{-}(s)\,,\quad\text{if}\quad s<4m_\phi^2\,.
    \end{cases}
\end{equation}
This is important to bear in mind in general, however since the amplitude $\cA_t$ has $t\leftrightarrow u$ crossing symmetry it is invariant under $\hip\leftrightarrow\him$ and so we may always take $\hip(s)$ as the momentum transfer ($t$ variable) of the forward limit $\phi\phi\xr\chi\chi$ amplitude and the differences between the merged and un-merged expressions do not make any difference to the final results.

\begin{figure}[htbp]\label{fig:hyperbola}
  \centering
  \begin{tikzpicture}
  
    \begin{axis}[
      axis lines = center,
      width=15cm,
      height=9cm,
      xlabel = $s$,
      ylabel = $h(s)$,
      xmin = -2, xmax = 2,
      ymin = -2, ymax = 4,
      samples=100,
      domain=-4:4,
      xtick=\empty,
      ytick=\empty
    ]
      % Left of zero
      \addplot[blue, thick, domain=-2:-0.03] {1-x-0.1/x};

      % Asymptote
      \addplot[orange!40, thick, domain=-2:2] {1-x};

      % Right of zero
      \addplot[blue, thick, domain=0.03:2] {1-x-0.1/x};

      % Add Delta label
      \node at (axis cs:0.316,0) [anchor=north] {\tiny $\Delta$};
      \node at (axis cs:-0.316,0) [anchor=north] {\tiny $-\Delta$};

      %Guidelines
      \draw [dotted, thick, gray] (axis cs:0.316,0) -- (axis cs:0.316,0.367);     \draw [dotted, thick, gray] (axis cs:0,0.38) -- (axis cs:0.316,0.38);

      %Labels of mass intercepts
      \node at (axis cs:0,0.38) [anchor=east] {\tiny $4m_\phi^2$};
      \node at (axis cs:0,1.6) [anchor= west] {\tiny $4m_\chi^2$};

      %Guidelines
      \draw [dotted, thick, gray] (axis cs:-0.316,0) -- (axis cs:-0.316,1.6);     \draw [dotted, thick, gray] (axis cs:0,1.6) -- (axis cs:-0.316,1.6);

      %Sigma label
      \draw [dotted, thick, gray] (axis cs:0,1) -- (axis cs:1,1);
      \node at (axis cs:1,1) [anchor=west] {\tiny $\Sigma=2m_\phi^2+2m_\chi^2$};

      %Guidelines for x intercepts
      \draw [dotted, thick, gray] (axis cs:0.113,0) -- (axis cs:0.2,-1);  
      \node at (axis cs:0.18,-0.85) [anchor=north west] {\tiny $(m_\phi-m_\chi)^2$};
      
      \draw [dotted, thick, gray] (axis cs:0.887,0) -- (axis cs:0.887,-0.5);
      \node at (axis cs:0.887,-0.4) [anchor=north ] {\tiny $(m_\phi+m_\chi)^2$};

    \end{axis}
  \end{tikzpicture}
  \caption{Plot of $h(s)=\Sigma-s-\frac{\Delta^2}{s}$. The asymptote shown in orange is given by $\Sigma-s$.}
  \label{fig:plot_h}
\end{figure}
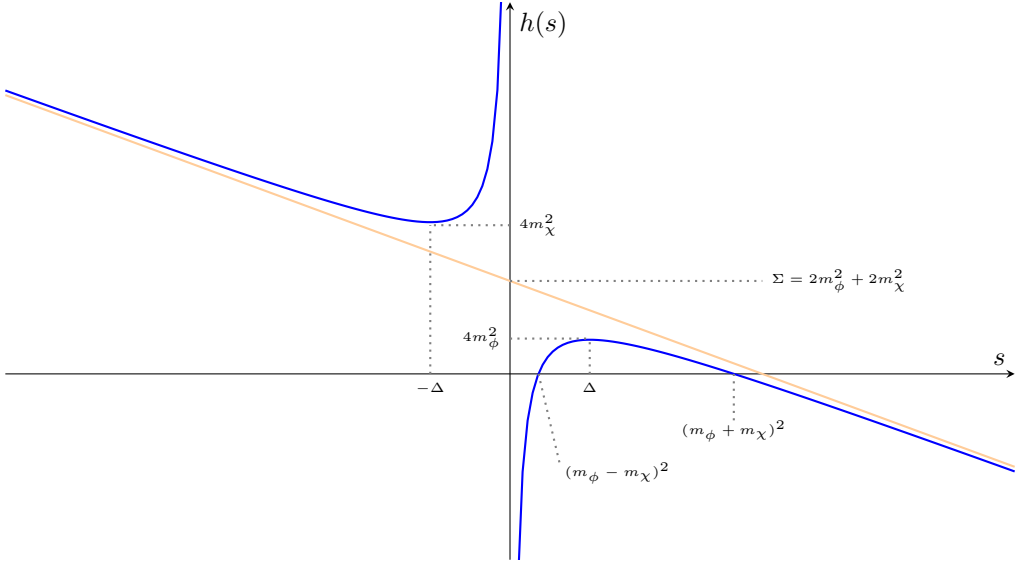

A plot of $h(s)$ is shown in figure \ref{fig:hyperbola}. Once more we assume maximal analyticity so that the branch cut structure of $\cA_t(s;1)=\cA_{t}(s,h^{-1}_+(s),h^{-1}_{-}(s))$ is given by values of $s$ in the complex plane for which any of the three Mandelstam variables exceed threshold values. For example, we expect a branch cut for $s>4m_\phi^2$ due to intermediate states with two $\phi$ particles, and a cut at $h^{-1}_+(s)>(m_\phi+m_\chi)^2$ due to intermediate states with one $\phi$ and one $\chi$ particle in the $s-t$ crossed channel. The latter inequality is satisfied for real $s$ in the region $s<0$. The threshold $h^{-1}_{-}(s)>(m_\phi+m_\chi)^2$ is not satisfied for any value of $s$ and so does not produce a branch cut. The result is that the $t$-channel amplitude has cuts along the real $s$ axis for $s<0$ and $s>4m_\phi^2$.

We can express the amplitude via a dispersion relation as usual (below we understand $\cA_t(s)$ to be the forward limit amplitude and omit its second argument) up to an integral along the portion of the contour at $|\mu|\xr\infty$:
\begin{eqnarray}
\cA_t(s) & = & \frac{1}{\pi} \int_{4 m_\phi^2}^{\infty}\rd\mu \; \frac{\operatorname{Im} \cA_t(\mu;1)}{(\mu-s)}+\frac{1}{2 \pi i} \int_{-\infty}^0\rd\mu \; \frac{\cA_t(\mu+\ri\epsilon)-\cA_t(\mu-\ri\epsilon)}{(\mu-s)} \label{tchan_Adisp} \\
& = & \frac{1}{\pi} \int_{4 m_\phi^2}^{\infty}\rd\mu \; \frac{\operatorname{Im} \cA_t(\mu;1)}{(\mu-s)} +\frac{1}{2 \pi i} \int_{\infty}^{\left(m_\phi+m_\chi\right)^2} \rd z\; \left(-1+\frac{\Delta^2}{z^2}\right) \frac{\cA_t(h(z)+\ri\epsilon)-\cA_t(h(z)-\ri\epsilon)}{(h(z)-s)} \notag
\end{eqnarray}
We have performed the variable change $\mu=h(z)$. Then using crossing we get (now including the integral at infinity explicitly),
\begin{equation}\label{eq:t_channel_disp}
\begin{aligned}
    \cA_t(s)=&\frac{1}{\pi} \int_{4 m_\phi^2}^{\infty}\rd\mu \; \frac{\operatorname{Im} \cA_t(\mu;1)}{(\mu-s)}+\frac{1}{\pi} \int_{\left(m_\phi+m_\chi\right)^2}^{\infty}\rd\mu\left(-1+\frac{\Delta^2}{\mu^2}\right) \frac{\operatorname{Im} \cA_{\phi \chi}(\mu;-1)}{(h(\mu)-s)}\\&+\frac{1}{2 \pi i} \int_{|\mu|\xr\infty}\rd\mu\;  \frac{\cA_t  (\mu)}{\mu-s}\,.
\end{aligned}
\end{equation}
%

%%%%%%%%%%%%%%%%%%%%%%%%%%%%%%%

\subsection{Backwards limit dispersion relation for \texorpdfstring{$\cA_{\phi\chi}$}{}}
\label{sec:backwards}

Dispersion relations for fixed angle elastic scattering were first derived by Hearn and Leader in \cite{Hearn:1962zz} (see also \cite{Atkinson:1962zz}), however we closely follow the presentation of Goldberg \cite{Goldberg:1968zza} with a variation that gives a new final form to the integrals.

\begin{figure}[h!]    
    \centering
    \begin{tikzpicture}[>=stealth,scale=1.7]
        % Draw x-axis
        \draw[->] (-4,0) -- (4,0) node[right] {$\operatorname{Re}\mu$};
        
        % Draw y-axis
        \draw[->] (0,-2.5) -- (0,2.5) node[above] {$\operatorname{Im}\mu$};

        % Draw branch cuts
        \draw[blue!50, line width = 1.2mm] (0,0) circle (1.9cm);
        \draw[blue!50, line width = 1.2mm] (-4,0) -- (0.9,0);
        \draw[blue!50, line width = 1.2mm] (2.7,0) -- (3.9,0);

        % Draw external s point
        \filldraw[black] (3,2) circle (1pt) node[below left, xshift=9pt] {$s$};

        % Draw contour around it
        \draw[thick,black!40, ->] (3.25,2) arc[start angle=0, end angle=359, radius=0.25];

        % Draw internal s point
        \filldraw[black] (1,-0.66) circle (1pt) node[below left, xshift=9pt,yshift=-12pt] {$\Delta^2/s$};

        % Draw contour around it
        \draw[thick,black!40, ->] (1.25,-0.66) arc[start angle=0, end angle=359, radius=0.25];

        % Draw label for delta
%        \node at (2.1,-0.1) {$\Delta$};
%       \draw[thin] (1.9,0) -- (2.03,-0.1);

        % Label C segment
        \node at (0.2,2.2) {$\mathtt{C}$};

        % Label F segment
        \node at (-0.2,-2.2) {$\mathtt{F}$};

        % Label D segment
        \node at (-3,0.3) {$\mathtt{D}$};

        % Label E segment
        \node at (-3,-0.3) {$\mathtt{E}$};

        % Label A segment
        \node at (3.2,0.3) {$\mathtt{A}$};

        % Label B segment
        \node at (3.2,-0.3) {$\mathtt{B}$};

        % Label K segment
        \node at (-1,-0.3) {$\mathtt{K}$};

        % Label H segment
        \node at (-1,0.3) {$\mathtt{H}$};

        % Label I segment
        \node at (0.5,0.3) {$\mathtt{I}$};

        % Label J segment
        \node at (0.5,-0.3) {$\mathtt{J}$};

        % Label G segment
        \node at (-0.2,1.57) {$\mathtt{G}$};

        % Label L segment
        \node at (0.2,-1.57) {$\mathtt{L}$};

        % Draw upper semicircular contour
        \draw[black, thick, postaction={decorate}, decoration={
            markings, mark=at position 0.5 with {\arrow{stealth}}}] 
            (-2.01,0.1) arc[start angle=177.152, end angle=2.8482, radius=2.0125];

        % Draw upper semicircular contour
        \draw[black, thick, postaction={decorate}, decoration={
            markings, mark=at position 0.5 with {\arrow{stealth}}}] 
            (1.77,0.1) arc[end angle=177.152, start angle=2.8482, radius=1.77282];

        \draw[thick] (1.77,-0.1) -- (1.77,0.1);

        % Draw lower semicircular contour
        \draw[black, thick, postaction={decorate}, decoration={
            markings, mark=at position 0.5 with {\arrow{stealth}}}] 
            (-1.77,-0.1) arc[start angle=-177.152, end angle=-2.8482, radius=1.77282];

        % Draw lower semicircular contour
        \draw[black, thick, postaction={decorate}, decoration={
            markings, mark=at position 0.5 with {\arrow{stealth}}}] 
            (2.01,-0.1) arc[end angle=-177.152, start angle=-2.8482, radius=2.0125];

        \draw[thick] (2.01,-0.1) -- (2.01,0.1);

        % Inner left contour

        % Left contour 
        \draw[thick, postaction={decorate},
        decoration={markings, mark=at position 0.5 with {\arrow{stealth}}}] 
        (-4,0.1) -- (-2,0.1);
        \draw[thick, postaction={decorate},
        decoration={markings, mark=at position 0.55 with {\arrow{stealth}}}] 
         (-2,-0.1) -- (-4,-0.1) ;

        % Right contour 
        \draw[thick, postaction={decorate},
        decoration={markings, mark=at position 0.5 with {\arrow{stealth}}}] 
        (2.7,0.1) -- (3.9,0.1);
        \draw[thick, postaction={decorate},
        decoration={markings, mark=at position 0.55 with {\arrow{stealth}}}] 
          (3.9,-0.1)-- (2.7,-0.1) ;

        % Internal straight contour 
        \draw[thick, postaction={decorate},
        decoration={markings, mark=at position 0.55 with {\arrow{stealth}}}] 
        (-1.77,0.1) --(0,0.1) ;
        \draw[thick, postaction={decorate},
        decoration={markings, mark=at position 0.5 with {\arrow{stealth}}}] 
         (0,-0.1) -- (-1.77,-0.1) ;

        % Internal straight contour 
        \draw[thick, postaction={decorate},
        decoration={markings, mark=at position 0.55 with {\arrow{stealth}}}] 
        (0,0.1) --(0.9,0.1) ;
        \draw[thick, postaction={decorate},
        decoration={markings, mark=at position 0.5 with {\arrow{stealth}}}] 
         (0.9,-0.1) -- (0,-0.1) ;

        % Little semi-circles
        \draw[thick] (0.9,0.1) arc[start angle=90, end angle=-90, radius=0.1];
        \draw[thick] (2.7,0.1) arc[start angle=90, end angle=270, radius=0.1];

        % Dashed lines
        \draw[dotted, thick, black!50] (2.7,0) -- (2.7,-0.5);
        \draw[dotted, thick, black!50] (0.9,0) -- (0.9,0.5);
        \draw[dotted, thick, black!50] (1.9,0) -- (1.9,-1.3);

        % Label thresholds
        \node at (0.9,0.65) {\small $ (m_{\phi}-m_\chi)^2$};
        \node at (2.7,-0.65) {\small $ (m_{\phi}+m_\chi)^2$};
        \node at (1.9,-1.45) {\small $ \Delta$};

        % Label small contours
        \node[black!70] at (2.55,2) {\small $\mathcal C_0$};
        \node[black!70] at (1.2,-0.3) {\small $\mathcal C_1$};
        
    \end{tikzpicture}
    \caption{Analytic structure of $B(\mu)\equiv \cA_{\phi\chi}(\mu;-1)$. Thick blue lines are branch cuts and each lettered segment of the contour is identified by an arrow at its mid-point. The radius of the circular contour is $\Delta \equiv m_\chi^2-m_\phi^2$.}
    \label{fig:contour-anatomy}
\end{figure}
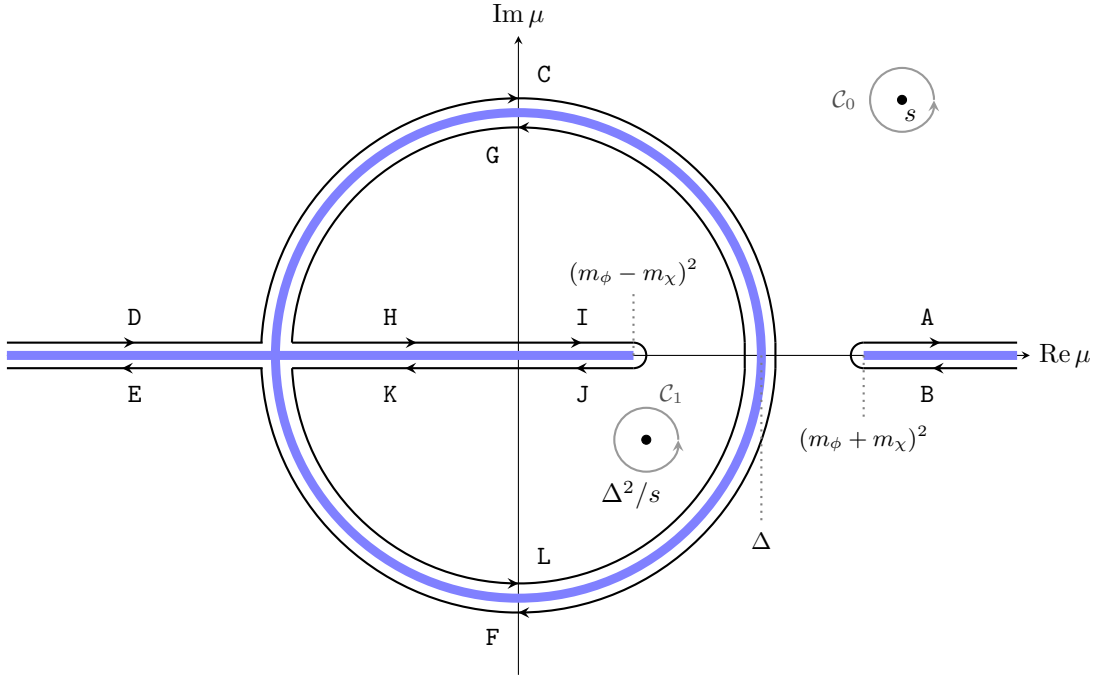

Scattering amplitudes at fixed scattering angle generally do not correspond to fixed-$t$ amplitudes but rather to amplitudes with $s$ dependent momentum transfer variables: $t=t(s)$. For example, we consider $\theta=\pi$ fixed angle elastic $\phi\chi\xr\phi\chi$ scattering, which has a momentum transfer variable given by the hyperbola $h(s)$ defined in Eq.~(\ref{hs_def}), 
\begin{equation}
    \cA_{\phi\chi}(s;-1)=\cA_{\phi\chi}\left(s,h(s)=\Sigma-s-\frac{\Delta^2}{s}\right)\,.
\end{equation}
For non-zero $\Delta>0$ this hyperbola diverges as $s\xr0$ indicating that the interpretation of the $s\xr0$ regime as `low-energy' is no longer accurate, as the momentum transfer is unbounded. From $h(s)$ we may derive the complex $s$-plane branch cut structure of $\cA_{\phi\chi}(s;-1)$ by determining the values of $s$ for which the three Mandelstam variables are real and exceed threshold values in each scattering channel. The result is summarised in figure \ref{fig:contour-anatomy}.\\

To declutter the presentation we shall define $B(s)$ to be the backwards limit of the $\apc$ scattering amplitude, i.e. $B(s)\equiv \apc(s;-1)$. Owing to the unusual branch cut structure of $B(s)$ which exhibits a circular branch cut that separates the complex plane into $|s|<\Delta$ and $|s|>\Delta$, a dispersion relation derived with $s$ taken in the former region may not be valid in the latter. However, the crossing symmetry of the amplitude implies the property,
\begin{equation}\label{eq:b_crossing}
    B(s)=B\left(\frac{\Delta^2}{s}\right)\qquad\text{(crossing symmetry)\,,}
\end{equation}
which implies that the amplitude in the region inside the circular contour is fully determined by the amplitude outside the contour. To take advantage of this crossing symmetry, we begin by re-writing the amplitude as
\begin{equation}
    \begin{aligned}
        B(s)&=\frac12 B(s)+\frac12 B(\Delta^2/s)\\
        &=\frac{1}{4\pi\ri}\oint_{\mathcal{C}_0}\rd\mu\,\frac{B(\mu)}{(\mu-s)}+\frac{1}{4\pi\ri}\oint_{\mathcal{C}_1}\rd\mu\,\frac{B(\mu)}{(\mu-\Delta^2/s)}\,,\quad\textrm{for}\quad|s|>\Delta\,,
    \end{aligned}
\end{equation}
where in the first line we have used crossing and in the second, Cauchy's integral formula and the contours of integration (which enclose the explicit poles of each integrand in an anti-clockwise sense) $\mathcal{C}_{0,1}$ are shown in figure \ref{fig:contour-anatomy}. This expression assumes that the variable $s$ takes values outside the region enclosed by the circular contour, otherwise it evaluates to zero as $\mathcal{C}_{0,1}$ will not enclose the poles of their respective integrands.

We could immediately proceed from this expression to re-write the contour integrals as integrals over physical centre-of-mass energy (squared) to obtain a dispersion relation, however we shall choose to make an additional step that will lead to a simpler final result with manifest crossing symmetry c.f. equation \eqref{eq:b_crossing}. We can `add zero' to both sides as
\begin{equation}
\begin{aligned} \label{Bdisp_C0C1}
    B(s)&=\frac{1}{4\pi\ri}\left[\oint_{\mathcal{C}_0}\rd\mu\,\left(\frac{1}{\mu-s}+\frac{1}{\mu-\frac{\Delta^2}{s}}\right)B(\mu)+\oint_{\mathcal{C}_1}\rd\mu\,\left(\frac{1}{\mu-s}+\frac{1}{\mu-\frac{\Delta^2}{s}}\right)B(\mu)\right]\\
    &=\frac{1}{4\pi\ri}\left[\oint_{\mathcal{C}_0}+\oint_{\mathcal{C}_1}\right]\rd\mu\,\left(\frac{1}{\mu-s}+\frac{1}{\mu-\frac{\Delta^2}{s}}\right)B(\mu)\,.
\end{aligned}
\end{equation}
Since the added poles are not inside the respective contours of integration, they do not affect the result and only serve to make the integrands crossing symmetric in the same fashion as the full amplitude.
As mentioned in the discussion on the assumed analytic structure of the sub-amplitudes, although the branch cut structure seems unusual, it is nevertheless reflecting the various physical multi-particle thresholds (which are assumed via maximal analyticity) that exist in amplitudes for different scattering channels between two $\phi$ and two $\chi$ particles. Hence, each of these integrals should be amenable to a change of variable and crossing operation that results in an integral over a physical region discontinuity of one of these definite species amplitudes\footnote{For example, in the standard forward limit $s-u$ symmetric dispersion relation for elastic scattering the left hand branch cut that runs over negative values of $s$ is transformed using a change of variable and $s-u$ crossing symmetry into an integral over positive $s$ but for the associated $u$-channel process.}.

To organise the derivation we split both contours $\mathcal C_{0,1}$ into several segments and label each with a letter $\mathtt{A,B,\ldots, L}$ as shown in figure \ref{fig:contour-anatomy}. Upon each lettered segment we perform a change of variable as given by the central column of Table \ref{tab:variable_change}, followed by a crossing operation to obtain the boundary value of a particular definite species amplitude, given in the final column of the table.
\begin{table}[htbp]
\centering
\begin{tabular}{c|c|l}
\hline
Segments & Change of variable & Crosses to\\
\hline
$\mathtt{A,B}$ & $\mu$ & $\cA_{\phi\chi}(\mu;-1)$\\
$\mathtt{C,G,H,K}$ & $h^{-1}_{+}(\mu)$&$\cA_{t}(\mu;\pm1)$\\
$\mathtt{D,E,F,L}$ & $h^{-1}_{-}(\mu)$&$\cA_{t}(\mu;\pm1)$\\
$\mathtt{I,J}$ & $\Delta^2/\mu$&$\cA_{\phi\chi}(\mu;-1)$\\
\hline
\end{tabular}
\caption{Change of variable required for each segment of the contour integral. For example, in segment $\mathtt{C}$ a new integration variable $\mu_{\text{new}}$ should be related to the original $\mu_{\text{old}}$ as $\mu_{\text{old}}=h^{-1}_{+}(\mu_{\text{new}})$ to obtain the desired result.\label{tab:variable_change}}
\end{table}

For example, analysis via the change of variables in the contour integrals above shows that the combination of integrals that enclose the circular branch cut (i.e. segments $\mathtt{C,G,L,F}$) give a very simple total:
\begin{equation}
    I_{\text{circular}} = c + \frac{1}{\pi}\int_{4 m_\phi^2}^{4 m_\chi^2}\rd \mu\,\frac{1}{\mu-h(s)}\operatorname{Im}\cA_{t}(\mu+\ri\epsilon;1)\,,
\end{equation}
where $c$ is a constant. This makes it clear that the circular branch cut appearing in the backwards $\phi\chi$ scattering amplitude is a manifestation of unitarity in the $t$-channel ($\phi\phi\xr\chi\chi$).

The remaining segments of the integral organise themselves in a similar manner to produce integrals over discontinuities in different scattering channels. We leave the details of these steps to Appendix \ref{app:backwards_dr} and state the final result of the contour integration:
\begin{equation}
\begin{aligned}\label{eq:backwards_dispersion}
    B(s) &=\int_{(m_\phi+m_\chi)^2}^{\infty}\frac{\rd\mu}{\pi}\, \operatorname{Im} B(\mu)\left(\frac{1}{\mu-s}+\frac{1}{\mu-\frac{\Delta^2}{s}}\right)+\int_{4m_\phi^2}^{\infty}\frac{\rd \mu}{\pi}\,\operatorname{Im} \cA_{t}(\mu;1)\left(\frac{1}{\mu-h(s)}\right)\\&+c+\frac{1}{2\pi\ri}\int_{|\mu|\xr\infty}\rd \mu\,\left(\frac{1}{\mu-s}+\frac{1}{\mu-\frac{\Delta^2}{s}}\right)B\left(\mu\right)\,,
\end{aligned}
\end{equation}
where we recall the function $h(s)\equiv \Sigma-s-\Delta^2/s$. The two terms on the second line are a generic constant $c$ (i.e. any constant terms arising from other segments are lumped into $c$) and an integral over arcs lying at $|\mu|\xr\infty$, both of which will not contribute to the final positivity bounds as they disappear upon taking a sufficient number of $s$ derivatives. Finally, note that while we initially assumed that $|s|>\Delta$ in deriving the dispersion relation, we could have equally started with $|s|<\Delta$ then defined $\tilde s=\Delta^2/s$ so that $|\tilde s|>\Delta$ and followed the same steps leading to  \eqref{eq:backwards_dispersion} but with $s\xr\tilde s$. However since the final result is manifestly symmetric under $s\leftrightarrow\Delta^2/s$, we end up with exactly the same dispersion relation as we'd obtained for $|s|>\Delta$, therefore the dispersion relation is valid for all $s$ in the complex plane away from non-analyticities.

%%%%%%%%%%%%%%%%%%%%%%%%%%%%%%%
\subsection{Superposition positivity bounds for \texorpdfstring{$\Delta>0$}{}}
\label{sec:superproof_1}

As we have obtained dispersion relations for each sub-amplitude on the RHS of \eqref{eq:eqenergysuper}, we may attempt to derive positivity bounds on $s$ derivatives of $\cA_{\rm S}$. Recall that derivatives are required to eliminate integrals along contours at $|\mu|\xr\infty$. Additionally, for $s-u$ symmetric amplitudes we require two $s$ derivatives to obtain a positive integrand, and so as a consistency check we begin by considering the second $s$ derivative of the superposition amplitude in the $\Delta\xr0$ limit:
\begin{equation}
\begin{aligned}
    \lim_{\Delta\xr0^+}\frac{\pi}{2}\del_s^2 \cA_{\rm S}(s) &=\int_{4m^2}^{\infty}\,\frac{\rd\mu}{(\mu-s)^3}\left(\sum_{i}\sfc_i\operatorname{Im}\cA_{i}(\mu;1) +\sfc_b\operatorname{Im}\cA_{\phi\chi}(\mu;-1)+\sfc_t \operatorname{Im} \cA_t(\mu;1)\right)\\
    &+\int_{4m^2}^{\infty}\,\frac{\rd\mu}{(\mu-u)^3}\left(\sum_{i}\sfc_i\operatorname{Im}\cA_{i}(\mu;1) +\sfc_t\operatorname{Im}\cA_{\phi\chi}(\mu;-1)+\sfc_b \operatorname{Im} \cA_t(\mu;1)\right)\,.
\end{aligned}
\end{equation}
Constant terms in the dispersion relation obviously disappear when derivatives are taken, however the integrals at infinity can only be assumed to vanish if the Froissart bound holds for $\cA_{\phi\chi}(s;-1)$ and $\cA_t(s;1)$. Unitarity applied to the superposition $S$-matrix element directly implies that the term in curved brackets above is positive, and hence the quantity on the left is positive for $s$ in the interval $0<s<4m^2$, leading to the positivity bound,
\begin{equation}\label{eq:equalmassbound}
    \lim_{\Delta\xr0}\del_s^2\cA_{\rm S}(s)>0\,,\quad 0<s<4m^2\,,
\end{equation}
which is in agreement with the analysis of $\cA_{\rm S}(s)$ presented in section \ref{sec:eqmass_proof}.\\

We now attempt to derive positivity bounds on $\del_s^2\cA_{\rm S}$ when $\Delta>0$. An immediate obstruction is the asymptotic behaviour of the integrand in the backwards dispersion relation \eqref{eq:backwards_dispersion}. At large $\mu$, all integrands in this dispersion relation have a weaker explicit $\mu$ suppression than all integrands in the forward limit elastic amplitudes $\del_s^2\cA_i(s;1)$ as well as the forward limit $\phi\phi\xr\chi\chi$ amplitude $\del_s^2\cA_t(s;1)$, {\it i.e.},
\begin{equation}
\hspace{-0.5cm}
\begin{aligned}\label{eq:derivatives_B}
    \del_s^2 \cA_{\phi\chi}(s;-1) &=\int_{(m_\phi+m_\chi)^2}^{\infty}\frac{\rd\mu}{\pi}\, \operatorname{Im} \cA_{\phi\chi}(\mu;-1)\left(\frac{2 \Delta ^2}{\mu ^2 s^3}+\ldots\right)+\int_{4m_\phi^2}^{\infty}\frac{\rd \mu}{\pi}\,\operatorname{Im} \cA_{t}(\mu;1)\left(-\frac{2 \Delta ^2}{\mu ^2 s^3}+\ldots\right) \hspace{-1cm} \\
    &+c+\frac{1}{2\pi\ri}\int_{|\mu|\xr\infty}\rd \mu\,\left(\frac{2 \Delta ^2}{\mu ^2 s^3}+\ldots\right)\cA_{\phi\chi}(\mu;-1)\,,
\end{aligned}
\end{equation}
where ellipses denote higher inverse powers of $\mu$. The elastic forward limit amplitudes on the other hand have dispersion relations with integrands that have explicit $\sim 1/\mu^3$ behaviour at large $\mu$. As a result, any attempt to prove a positivity bound will fail because at sufficiently large $\mu$ the sign indefinite imaginary part of $\cA_{\phi\chi}(s;-1)$ will begin dominating the integrals in the dispersion relation for $\del_s^2\cA_{\rm S}(s)$. Additionally, the assumption of Froissart boundedness is no longer sufficient to guarantee that the integral at $\infty$ evaluates to zero. It can clearly be seen that these terms vanish when $\Delta\xr0$ as expected, indicating that this is a novel issue encountered only when the difference between the two masses is taken into account. 

The expectation of $\del_s^2\cA_{\rm S}$ to be a positive quantity is motivated from the example of forward limit definite species elastic scattering and for the case where $\Delta=0$ as demonstrated above, however there is no inherent reason it must be true when $\Delta>0$. Given the clear obstruction explained above it is natural to propose that an alternative EFT observable should be considered. If we are motivated by constructing an EFT quantity that has a dispersion relation with all integrands behaving as $\operatorname{Im}\cA\times\mu^{-3}$ (for some definite species amplitude $\cA$) at large $\mu$ then we need to add a term to $\del_s^2\cA_{\rm S}$
 that would cancel the $\operatorname{Im}\cA\times\mu^{-2}$ terms on the RHS of \eqref{eq:derivatives_B} without introducing any new such terms. Consider the following ansatz for such a quantity and its dispersion relation:
\begin{equation}
\hspace{-0.5cm}
\begin{aligned}\label{eq:ansatz}
    \left[\del_s^2 +f(s)\del_s\right] \cA_{\phi\chi}(s;-1)&=\int_{(m_\phi+m_\chi)^2}^{\infty}\frac{\rd\mu}{\pi}\, \operatorname{Im} \cA_{\phi\chi}(\mu;-1)\left(\frac{\frac{2 \Delta ^2}{s^3}+f(s)\left(1-\frac{\Delta ^2}{s^2}\right)}{\mu ^2}+\cO(\mu^{-3})\right)\hspace{-1cm}\\&+\int_{4m_\phi^2}^{\infty}\frac{\rd \mu}{\pi}\,\operatorname{Im} \cA_{t}(\mu;1)\left(\frac{-\frac{2 \Delta ^2}{s^3}-f(s) \left(1-\frac{\Delta ^2}{s^2}\right)}{\mu ^2}+\cO(\mu^{-3})\right)\\&+c+\frac{1}{2\pi\ri}\int_{|\mu|\xr\infty}\rd \mu\,\left(\frac{\frac{2 \Delta ^2}{s^3}+f(s)\left(1-\frac{\Delta ^2}{s^2}\right)}{\mu ^2}+\cO(\mu^{-3})\right)\cA_{\phi\chi}(\mu;-1)\,,\hspace{-1cm}
\end{aligned}
\end{equation}
from which we can deduce that if $f(s)$ is chosen so that $\frac{2 \Delta ^2}{s^3}+f(s)\left(1-\frac{\Delta ^2}{s^2}\right)=0$, all terms of the form $\operatorname{Im}\cA\times\mu^{-2}$ at large $\mu$ in the dispersion relation are cancelled. Hence we continue by attempting to derive positivity bounds on the modified EFT quantity,
\begin{equation}
    \cO_1(s)\equiv\del_s^2\cA_{\rm S}(s)-\frac{2 \Delta ^2\sfc_b}{s \left(s^2-\Delta ^2\right)}\del_s \cA_{\phi\chi}(s;-1)=\del_s^2\cA_{\rm S}(s)-\sfc_b\frac{h''}{h'}\del_s \cA_{\phi\chi}(s;-1)\,. \label{A2_subtract1d}
\end{equation}
Note that this new quantity reduces to $\del_s^2\cA_{\rm S}$ in the limit $\Delta\xr0$, indicating that any bounds we derive on this quantity should smoothly reduce to those derived above for the $\Delta=0$ case.

The integral expression for $\at$ is cumbersome but can be organised into two terms $\at=I_{\rm RH}+I_{\rm LH}$, where $I_{\rm RH}$ contains integrands that have a denominator $(\mu-s)^3$ in the limit $\Delta\xr0$, and $I_{\rm LH}$ contains integrands that have denominator $(\mu-u)^3$ in the same limit:
\begin{equation}
    \begin{aligned}
        \frac{\pi}{2} I_{\rm RH} = \sum_{i=1}^{3}\,&\sfc_i \int_{\Lambda_i}^{\infty}\rd\mu\,\operatorname{Im} \cA_{i}(\mu;1)\,\left[\frac{1}{(\mu-s)^3}\right]\\
        +&\sfc_b\int_{\Lambda_2}^{\infty}\rd\mu\,\operatorname{Im} \cA_{\phi\chi}(\mu;-1)\, \left[\left(1-\frac{\Delta^2}{s^2}\right)^2\frac{ \mu^2  \left(\mu-\frac{\Delta ^2}{\mu}\right) }{  \left(\mu  -\frac{\Delta ^2}{s}\right)^3}\right]\frac{1}{(\mu-s)^3}\\
    +&\sfc_t\int_{\Lambda_1}^{\infty}\rd\mu\,\operatorname{Im}\cA_t(\mu;1)\left[\frac{1}{(\mu-s)^3}\right]\\
    \frac{\pi}{2} I_{\rm LH} = \sum_{i=1}^{3}\,&\sfc_i \int_{\Lambda_i}^{\infty}\rd\mu\,\operatorname{Im} \cA_{i}(\mu;1)\,\left[\frac{1}{(\mu-u_i)^3}\right]\\
        +&\sfc_b\int_{\Lambda_1}^{\infty}\rd\mu\,\operatorname{Im} \cA_{t}(\mu;1)\,\left[\left(1-\frac{\Delta^2}{s^2}\right)^2\frac{1}{(\mu -h(s))^3}\right]\\ +&\sfc_t\int_{\Lambda_2}^{\infty}\rd\mu\,\operatorname{Im}\cA_{\phi\chi}(\mu;-1)\left[\left(1-\frac{\Delta^2}{\mu^2}\right)\frac{ 1}{(s-h(\mu ))^3}\right]\,,
    \end{aligned}\label{eq:LHRHdr}
\end{equation}
where $\Lambda_{i}=(4m_\phi^2,(m_\phi+m_\chi)^2,4m_\phi^2)_i$ and $u_i=(4m_\phi^2-s,\Sigma-s,4m_\chi^2-s)_i$. Our strategy will be to prove that the above integrands have a positive sum over the range of integration and thereby conclude that the integral itself is positive. We shall find that $I_{\rm RH}$ is positive, however the sign of $I_{\rm LH}$ is unclear. Although there are differing lower limits for the integrals, we can safely assume that the imaginary part of $\cA_{\phi\chi}(\mu;\pm1)$ vanishes\footnote{The forward limit is expected to vanish due to lack of on-shell intermediate states due to the assumed $\mathbb{Z}_2\times \mathbb{Z}_2$ symmetry, and the backwards limit will hence vanish as a result of the partial wave expansion of the imaginary part.} between the thresholds $(m_\phi+m_\chi)^2$ and $4m_\phi^2$ and so we can lower all limits to $4m_\phi^2$ without affecting the answer. \\

\paragraph{Positivity of $I_{\mathrm{RH}}:$}To prove that $I_{\rm RH}$ is positive it is sufficient to show that the function in square brackets multiplying $\operatorname{Im}\cA_{\phi\chi}(\mu;-1)$ has absolute value less than unity over the range of integration and for real values of $s$ where the superposition amplitude is analytic: $4\Delta<s<4m_\phi^2$. This is sufficient because we know from unitarity that
\begin{equation}\label{eq:im_super}
    \operatorname{Im}\cA_{\rm S}(s)=\sum_{i=1}^{3}\sfc_i \operatorname{Im}\cA_{i}(s;1)+\sfc_b\operatorname{Im} \cA_{\phi\chi}(s;-1)+\sfc_t \operatorname{Im}\cA_{\phi\phi\xr\chi\chi}(s;1)>0\,,
\end{equation}
and $\sfc_b$ can freely be taken to zero from above or below by a choice of mixing angle without affecting $\sfc_i$ and $\sfc_t$, therefore
\begin{equation}
    \sum_{i=1}^{3}\sfc_i \operatorname{Im}\cA_{i}(s;1)+p(s)\sfc_b\operatorname{Im} \cA_{\phi\chi}(s;-1)+\sfc_t \operatorname{Im}\cA_{\phi\phi\xr\chi\chi}(s;1)>0\,,\quad \forall\, |p(s)|\leq1\,.
\end{equation}
In our case, for large positive or negative values of $\mu$ the function $p(\mu)$ approaches a constant value which for $\Delta<s$ is positive and less than unity:
\begin{equation}
    p(\mu)\equiv\left(1-\frac{\Delta^2}{s^2}\right)^2\frac{ \mu^2  \left(\mu-\frac{\Delta ^2}{\mu}\right) }{  \left(\mu  -\frac{\Delta ^2}{s}\right)^3} = \left(1-\frac{\Delta ^2}{s^2}\right)^2+\cO\left(\frac{1}{\mu }\right)\,.
\end{equation}
We can solve the equation $p(\mu_*)=1$ to find three solutions,
\begin{equation}
    \mu_* = \begin{cases}
        s\\[5pt]
        \frac{s^3-\left(s^2-\Delta ^2\right) \sqrt{4 \Delta ^2+s^2}+\Delta ^2 s}{2 \left(2 s^2-\Delta ^2\right)}\equiv r_1
        \\[5pt]
        \frac{s^3+(s^2-\Delta^2 ) \sqrt{4 \Delta ^2+s^2}+\Delta ^2 s}{2 \left(2 s^2-\Delta ^2\right)}\equiv r_2\,.
    \end{cases}
\end{equation}
Given the constraints, $0<\Delta<m_\phi^2$ (which follows from our assumption $m_\chi<\sqrt{2}m_\phi$) and $4\Delta<s<4m_\phi^2$ (the analytic region of $\cA_{\rm S}$), one finds that $r_1<r_2<s$. Hence the largest value of $\mu$ for which $p(\mu) = 1$ is $\mu=s$, which is not within the range of integration of the dispersion relation. Additionally, the largest value of $\mu$ for which $p(\mu)=0$ is $\mu=\Delta$, and hence $p(\mu)$ is also positive throughout the region of integration. Therefore,
\begin{equation}
    0 < p(\mu) < 1 \quad \text{for}\quad \mu>s\quad\text{and}\quad 4\Delta<s<4m_\phi^2\,,
\end{equation}
and so the sum of integrands in $I_{\rm RH}$ is positive over the entire region of integration for $4\Delta<s<4m_\phi^2$ and hence,
\begin{equation}\label{eq:s_positivity}
    I_{\rm RH}>0 \quad \text{for}\quad 4\Delta<s<4m_\phi^2\,.
\end{equation}
\paragraph{Non-positivity of $I_{\text{LH}}$:}The difficulty with determining the sign of $I_{\text{LH}}$ is that the functions multiplying the amplitudes' imaginary parts are now \textit{all} different and so using \eqref{eq:im_super} to prove positivity is not immediately possible. The sign indefinite imaginary parts: $\operatorname{Im}\apc(\mu;-1)$ and $\operatorname{Im}\cA_t(\mu;1)$ are multiplied against the two functions  $p_b$ and $p_t$ respectively, defined:
\begin{equation}
\begin{aligned}
    p_b(\mu,s) &\equiv\left(1-\frac{\Delta^2}{\mu^2}\right)\frac{ 1}{(s-h(\mu ))^3}\,,\\  
    p_t(\mu,s) &\equiv \left(1-\frac{\Delta^2}{s^2}\right)^2\frac{1}{(\mu -h(s))^3}\,.
\end{aligned}
\end{equation}
If the absolute values of these two functions are bounded by $(\mu-u_i)^{-3}$ for all three $u_i$ then we may conclude that $I_{\rm LH}$ is positive, since positivity of the three $\operatorname{Im}\cA_i$ terms will compensate for the sign indefinite terms according to \eqref{eq:im_super}. Noting that for values of $\mu$ and $s$ in the analytic region we have,
\begin{equation}
    \frac{1}{(\mu-4m_\phi^2+s)^3}<\frac{1}{(\mu-\Sigma+s)^3}<\frac{1}{(\mu-4m_\chi^2+s)^3}\quad\text{for}\quad 4\Delta<s<4m_\phi^2\,,
\end{equation}
so if we can prove that $|p_{b,t}|<(\mu-4m_\phi^2+s)^{-3}$ within the integrals, we can conclude that the total integral is positive. Both $p_b$ and $p_t$ are positive within the limits of integration and for $4\Delta<s<\mu$. It is straightforward to see by inspection\footnote{For example taking the expression for $p_b$, $$p_b=\left(1-\frac{\Delta^2}{\mu^2}\right)\frac{ 1}{(s-h(\mu ))^3}$$ we know that the first factor in brackets is valued between $0$ and $1$ since $\mu>\Delta>0$ and the denominator $s-h(\mu)=\mu+s-\Sigma+\Delta^2/\mu$ is positive and greater than $\mu+s-\Sigma$ since $\mu>0$. Therefore,
$$p_b<\frac{1}{(\mu+s-\Sigma)^3}\,.$$ Similar reasoning holds for $p_t$.} that $p_t$ and $p_b$ are less than $(\mu-\Sigma+s)^{-3}$, and therefore also less than $(\mu-4m_\chi^2+s)^{-3}$, however they are \textit{not} less than $(\mu-4m_\phi^2+s)^{-3}$ between the limits of integration. One way to see this is to expand at large $\mu$:
$$\frac{1}{(\mu-4m_\phi^2+s)^{3}}-p_b(\mu,s)=-\frac{6\Delta}{\mu^4}+\cO\left(\frac{1}{\mu^5}\right),$$
revealing that the pre-factor of $\operatorname{Im}\cA_{\phi\chi}(\mu;-1)$ within the integrals will always be larger than that of $\operatorname{Im}\cA_{\phi\phi}(\mu;1)$ at sufficiently large $\mu$. Due to this, we \textbf{cannot} conclude that $I_{\rm LH}$ is positive as in the case of $I_{\rm RH}$.

\paragraph{Constructing a positivity bound:}It is clear however that the obstruction to proving positivity is the `smallness' of the function $(\mu+s-4m_\phi^2)^{-3}$ multiplying $\operatorname{Im}\cA_{\phi\phi}(\mu;1)$, as compared to $p_b$ and $p_t$ which are multiplying the sign indefinite imaginary parts.  Therefore we can straightforwardly construct a positive quantity from $I_{\rm LH}$ by adding a large enough positive multiple, which we denote $x(s)$, of $\del_s^2\cA_{\phi\phi}(s;1)$ such that,
\begin{equation}\label{eq:modification_1}
    I_{\rm LH} +x(s)\del_s^2\cA_{\phi\phi}(s;1)>0\,.
\end{equation}
Since we are adding $\del_s^2\cA_{\phi\phi}(s;1)$ which is a positive quantity the strongest bound will correspond to the minimum possible value of $x(s)$. In other words, we need to find the smallest value of $x(s)$ such that:
\begin{equation} \label{firstxs}
\begin{aligned}
    \frac{1}{(\mu-4m_\phi^2+s)^3}+x(s)\left(\frac{1}{(\mu-s)^3}+\frac{1}{(\mu-4m_\phi^2+s)^3}\right)&\geq p_b(\mu,s)\,,\quad\forall\mu>(m_\phi+m_\chi)^2\\
    \textbf{and}\quad\frac{1}{(\mu-4m_\phi^2+s)^3}+x(s)\left(\frac{1}{(\mu-s)^3}+\frac{1}{(\mu-4m_\phi^2+s)^3}\right)&\geq p_t(\mu,s)\,,\quad\forall\mu>4m_\phi^2\,.
\end{aligned}
\end{equation}
Note the different ranges of $\mu$ arising due to the different lower limits of the integrals in which $p_b$ and $p_t$ appear.
% As $p_b$ and $p_t$ are bounded in magnitude by $(\mu-\Sigma+s)^{-3}$, one way to obtain the multiplier is to find $x(s)$ such that 
% %
% \begin{equation} \label{firstxs}
%     \frac{1}{(\mu-4m_\phi^2+s)^3}+x(s)\left(\frac{1}{(\mu-s)^3}+\frac{1}{(\mu-4m_\phi^2+s)^3}\right)\geq\frac{1}{(\mu-\Sigma+s)^3}\,.
% \end{equation}
% %
% for all $\mu$ within the limits of integration of the .

These inequalities can be re-arranged\footnote{When rearranging one must use the fact that $s$ takes values in the range $4\Delta<s<4m_\phi^2$ so that the term in curved brackets multiplying $x(s)$ is positive and so the direction of the inequality does not flip when dividing by it.} to the form $x(s)\geq r_b(\mu;s,\Delta,\Sigma)$ and $x(s)\geq r_t(\mu;s,\Delta,\Sigma)$ respectively where $r_{b,t}$ are  rational functions in $\mu,s,\Delta$ and $\Sigma$. Finding a closed form for the maximum value of these rational functions over $\mu$ for generic values of $s,\Delta$ and $\Sigma$ is straightforward in principle but difficult in practice as the polynomials involved are quartic order and higher leading to extremely complicated expressions. However practically speaking the masses of the low energy particles will be known and hence $\Sigma$ and $\Delta$ can be fixed to their respective values. As we want $x(s)$ to be as small as possible whilst still satisfying both inequalities, we can write it as
\begin{equation}\label{eq:x_maxx}
    x(s) = \operatorname{max}(x_b(s),x_t(s))\,,\qquad\text{where}\qquad \begin{array}{ll}
        x_b(s)\equiv &\max_{\mu>(m_\phi+m_\chi)^2}r_b\\
        x_t(s)\equiv &\max_{\mu>4m_\phi^2}r_t
    \end{array}
\end{equation}
Combining equations \eqref{eq:modification_1}, \eqref{eq:s_positivity} and \eqref{eq:x_maxx} leads immediately to the non-perturbative positivity bound:
\begin{equation}\label{eq:Unequal_bound1}
\boxed{
     \del_s^2\cA_{\rm S}(s)-\frac{2 \Delta ^2\sfc_b\del_s \cA_{\phi\chi}(s;-1)}{s \left(s^2-\Delta ^2\right)}+x(s)\,\sfc_1\,\del_s^2\cA_{\phi\phi}(s;1) > 0 \,,\quad\text{for}\quad 4\Delta<s<4m_\phi^2}
\end{equation}
Recall that we have made the assumption that $\Delta>0$ and $m_\chi<\sqrt{2}m_\phi$. The quantity on the left hand side can be evaluated in an EFT, leading to a restriction on the space of Wilson coefficients. As the bound holds for all values of the mixing angles $\theta_{A,B},\varphi_{A,B}$, one must vary over these to obtain the strongest restriction on the parameter space of the EFT. As $\Delta\xr0$ the functions $r_{b,t}\xr0$ and the superposition amplitude $\at$ equals the superposition amplitude derived for the equal mass case given in \eqref{eq:eqenergysuper}, hence the bound \eqref{eq:Unequal_bound1} becomes \eqref{eq:eqmass_superbound} in the equal mass limit as expected.
%%%%%%%%%%%%%%%%%%%%%%%%%%%%%%%%%%%%%%%%%%%%%%%%%%%%%%%%
\subsection{Probing the size of corrections to the bound}
\label{sec:EFTprobe1}

In arriving at the bound above we have made two ‘corrections' to the EFT quantity being bounded: adding a multiple of $\del_s^2\cA_{\phi\phi}(s;1)$, and subtracting $\frac{2 \Delta ^2\sfc_b}{s \left(s^2-\Delta ^2\right)}\del_s B(s)$ as in \eqref{A2_subtract1d}; both of these vanish as $\Delta\xr0$. The former correction weakens the bound the larger the multiple is (we are adding a quantity that is known to be positive to the left hand side of \eqref{eq:Unequal_bound1} without changing the right-hand side) and so should be minimised over $s$ when searching for the strongest constraints. The latter correction involves the addition of a sign indefinite quantity and so it is not immediately clear if it strengthens or weakens the bound. We shall examine the size of the former correction in a theory independent way and for the latter we shall use our toy model EFT in \eqref{EFTaction}.

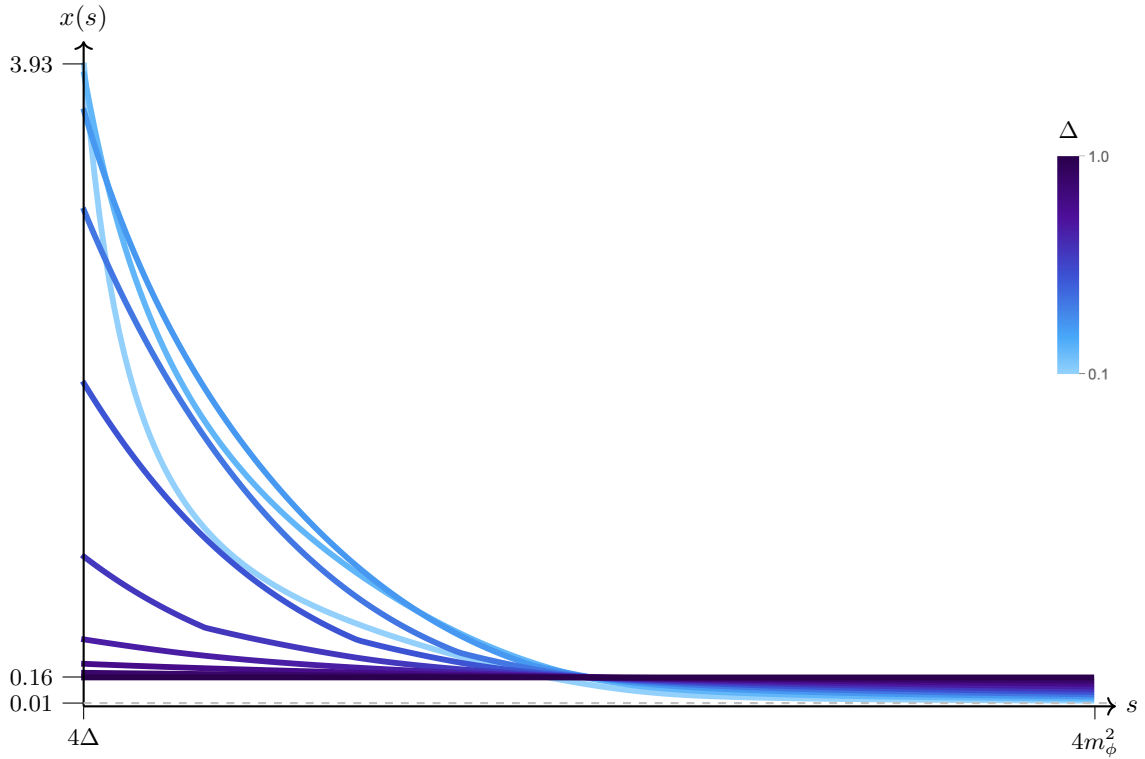
\begin{figure}[h]
    \centering
    \begin{tikzpicture}[]
        \node[anchor=south west,inner sep=0] (image) at (0,0) {\includegraphics[width=0.9\textwidth]{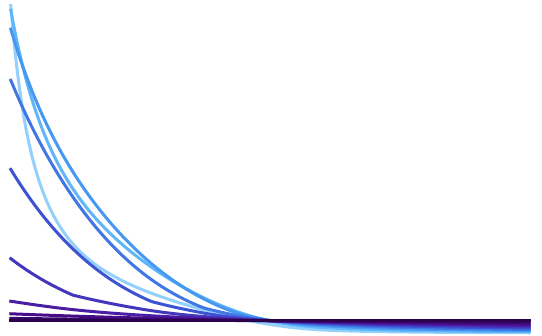}};
        \begin{scope}[x={(image.south east)},y={(image.north west)}]
            % Draw axes starting at bottom-left (0,0)
            \draw[->,thick] (0.02,0) -- (1,0) node[right] {$s$};
            \draw[->,thick] (0.02,0) -- (0.02,1.02) node[above] {$x(s)$};

            \draw (0.02,0) -- (0.02,-0.02) node[below] {\small $4\Delta$};

            \draw (0.955,0.91) node[below] {\small $\Delta$};

            \draw (0.98,0) -- (0.98,-0.02) node[below] {\small $4m_\phi^2$};

            \draw (0.02,0.045) -- (0.0,0.045) node[left] {\small $0.16$};
            \draw (0.02,0.005) -- (0.0,0.005) node[left] {\small $0.01$};
            \draw (0.02,0.985) -- (0.0,0.985) node[left] {\small $3.93$};

            %draw horizontal dashed line
            \draw[gray!50, thick, dashed] (0.02,0.005) -- (1,0.005);
        \end{scope}
        \node[anchor=south west,inner sep=0] (image) at (13,4) {\includegraphics[width=0.065\textwidth]{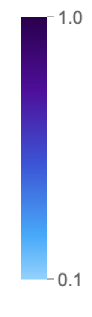}};
    \end{tikzpicture}
    \caption{As $s$ (horizontal axis) increases between $4\Delta$ and $4$ in units of $m_\phi^2$, the value of $x(s)$ (vertical axis) decreases monotonically. When $\Delta=0$ the curve lies flat along the $s$-axis. These plots are in fact discrete and are generated by fixing evenly spaced values of $s$ between $4\Delta$ and $4m_\phi^2$ and numerically maximising the functions $r_{b,t}$ over the range of the integral. The plot points are joined by a curve that we expect to be piecewise smooth since the underlying functions $r_{b,t}$ are smooth in $\mu$ -- however due to the discrete maximisation in Equation \eqref{eq:x_maxx} there is a visible kink in the curves where $x_t$ grows larger than $x_b$.}
    \label{fig:maxrcurves}
\end{figure}

\paragraph{First correction:}This correction depends on the value of $s$ as well as the values of the parameters $\Delta$ and $\Sigma$. To investigate this we first work in units where $m_\phi=1$ which fixes $\Sigma=4+2\Delta$, and then plot $x(s)$ as a function of $s$ for various values of $\Delta$ in the allowed range $0<\Delta<1$, as shown in figure \ref{fig:maxrcurves}.
We observe that $x(s)$ decreases monotonically as $s$ increases, and so to minimise it we should take the largest value of $s$ in the allowed range.

\paragraph{Second correction:}As mentioned above, it is not immediately clear if the addition of the term proportional to $\del_s B(s)$ improves or weakens the bound. To probe this question we turn back to the toy model EFT \eqref{EFTaction} in which the superposition bound reads\footnote{The absolute value of the terms on the RHS can be taken since there is complete freedom of choosing the signs of $\sfc_{b,t}$ by varying the mixing angles $\varphi_{A,B}$.},
\begin{equation}\label{eq:eft_superbound}
    \begin{aligned}
        8\left(1+x(s)\right)&\sfc_1 \lambda_\phi+8 \sfc_3 \lambda_\chi+2 \lambda_1 \sfc_2>\left|\sfc_t\left(\lambda_1+2 \lambda_2\right)\right|+\left|\sfc_b\left(1-\frac{\Delta^2}{s^2}\right)^2\left(\lambda_1+2 \lambda_2\right)\right| \,.
    \end{aligned}
\end{equation}
The second correction can be identified as the $\Delta$ dependent term on the RHS and so the bound will be strongest if $(1-\Delta^2/s^2)^2$ is maximised. For a given $0<\Delta<1$ and $s$ in the range $4\Delta<s<4$, this quantity is maximised at the largest possible value of $s$.\\

\begin{figure}[h]
\begin{center}
\includegraphics[width=0.9\linewidth]{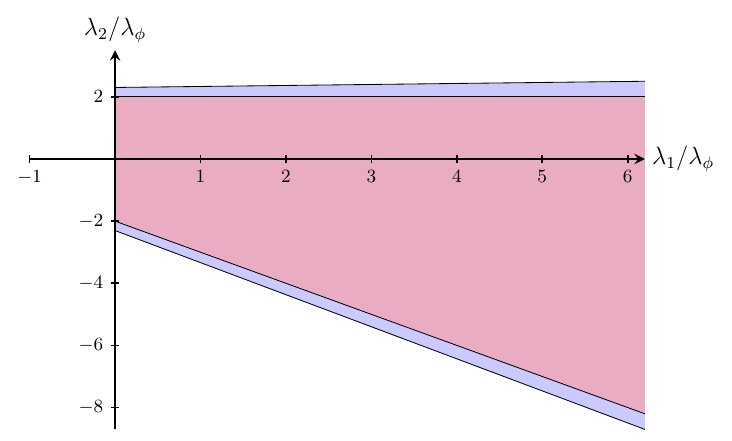}
\caption{\small The allowed parameter spaces according to the rigorous unequal mass bound (\ref{boundEFTresult_1}) (in purple) compared to that predicted by the equal mass bound (\ref{tightestsec3}) (in red), for the choice $\lambda_{\chi} = \lambda_\phi$ (also using $s = 4m_\phi^2$ and $\Delta = m_\phi^2$ and $x(s) \simeq 0.16$) applied to the tree-level EFT amplitude. The unequal mass bound is slightly weaker than the equal mass bound however requires no weak-coupling assumptions.} \label{Fig:Compare1}
\end{center}
\end{figure}

Extremizing over all the angles in the $\mathtt{c}$ variables (see Appendix \ref{App:EqMass}) one finds that the tightest bound is $\lambda_{\phi}, \lambda_1, \lambda_\chi> 0$, along with 
\begin{small}
\begin{equation} \label{boundEFTresult_1}
-\lambda _1 \left( \frac{1}{1+(1 - \frac{\Delta^2}{s^2})^2}+\frac{1}{2}\right)-\frac{4 \sqrt{(1+x(s)) \lambda _{\chi } \lambda _{\phi }}}{1+(1 - \frac{\Delta^2}{s^2})^2} < \lambda_2 <  \left(\frac{1}{1+(1 - \frac{\Delta^2}{s^2})^2}-\frac{1}{2}\right) \lambda_1+\frac{4 \sqrt{(1+x(s)) \lambda _{\chi } \lambda _{\phi }}}{1+(1 - \frac{\Delta^2}{s^2})^2} \ .
\end{equation}
\end{small}\ignorespaces
{\it c.f.} Eq.~(\ref{tightestsec3}) in the equal mass limit. See Figure \ref{Fig:Compare1}.

Hence in this example we see that the largest allowed value of $s$ leads to the strongest positivity bound with respect to both corrections simultaneously. Inserting this optimal value of $s=4$ into the bound, we can evaluate the size of the corrections for different values of $\Delta$. The corrections both vanish at $\Delta=0$ and grow larger as $\Delta$ increases, so we take the largest allowed value of $\Delta=1$ so as to probe the worst case scenario, and find the size of the two corrections to be
\begin{equation}
    1-\left(1-\frac{1}{4^2}\right)^2=\frac{31}{256}\approx0.12\,,\quad\&\quad x(4)|_{\Delta=1}\approx 0.16\,.
\end{equation}
Strictly speaking this is at the boundary of the region of validity of our dispersion relations as $s=4$ is a branch point. The above analysis of $x(s)$ applies independently of the EFT under consideration as standard forward limit positivity bounds demand that  $\del_s^2\cA_{\phi\phi}(s;1)$ be positive, hence the smallest value of $x(s)$ will always give the best bound. On the other hand, the evaluation of the size of the second correction has been done in an EFT example. We have found that the differences in the bound between the equal mass case \eqref{eq:equalmassbound} are $\cO(10\%)$ in this case. 

%%%%%%%%%%%%%%%%%%%%%%%%%%%%%%%%%%%%%%%%%%%%%%%%%%%%%%%%

\section{Bounds from generalized superposition amplitudes}
\label{sec:gen}

The choice of superposition state leading to the superposition amplitude $\cA_{\mathrm{S}}$ defined in Eq.~(\ref{eq:eqenergysuper}) was chosen for simplicity in order to give all definite species amplitudes the same Mandelstam $s$. We can consider a more general construction with the state ({\it c.f.} Eq.~(\ref{eq:2pstate}))
\begin{eqnarray} \label{genstate}
| \psi \rangle & = & \cos \theta_{A} \cos \theta_{B}\; \big| \big[ \mathbf{k} \big]^{\phi} \big[ - \mathbf{k} \big]^{\phi} \rangle  + \cos \theta_A \sin \theta_B e^{i \varphi_{B}} \; \big| \big[ \mathbf{p} \big]^{\phi} \big[ - \mathbf{p} \big]^{\chi} \rangle  \\
&& + \cos \theta_{B} \sin \theta_{A} e^{i \varphi_{A}} \; \big| \big[ \boldsymbol{\ell} \big]^{\chi} \big[ -  \boldsymbol{\ell} \big]^{\phi} \rangle+  \sin \theta_{A} \sin \theta_{B} e^{i (\varphi_{A} + \varphi_{B})} \;  \big| \big[ \mathbf{q} \big]^{\chi} \big[ - \mathbf{q} \big]^{\chi} \rangle \notag 
\end{eqnarray}
where $[\mathbf{k}]^i$ denotes a particle of species $i$ and 3-momentum $\mathbf{k}$. Each individual state in the sum has zero total 3-momentum but now have independent centre-of-mass energies.

One can naturally build a superposition amplitude analogous to Eq.~(\ref{eq:eqenergysuper}), although for all five of the sub-amplitudes to appear they all must satisfy momentum conservation simultaneously which imposes the constraints: 
\begin{equation} \label{distinctmom}
\mathbf{q}^2 = \mathbf{k}^2 - \Delta \qquad \mathrm{and} \qquad \boldsymbol{\ell}^2 = \mathbf{p}^2  \ .
\end{equation}
This leaves a generalized superposition amplitude which depends on two {\it independent} centre-of-mass energies which for physical kinematics are given by $s=4 \mathbf{k}^2 + 4m_\phi^2$ and $s'=h_{+}^{-1}(-4\mathbf{p}^2)$, given by $\langle \psi | T | \psi \rangle = (2\pi)^4 \delta^{(4)}( 0 )  \; \mathscr{G}(s,s')$ where ({\it c.f.}~Eq.~(\ref{eq:eqenergysuper}))
\begin{eqnarray} \label{Gdef}
\mathscr{G}(s,s') & \equiv &  \sfc_1 \; \mathcal{A}_{\phi\phi}( s ; 1  ) + \sfc_3  \; \mathcal{A}_{\chi\chi}\big( s;1 ) + \sfc_t \;  \mathcal{A}_{t}( s;1) + \sfc_2 \; \mathcal{A}_{\phi\chi}( s' ; 1) + \sfc_b \; \mathcal{A}_{\phi\chi}( s'; -1 )  \ .
\end{eqnarray}
Notice that the original amplitude $\cA_{\mathrm{S}}(s)$ from Eq.~(\ref{eq:eqenergysuper}) is a special case of the generalized superposition amplitude with $\cA_{\mathrm{S}}(s) = \mathscr{G}( s, s)$.
% %
% \begin{equation} \label{AS_fromG}
% \cA_{\mathrm{S}}(s) = \mathscr{G}( s, s) \ .
% \end{equation}
% %
The fact that $\mathscr{G}$ arises from a matrix element of the form $\langle \psi | T | \psi \rangle$ implies that it has a positive imaginary part for physical values $s$ and $s'$:
\begin{equation} \label{general_Imsuperpos}
\mathrm{Im}[ \mathscr{G}(s,s') ] > 0\quad\text{for}\quad s>4m_\phi^2\text{ and }s'>(m_\phi+m_\chi)^2\,.
\end{equation}
As before the above assumes that unitarity holds for the $\cA_{\chi\chi}(s;1)$ amplitude in the extended region $4m_\phi^2<s<4m_\chi^2$. In principle there is no restriction on $s'$ however in the next sub-section we shall choose a physically motivated function $s'=s'(s)$ so that the generalised superposition amplitude is a function of a single complex variable, and that we might attempt to derive positivity bounds on it and its derivatives.

%%%%%%%%%%%%%%%%%%%%%%%%%%%%%%%%%%%%%%%%%%
\subsection{Positivity bound: equal 3-momenta}\label{sec:gen_bound_1}
One choice for the curve $s'(s)$ is obtained by enforcing the 3-momenta of the $\phi\chi$ states to obey $|\boldsymbol{\ell}|=|\mathbf{k}|$ which is equivalent to setting,
\begin{equation}
    s'(s)=\frac{s}{2}+\Delta+\frac{1}{2} \sqrt{s} \sqrt{s+4 \Delta}\,.
\end{equation}
To derive a positivity bound involving $\scrg(s,s'(s))$ we follow the same steps as in the previous section, starting by looking at the integral expression for $\del_s^2\left[\scrg(s,s'(s))\right]$ using dispersion relations for each term on the right-hand side of \eqref{Gdef}. For the first three terms (with coefficients $\sfc_1$, $\sfc_3$ and $\sfc_t$) we can use the standard twice subtracted forward limit dispersion relations. 

For the $\sfc_2$ term we can also use the standard twice subtracted dispersion relation and simply insert $s'(s)$ in place of $s$; this is valid provided $s'$ does not lie on any branch cuts: $(m_\phi-m_\chi)^2<s'(s)<(m_\phi+m_\chi)^2$, which is satisfied if $0<s<4m_\phi^2$. For the $\sfc_b$ term we may use the dispersion relation for the backwards amplitude given in \eqref{eq:backwards_dispersion} and again simply insert $s'(s)$ in place of $s$ provided that we do not take $s$ values for which $s'$ is on top of any branch cuts of $\cA_{\phi\chi}(s'(s);-1)$, which is again ensured if $0<s<4m_\phi^2$. For future convenience we define the following notation: 
\begin{equation}\label{eq:ghs_def}
    \cG(s)\equiv \scrg\left(s,H(s)\right)\,,\quad H(s)\equiv \hip(\Sigma-2\Delta-s)\,,\quad \hat{S}\equiv \sqrt{s(s+4\Delta)}\,,
\end{equation}
from which it follows that $2H(s)=s+2\Delta+\hat{S}$. We will often suppress the argument of $H$ for compactness.

It may seem concerning that we have started with the amplitude $\mathscr{G}(s,s'(s))$ and are now considering $\mathscr{G}(s,H(s))$ despite the fact that $s'$ and $H$ are distinct as complex functions. However, since we will only ever evaluate the bound at positive $s$ where it is true that $s'(s)=H(s)$ we can use $H(s)$ in place of $s'(s)$ freely.

\paragraph{Convergence of the dispersion relations:}An immediate issue arises, similar to that highlighted in \eqref{eq:derivatives_B}. If we take the standard dispersion relation for $\cA_{\phi\chi}(s;1)$ and insert $H(s)$ in place of $s$ the integral at infinity does not converge after two subtractions. Taking two derivatives of the asymptotic UV integral used to close the contour in deriving the dispersion relation results in,
\begin{equation}
\begin{aligned}
    \del_s^2I_{\infty}(s)&=\del_s^2\left[\frac{1}{2 \pi \mathrm{i}} \int_{|\mu| \rightarrow \infty} \mathrm{d} \mu \frac{\mathcal{A}_{\phi \chi}(\mu ; 1)}{\left(\mu-H(s)\right)}\right]\\&=-\frac{1}{2 \pi \mathrm{i}} \int_{|\mu| \rightarrow \infty} \mathrm{d} \mu\left\{\frac{2 \Delta^2}{ \hat{S}^3}\frac{1}{\mu^2}+\cO\left(\frac{1}{\mu^3}\right)\right\}\cA_{\phi\chi}(\mu;1)\,.
\end{aligned}
\end{equation}
As the suppression at large energies is $1/\mu^2$ rather than $1/\mu^3$ we cannot conclude that this contribution vanishes via the Froissart bound. Nevertheless, just as in  \eqref{A2_subtract1d} we can construct a unique quantity involving both first and second $s$ derivatives that satisfies the usual convergence in the UV. This quantity is given by,
\begin{equation}
    \partial_s^2\left[\mathcal{A}_{\phi \chi}\left(H ; 1\right)\right]+\frac{2 \Delta^2 }{\hat{S}^2 H}\partial_s\left[\mathcal{A}_{\phi \chi}\left(H ; 1\right)\right]=\partial_s^2\left[\mathcal{A}_{\phi \chi}\left(H ; 1\right)\right]-\frac{H''}{H'}\partial_s\left[\mathcal{A}_{\phi \chi}\left(H ; 1\right)\right]\,.
\end{equation}
Interestingly the form of the coefficient of the second term above is exactly the same as in \eqref{A2_subtract1d} but with $H(s)$ instead of $h(s)$. The UV integral now has sufficient $\mu$ suppression for us to assume that it vanishes, given the Froissart bound:
\begin{equation}
    \partial_s^2\left[I_\infty\right]-\frac{H''}{H'}\partial_s\left[I_\infty\right]=\frac{1}{2 \pi \mathrm{i}} \int_{|\mu| \rightarrow \infty} \mathrm{d} \mu\left\{ \frac{(s+2\Delta+\hat{S})^2}{2\hat{S}^2}\frac{1}{\mu^3}+\cO\left(\frac{1}{\mu^4}\right)\right\}\cA_{\phi\chi}(\mu;1)=0\,.
\end{equation}
% Given that the integral at infinity now vanishes we have the dispersion relation:
% \begin{small}
% \begin{equation}
% \begin{aligned}
% \partial_s^2\left[\mathcal{A}_{\phi \chi}\left(H ; 1\right)\right]+\frac{2 \Delta^2 }{\hat{S}^2 H}\partial_s\left[\mathcal{A}_{\phi \chi}\left(H ; 1\right)\right] & =\frac{2}{\pi} \int_{\left(m_\phi+m_\chi\right)^2}^{\infty} \mathrm{d} \mu\left(\frac{-\Delta^2+(s+2 \Delta) H}{\hat{S}^2}\right) \frac{\operatorname{Im} \mathcal{A}_{\phi \chi}(\mu ; 1)}{(\mu-H)^3} \\
% & +\frac{2}{\pi} \int_{\left(m_\phi+m_\chi\right)^2}^{\infty} \mathrm{d} \mu\left(\frac{-\Delta^2(s+2 \Delta)+\left(\hat{S}^2+2 \Delta^2\right) H}{\hat{S}^3}\right) \frac{\operatorname{Im} \mathcal{A}_{\phi \chi}(\mu ; 1)}{(\mu+H-\Sigma)^3} .
% \end{aligned}
% \end{equation}
% \end{small}
One might worry that the same issue arises when $H(s)$ is inserted into the dispersion relation for $\cA_{\phi\chi}(s;-1)$ given in \eqref{eq:backwards_dispersion} however this is not the case as the second $s$ derivative of the integrand already has $1/\mu^3$ behaviour at large $\mu$ and so no modification is needed:
\begin{equation}
    \del_s^2\bigg(\frac{1}{\mu-H}+\frac{1}{\mu-\frac{\Delta^2}{H}}\bigg)=\frac{-\Delta^2(s+2\Delta)+H(\hat{S}^2+2\Delta^2)}{\hat{S} H^2}\frac{2}{\mu^3}+\cO\left(\frac{1}{\mu^4}\right)\,.
\end{equation}

\begin{figure}[h]
    \centering
    \begin{tikzpicture}[]
        \node[anchor=south west,inner sep=0] (image) at (0,0) {\includegraphics[width=0.9\textwidth]{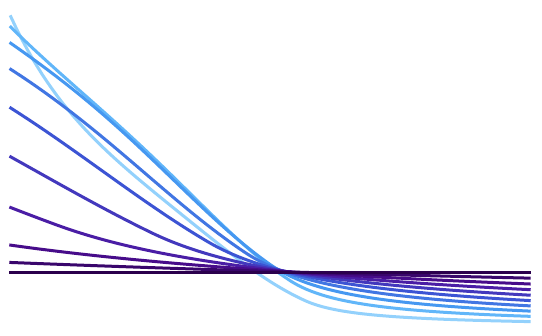}};
        \begin{scope}[x={(image.south east)},y={(image.north west)}]
            % Draw axes starting at bottom-left (0,0)
            \draw[->,thick] (0.020,0) -- (1,0) node[right] {$s$};
            \draw[->,thick] (0.020,0) -- (0.02,1) node[above] {$\text{max}_\mu\{R\}$};

            \draw (0.020,0) -- (0.020,-0.02) node[below] {\small $4\Delta$};

            \draw (0.95,0.9) node[below] {\small $\Delta$};

            \draw (0.98,0) -- (0.98,-0.02) node[below] {\small $4m_\phi^2$};

            \draw (0.020,0.19) -- (0.001,0.19) node[left] {\small $0.11$};
            \draw (0.020,0.95) -- (0.001,0.95) node[left] {\small $0.64$};
            \draw (0.020,0.04) -- (0.001,0.04) node[left] {\small $0.01$};

            %draw horizontal dashed line
            \draw[gray!50, thick, dashed] (0.02,0.04) -- (0.95,0.04);
        \end{scope}
        \node[anchor=south west,inner sep=0] (image) at (12.9,3.95) {\includegraphics[width=0.065\textwidth]{DeepSeaLegend.pdf}};
    \end{tikzpicture}
    \caption{As $s$ (horizontal axis) increases between $4\Delta$ and $4$ in units of $m_\phi^2$, the value of $\text{max}_\mu\{R\}$ (vertical axis) decreases monotonically. When $\Delta=0$ the curve lies flat along the $s$-axis. These plots are in fact discrete and are generated by fixing evenly spaced values of $s$ between $4\Delta$ and $4m_\phi^2$ and numerically maximising the function $R$ over the range of the integral. The plot points are joined by a curve that we expect to be smooth.}
    \label{fig:maxrcurves_2}
\end{figure}

\paragraph{Positivity bound:}We can therefore write down an integral expression for the following quantity:
\begin{equation}
    \begin{aligned}
    \cO_2&\equiv\partial_s^2[\cG(s)]+\sfc_2\frac{2 \Delta^2}{\hat{S}^2 H} \partial_s \left[\mathcal{A}_{\phi \chi}(H ; 1)\right]\\
    &=\sfc_1 \partial_s^2\left[\mathcal{A}_{\phi \phi}(s ; 1)\right]+\sfc_3 \partial_s^2\left[\mathcal{A}_{\chi \chi}(s ; 1)\right]+\sfc_t \partial_s^2\left[\mathcal{A}_t(s ; 1)\right] \\
    &+\sfc_2\left\{\partial_s^2\left[\mathcal{A}_{\phi \chi}(H ; 1)\right]+\frac{2 \Delta^2}{\hat{S}^2 H} \partial_s\left[\mathcal{A}_{\phi \chi}(H ; 1)\right]\right\}+\sfc_b \partial_s^2\left[\mathcal{A}_{\phi \chi}(H ;-1)\right] .
\end{aligned}
\end{equation}
The method of proof is very similar to that of \eqref{eq:Unequal_bound1} and so we have placed the details in Appendix \ref{app:gen_bound_proof}. One finds that it is not evident that $\cO_2$ is positive on its own and so one must add a positive multiple of another positive quantity (in this case $\del_s^2\cA_{\phi\chi}(s,0)$) to obtain something positive. The other difference is that in proving the bound one must use partial wave unitarity of the backwards limit $\phi\chi$ amplitude rather than the basic positivity of the imaginary part. We find the bound on the generalised superposition amplitude:
\begin{equation}\label{eq:Unequal_bound}
\boxed{
     \partial_s^2[\cG(s)]+\sfc_2\frac{2 \Delta^2}{\hat{S}^2 H} \partial_s \left[\mathcal{A}_{\phi \chi}(H ; 1)\right]+\max_{\mu>(m_\phi+m_\chi)^2}\left\{R(\mu;s,\Delta,\Sigma)\right\}\sfc_2\,\del_s^2\cA_{\phi\chi}(s;1) > 0 }
\end{equation}
where 
    \begin{equation}
         R(\mu;s,\Delta,\Sigma)\equiv \frac{\left(1-\frac{\Delta^2}{\mu^2}\right) \frac{1}{(s-h(\mu))^3}-\left(\frac{-\Delta^2(s+2\Delta)+(\hat{S}^2+2\Delta^2)H}{\hat{S}^3}\right)\frac{1}{(\mu+H-\Sigma)^3}}{\frac{1}{(\mu-s)^3}+\frac{1}{(\mu+s-\Sigma)^3}}\,.
    \end{equation}
\paragraph{Probing the size of corrections:}The function $R$ is a ratio of two order $12$ polynomials in $\mu$ and involves square-roots in $s$ that come from $H(s)$, hence it is not straightforward to obtain an analytic expression for $x(s,\Delta)$. As we have assumed a finite region in $(s,\Delta)$ space (i.e. $4\Delta<s<4$ and $0<\Delta<1$) we can nonetheless numerically evaluate the smallest allowed value of $x(s,\Delta)$ by maximising $R(\mu;s,\Delta,\Sigma)$ over $\mu>\Lambda_2$, for different points in this space. We use a combination of Mathematica's \texttt{FindMaxmimum} and \texttt{(N)Maximize} functions to discretely evaluate the maximum of $R$ and plot the results in Figure \ref{fig:maxrcurves_2}.

From this we see once again that the curves are monotonically decreasing and so evaluating the bound at the largest allowed value of $s=4$ gives the smallest correction factor $x(s,\Delta)$. For this value of $s$ the correction factor lies between 0 and approximately $0.11$ as $\Delta$ ranges from $0$ to $1$. Note that the only dependence of this bound so far on low energy data is via the value of $\Delta$, {\it i.e.} the low energy masses.

\begin{figure}[h]
\begin{center}
\includegraphics[width=0.9\linewidth]{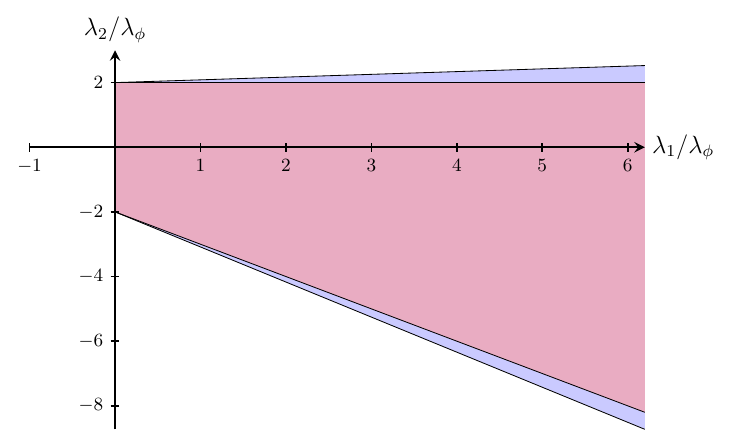}
\caption{\small Region spanned by the bound (\ref{boundEFTresult_2}) (in purple) compared to that predicted by the equal mass bound (\ref{tightestsec3}) (in red), for the choice $\lambda_{\chi} = \lambda_\phi$ (also $\mathfrak{C} \simeq 0.17$) and $\Delta=m_\phi^2$. The unequal mass bound is slightly weaker than the equal mass bound again, but takes a slightly different shape than the unequal mass bound from Figure \ref{Fig:Compare1}. Remarkably, we see that when $\lambda_1=0$, the unequal mass bounds reduce to the equal mass ones.} \label{Fig:Compare2}
\end{center}
\end{figure}

\begin{figure}[h]
    \centering
    \begin{tikzpicture}[]
        \node[anchor=south west,inner sep=0] (image) at (0,0) {\includegraphics[width=0.9\textwidth]{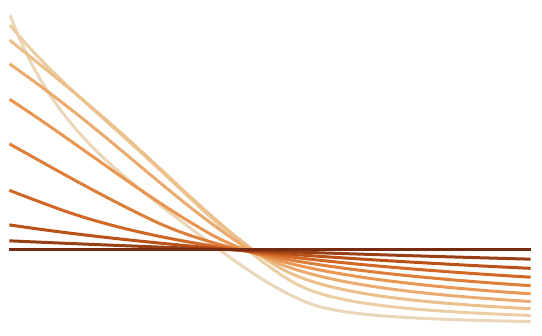}};
        \begin{scope}[x={(image.south east)},y={(image.north west)}]
            % Draw axes starting at bottom-left (0,0)
            \draw[->,thick] (0.020,0) -- (1,0) node[right] {$s$};
            \draw[->,thick] (0.020,0) -- (0.02,1.02) node[above] {$\mathfrak{C}(s,\Delta)$};

            \draw (0.020,0) -- (0.020,-0.02) node[below] {\small $4\Delta$};

            \draw (0.95,0.9) node[below] {\small $\Delta$};

            \draw (0.98,0) -- (0.98,-0.02) node[below] {\small $4m_\phi^2$};

            \draw (0.020,0.255) -- (0.001,0.255) node[left] {\small $0.17$};
            \draw (0.020,0.95) -- (0.001,0.95) node[left] {\small $0.70$};
            \draw (0.020,0.04) -- (0.001,0.04) node[left] {\small $0.01$};

            %draw horizontal dashed line
            \draw[gray!50, thick, dashed] (0.02,0.04) -- (0.98,0.04);
        \end{scope}
        \node[anchor=south west,inner sep=0] (image) at (12.9,3.95) {\includegraphics[width=0.065\textwidth]{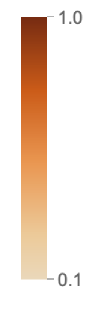}};
    \end{tikzpicture}
    \caption{As $s$ (horizontal axis) increases between $4\Delta$ and $4$ in units of $m_\phi^2$, the value of $\mathfrak C(s,\Delta)$ (vertical axis) decreases monotonically. When $\Delta=0$ the curve lies flat along the $s$-axis. These plots are in fact discrete and are generated by fixing evenly spaced values of $s$ between $4\Delta$ and $4m_\phi^2$ and evaluating $\mathfrak C(s,\Delta)$. The plot points are joined by a curve that we expect to be smooth. The quantity $\mathfrak C$ captures the total correction terms to the positivity bound in the toy model EFT specifically whilst the maximisation of $R$ shown in Figure \eqref{fig:maxrcurves_2} is completely EFT independent.}
    \label{fig:maxrcurves_3}
\end{figure}
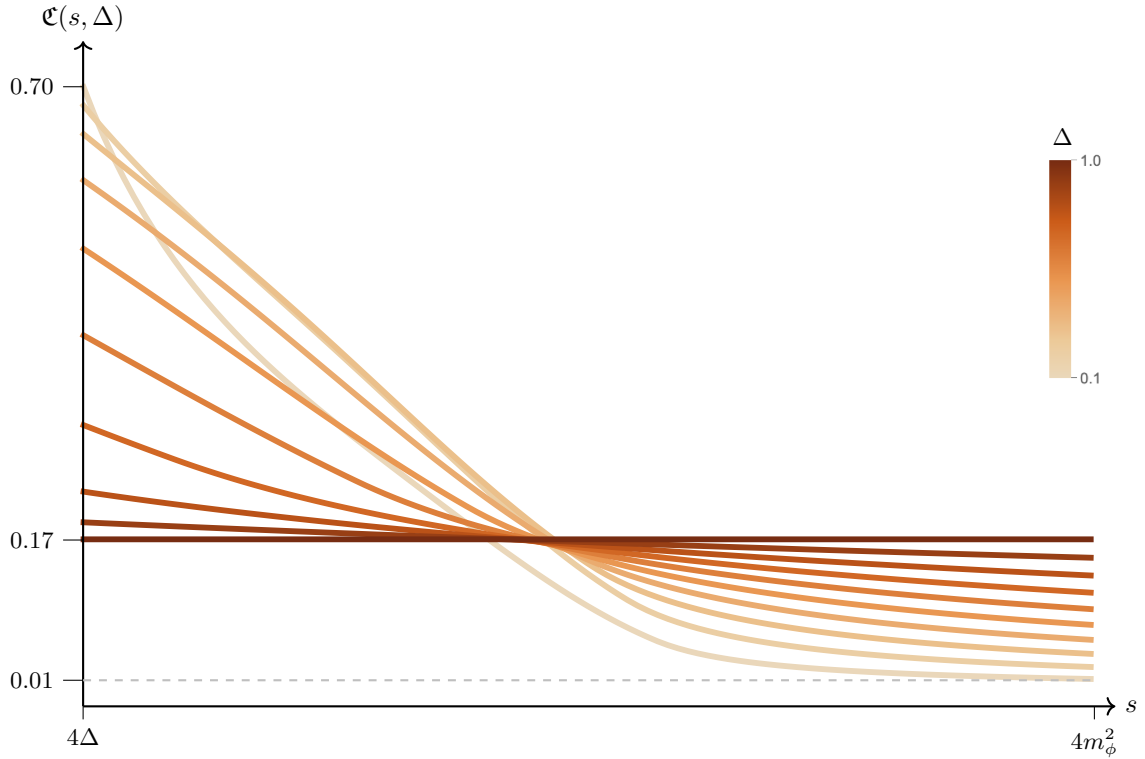

\paragraph{EFT toy-model:} If we return to our toy model EFT and insert the tree-level amplitudes into the above positivity bound, the inequality reads,
\begin{equation}
    \begin{aligned} \label{eq:eft_superbound2}
        (8\lambda_\phi) \sfc_1+(8\lambda_\chi) \sfc_3+(2\lambda_1)\sfc_2\left(1+\mathfrak{C}(s,\Delta)\right)>-(\sfc_t+\sfc_b)\left(\lambda_1+2\lambda_2\right)\,,
    \end{aligned}
\end{equation}
where the correction factor $\mathfrak{C}$ is defined by grouping together any terms that vanish when the $\Delta\xr0$ limit is taken,
\begin{equation}
    \mathfrak C(s,\Delta)\equiv -1 +\frac{H \hat{S}+\Delta^2}{\hat{S}^2}+\max_{\mu>(m_\phi+m_\chi)^2}\left\{R(\mu;s,\Delta,\Sigma)\right\}\,,
\end{equation}
with $\mathfrak{C}(s,0)=0$. In the EFT the bound only differs from the equal mass bound by an amplification of the $\sfc_2\lambda_1$ term and from plotting this correction at different values of $s$ and $\Delta$ we again see that the correction is minimised at the largest value of $s$ for a given $\Delta$ and that if we take this value, the correction is at most $\sim17\%$ (see Figure \ref{fig:maxrcurves_3}). This bound again predicts that $\lambda_\phi,\lambda_\chi,\lambda_1>0$ but instead says
\begin{equation} \label{boundEFTresult_2}
- \left( 1 + \frac{\mathfrak{C}}{2}\right) \lambda _1 - 2 \sqrt{ \lambda _{\chi } \lambda _{\phi }} < \lambda_2 <  \frac{\mathfrak{C}}{2} \lambda_1+2 \sqrt{\lambda _{\chi } \lambda _{\phi }} \ .
\end{equation}
where we use at worst $\mathfrak{C} \simeq 0.17$. See Figure \ref{Fig:Compare2}. Note that in this case, the correction arising from the unequal masses dress the $\cA_{\phi \chi}$ contribution to the bound \eqref{eq:Unequal_bound}.  This implies that when that part of the amplitude vanishes, {\it i.e.} when $\lambda_1=0$, there are no corrections to the unequal mass bounds we obtain as compared to the equal mass case. 
In principle, when considering a particular EFT, one may consider a new version of the generalized bound presented here with a specific choice of function $s'(s)$ specifically engineered so that it gives the strongest possible constraint on EFT coefficients. 

%%%%%%%%%%%%%%%%%%%%%%%%%%%%%%%%%%%%%%%%%%

%%%

\subsection{Further generalised bounds}
\label{sec:furthergen}

There exist many possible families of bounds that can be constructed using variations of the procedure described above. Another related example is presented in Appendix~\ref{App:Alt_Bound1} where generally the first derivative terms appearing in the observables $\cO_i$ in the above get replaced by integrals in the IR, which can always be computed in the EFT. Bounds depending on IR integrals (like the one in Appendix~\ref{App:Alt_Bound1}) rely on the assumption of weak coupling.

There is also a functional degree of freedom in choosing the function $s'(s)$ that sits inside $\mathscr{G}(s,s')$. We have chosen a physically motivated function for $s'$, however in principle there is no reason to place such a restriction and it is possible that by considering other functions or even just considering $s'$ as a totally independent complex variable will lead to stronger bounds. Systematically exploring all possible bounds for $\Delta > 0$ is a challenging problem. Our observation is that all unequal-mass bounds we have examined inevitably exhibit mass dependence of order $\mathcal{O}(\Delta/\Sigma)$, though the precise placement of this dependence within the bound depends on the choice of the generalized superposition amplitude $\mathscr{G}$. Determining the strongest and most general bound that incorporates all such possibilities is non-trivial, and we leave a detailed exploration of this question for future work.

%%%%%%%%%%%

\section{Improved positivity bounds}
\label{sec:Improved_Bounds}

Whilst the bounds derived above are valid even in the presence of low-energy branch cuts, one might worry that the assumptions used to derive them are rather restrictive. In particular the restriction on the mass differences between the particles $m_\chi<\sqrt{2}m_\phi$, which was necessary to ensure a region free of cuts on the real $s$ axis where the amplitude's derivatives could be evaluated. In the case where this is not true for the particles under consideration, provided that an accurate approximation of the imaginary parts of the amplitude is available via an EFT, we may use the dispersion relations derived earlier to explicitly subtract off the low energy branch cuts of the amplitude following the logic of ‘improved positivity bounds'. Once this has been done, the ‘improved' amplitude enjoys a larger region of analyticity (though now in an EFT approximation) and we can ignore the upper bound on the heavy mass $m_\chi$.\\
As an example let's take the dispersion relation for the observable $\cO_1$ defined in equation \eqref{A2_subtract1d}. The portions of the integrals above the EFT cut-off can be approximated using the fact that the integration variable $\mu$ takes values much larger than $s,\Delta$ or any individual masses of particles in the EFT. So assuming that we are not evaluating the bound at $s$ values very near the cut-off scale $\varepsilon^2\Lambda^2$, the integrals above the cut-off scale are approximately
\begin{equation}
    \begin{aligned}
        I_{\rm UV} \approx & \int_{\varepsilon^2\Lambda^2}^{\infty}\frac{\rd\mu}{\mu^3}\,\left[\sum_{i=1}^{3}\,\sfc_i\operatorname{Im} \cA_{i}(\mu;1)+\sfc_b\left(1-\frac{\Delta^2}{s^2}\right)^2\operatorname{Im} \cA_{\phi\chi}(\mu;-1)+\sfc_t\operatorname{Im}\cA_t(\mu;1)\right]\\
    +&\int_{\varepsilon^2\Lambda^2}^{\infty}\frac{\rd\mu}{\mu^3}\,\left[\sum_{i=1}^{3}\,\sfc_i\operatorname{Im} \cA_{i}(\mu;1)+\sfc_b\left(1-\frac{\Delta^2}{s^2}\right)^2\operatorname{Im}\cA_t(\mu;1)+\sfc_t\operatorname{Im} \cA_{\phi\chi}(\mu;-1)\right]\,.
    \end{aligned}
\end{equation}
Using unitarity (positivity) in the form \eqref{eq:im_super} combined with the fact that the correction factor multiplying $\sfc_b$ above satisfies $0<\left(1-{\Delta^2}/{s^2}\right)^2<1$ for $s>\Delta$, we can conclude that the UV integrals in this approximation are positive. This immediately leads to the improved positivity bound,
\begin{equation}
    \cO_1(s)-I_{\rm LH;\epsilon}-I_{\rm RH;\epsilon}>0\,,\qquad\text{for}\qquad \Delta<s\ll\varepsilon^2\Lambda^2\,,
\end{equation}
where $I_{\rm LH,RH;\epsilon}$ are defined to be the expressions in equation \eqref{eq:LHRHdr} but with the upper limit of every integral set to $\varepsilon^2\Lambda^2$. Similar bounds can be derived for the observable $\cO_2$.

The dispersive integrals in $I_{\rm LH,RH}$ reproduce the low energy branch cuts of the amplitude meaning that the above subtraction results in a function that is analytic in the low energy region of the $s$ plane and so no restriction regarding the upper limit on the heavy mass $m_\chi$ -- which came from demanding that there was at least some region of analyticity on the real $s$ axis -- is required. Since we have obtained explicit expressions for the dispersive integrals the evaluation of improved bounds within a given EFT is straightforward and the improved bounds method should provide stronger constraints than those obtained by ignoring low energy branch cuts altogether. Once low energy cuts have been subtracted, the resulting improved amplitude can be optimally constrained using powerful techniques developed in \cite{Du:2021byy} since we have analyticity up to the EFT cut-off and simpler dispersive integrals to manipulate (as in $I_{\rm UV}$).

One final word of caution: it may seem that by subtracting the low energy branch cuts from the superposition amplitude that the entire region $|s|<\varepsilon^2\Lambda^2$ is analytic, however this is not the case. Due to the inclusion of the backwards limit scattering amplitude $\cA_{\phi\chi}(s;-1)$ within the superposition amplitude the region $|s|<\Delta$ is not solely IR. The peculiar crossing property of this amplitude \eqref{eq:b_crossing}, which involves $s\leftrightarrow\Delta^2/s$ means that the $|s|<\Delta^2/(\varepsilon^2\Lambda^2)$ is also ‘UV' and still contains un-subtractable branch cuts.

%%%%%%%%%%%%%%%%%%%%%%%%%%%%%%%

\section{Conclusions and further directions}
\label{sec:conclusion}

EFTs involving different species are ubiquitous in nature. As the number of fields and their interactions increases, so too does the number of positivity bounds, particularly when one can leverage superposition states to strengthen parameter space constraints. These bounds provide powerful constraints on the space of consistent EFTs and have proven essential in taming the complexity of multi-field theory parameter space.

When the light fields are much lighter than the scale of new physics, one might expect that differences in mass can be ignored, allowing indefinite bounds to be applied without concern for non-analytic structures from fixed angle scattering that peel off the real axis into the complex plane. However, in practice, even the smallest mass difference implies that indefinite species states are no longer momentum eigenstates, complicating the derivation of standard positivity bounds. For instance a consequence of this mass difference is the non-trivial relationships between Mandelstam variables in the backwards limit, which imprints onto the complex $s$ plane analytic structure.
\begin{figure}[!h]
    \centering
    \includegraphics[width=0.9\linewidth]{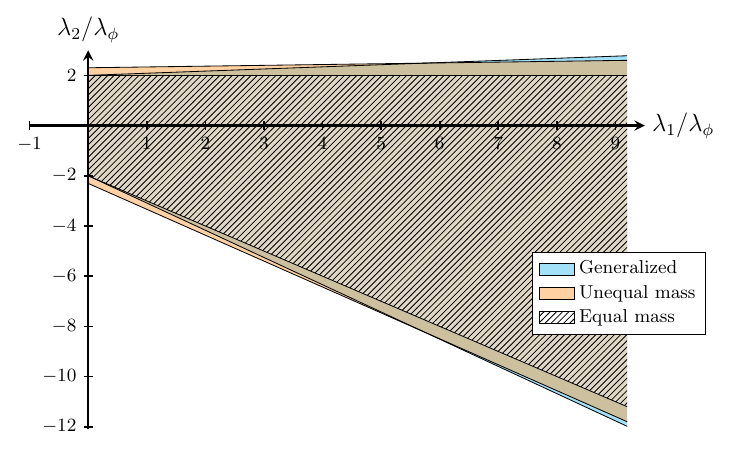}
    \caption{A comparison of the equal-mass/weak-coupling bound, unequal mass and generalised unequal mass bounds in the EFT \eqref{EFTaction} for $\lambda_\phi=\lambda_\chi>0$ and $s=4m_\phi^2$ with $\Delta=m_\phi^2$.}
    \label{fig:combined_bound}
\end{figure}

In this work, we have carefully examined the analytic structure of superposition amplitudes involving different mass particles -- making no assumption of weak coupling, and presented dispersion relations for these amplitudes. Following this we use positivity of the imaginary part of the superposition amplitude to derive constraints on combinations of its derivatives. In addition we have presented a generalised class of superposition amplitudes which depend on two invariant center-of-mass energies $s$ and $s'$, in which case we choose a relation between the two $s'(s)$ to obtain a complex function in a single variable. Bounds on these generalised amplitudes are shown in the context of a toy EFT to be complementary to the original unequal mass superposition bounds (un-generalised case) as shown in Figure \ref{fig:combined_bound}. This suggests that considering more and more functions $s'(s)$ and their associated positivity bounds will continue restricting the allowed region -- perhaps even converging to the equal-mass/weak coupling region. 

While making no weak coupling assumption, the first part of our analysis still assumes a gapped theory, requiring relatively tight constraints on the ratio between the two masses. However we then lay down the steps for a procedure in which the low-energy branch cut can be subtracted up to the EFT cut-off, allowing applications to more generic (and physically realistic) situations as well as the use of techniques originally derived in the weak coupling approximation. In our results the mass difference enters as a small, analytic, and trackable correction to the standard equal-mass bounds, as expected, though the size of the correction depends on the specific EFT under consideration.

The formalism we have developed offers explicit dispersion relations which can be used as foundation for generalizations that incorporate full positivity as formulated in \cite{Li:2021lpe}. Since our unequal mass bounds are directly relatable to the equal mass indefinite bounds, we only expect additional bounds from full positivity when considering three fields or more. 
With the use of the dispersion relations we have derived, one can also be in principle pick a specific choice of indefinite states where the amplitude enjoys full crossing symmetry and further non-linear bounds can be inferred from the null constraints, potentially generalizing the results of \cite{Du:2021byy} to the case of unequal mass. 

Although the present analysis is not directly applicable to the Standard Model, which (to date) only counts  one scalar field, the Higgs, it can be applied to amplitudes involving composite particles, of which there are numerous scalar examples with mass ratios in the range considered here. Our bounds thus offer precise, analytic constraints on any EFT that includes such composites in its low-energy spectrum. Our results should also be generalizable to particles with spin, although developing the precise formalism to account for the particle's respective spins is beyond the remit of the work presented here. 

Interestingly, our work also lays down the foundations for a more systematic exploration of generalized superposition amplitudes for both equal and unequal mass cases, and the analytic bounds they imply. Of particular interest, is the application of our generalized formalism to the equal-mass  $m_\phi = m_\chi = m$, where the generalized superposition amplitude $\mathscr{G}$ reduces to
\begin{small}
\begin{eqnarray} 
\mathscr{G}(s,s') |_{m_\phi = m_\chi = m} & = &  \sfc_1 \mathcal{A}_{\phi\phi}( s ;1   ) + \sfc_3  \mathcal{A}_{\chi\chi}( s ;  1 ) + \sfc_t  \mathcal{A}_{t}( s ; 1 ) + \sfc_2  \mathcal{A}_{\phi\chi}( s' ;  1) + \sfc_b \mathcal{A}_{\phi\chi}( s', -1 ) \ ,
\end{eqnarray}
\end{small}\ignorespaces
providing us with the freedom to pick different curves of $s'(s)$ to leverage this to derive different bounds. The full implications of this generalized amplitude to the same (and unequal) mass case deserves its own dedicated investigation. 

\paragraph{Acknowledgements:} We thank Andrew J. Tolley for pointing us to the existence of hyperbolic dispersion relations, Miguel Correia for helpful correspondence regarding anomalous thresholds, as well as Martin Kruczenski, Brian McPeak and Shuang-Yong Zhou and Zhuo-Hui Wang for useful discussions and comments on our work. 
The research of
SJ is supported by the ERC (NOTIMEFORCOSMO, 101126304) funded by the European
Union, as well as the Simons Investigator award 690508. The work of CdR is supported by STFC Consolidated Grant ST/X000575/1. CdR is also supported by a Simons Investigator award 690508.  Views and opinions expressed are
however those of the author(s) only and do not necessarily reflect those of the European
Union or the European Research Council Executive Agency. Neither the European Union
nor the granting authority can be held responsible for them.
Discussions during the long-term workshop Progress of Theoretical Bootstrap at Yukawa Institute for Theoretical Physics (YITP-T-25-01) were useful as we finished this work.

%%%%%%%%%%%%%%%%%%%%%%%%%%%%%%%
\appendix
%\newpage

\section{Kinematics in the centre-of-mass frame}
\label{App:kinematics}
We here briefly summarize our conventions for $2 \to 2$ scattering in the centre-of-mass  frame \cite{Melville:2019gux}, focusing on the peculiar $s$-dependence of the inelastic amplitudes discussed in \S\ref{sec:tchan} and \S\ref{sec:backwards}. For each sub-amplitude, we assume a frame where the first and second 3-momenta sum to zero (and by momentum conservation so do the third and fourth), and rotate so that all particles lie along the $z$-axis and scatter within the $xz$-plane. In the all-incoming convention, the momenta are:
\begin{eqnarray} \label{CoM}
    k_{1} &=& ( \omega_1, 0, 0 , k ) \\
    k_{2} &=& ( \omega_2, 0, 0 , -k )  \\
    k_{3} &=& ( - \omega_3, - k' \sin\theta , 0 , - k' \cos\theta ) \\
    k_{4} &=& ( - \omega_4, k' \sin\theta, 0 , k' \cos\theta ) 
\end{eqnarray}
The above parameters may be related to the Mandelstam $s$ and $t$ variables through the relations
\begin{equation} \label{CoM_tos}
  \begin{split}
  \omega_{1} & = \tfrac{1}{2\sqrt{ s }} ( s + m_\phi^2 - m_\chi^2 ) \\
  \omega_{2} & = \tfrac{1}{2\sqrt{ s }} ( s + m_\chi^2 - m_\phi^2 ) \\
  \omega_{3} & = \tfrac{1}{2\sqrt{ s }} ( s + m_3^2 - m_4^2 ) \\
  \omega_{4} & = \tfrac{1}{2\sqrt{ s }} ( s + m_4^2 - m_3^2 )
  \end{split}
\hspace{20mm}
  \begin{split}
    k^2 & = \frac{\Lambda(s,m_\phi^2,m_\chi^2)}{4 s} \\
    k^{\prime 2} & = \frac{\Lambda(s,m_3^2,m_4^2)}{4 s} \\
    \cos \theta & = \frac{ s^2 + s ( 2 t - \Sigma_i m_i^2) + (m_\phi^2 - m_\chi^2)(m_3^2 - m_4^2)  }{\sqrt{ \Lambda\mathtt(s,m_\phi^2,m_\chi^2) \Lambda\mathtt(s,m_3^2,m_4^2) }}
  \end{split}
\end{equation}
where $\Lambda(x,y,z) = x^2 + y^2 + z^2 - 2 x y - 2 y z - 2 x z$ is the triangle function. We can now immediately apply the above formulae to find the Mandelstam variables for various fixed angle scattering processes. Recall the definitions, $\Sigma=2m_\phi^2+2m_\chi^2$ and $\Delta\equiv m_\chi^2-m_\phi^2$.
 \begin{itemize}
    \item For the scattering process $\cA_{\phi\chi\xr\phi\chi}(s,t)$ in the forward limit we have,
    \begin{equation}
         \cos\theta=1=\frac{ s^2 + s ( 2 t -2 m_\phi^2-2m_\chi^2) + (m_\phi^2 - m_\chi^2)^2 }{ \Lambda\mathtt(s,m_\phi^2,m_\chi^2)  } = 1 - \frac{2t}{h(s)} \quad\implies\quad t(s)=0
    \end{equation}
    derived for physical kinematics such that $s > (m_\phi + m_\chi)^2$ (equivalently $k^2 > 0$) and we have used the definition of $h(s)$ from Eq.~(\ref{hs_def}). This shows that in the elastic case, one recovers the usual forward limit $t \to 0$. 
    
    In contrast when we take the backwards limit where $\cos\theta=-1$, one finds the $t$ variable has the following dependence on $s$:
    \begin{equation}
    \cos\theta=-1 =   1 - \frac{2t}{h(s)} \qquad \implies \qquad t(s) = h(s) = \Sigma-s-\frac{\Delta^2}{s} \ .
    \end{equation}
    which again assumes $s > (m_\phi + m_\chi)^2$. 
    
    \item Finally for the inelastic $t$-channel amplitude $\mathcal{A}_{t} = \mathcal{A}_{\phi\phi\to\chi\chi}$ one finds that in the forward and backward limits the $t$ variable depends on $s$ as:
    \begin{equation} \label{t_tcurve}
    \begin{aligned}
    \cos\theta_t=1 = \frac{s + 2 t(s) - \Sigma }{ \sigma(s) } \qquad \implies \qquad t(s) = h_{+}^{-1}(s)\\
    \cos\theta_t=-1 = \frac{s + 2 t(s) - \Sigma }{ \sigma(s) } \qquad \implies \qquad t(s) = h_{-}^{-1}(s)
    \end{aligned}
    \end{equation}
    which assumes $s > 4m_\phi^2$, and uses the definition $\sigma(s)$ from Eq.~(\ref{sigmatau_def}), and the inverses of $h(s)$ are given in Eq.~\eqref{eq:h_inverse}.
\end{itemize}

%%%%%%%%%%%%%%%%%%%%%%%%%%%%%%
\section{Derivation of the backwards limit dispersion relation}\label{app:backwards_dr}
In this appendix we give the remainder of the details regarding the derivation of the backwards limit dispersion relation, Equation \eqref{eq:backwards_dispersion}. The labels of segments refer to Figure \ref{fig:contour-anatomy}.

\paragraph{Circular contour integrals:}Consider segment $\mathtt{C}$ and perform the given change of variable and use $s-t$ crossing symmetry to find,
\begin{equation}
\begin{aligned}
    I_{\mathtt{C}} &= \frac{1}{4\pi\ri}\int_{\mathtt{C}}\rd\mu\left(\frac{1}{\mu-s}+\frac{1}{\mu-\frac{\Delta^2}{s}}\right)B(\mu)
    \\&= \frac{1}{4\pi\ri^2}\int^{4m_\chi^2}_{4m_\phi^2}\rd z\frac{h_{+}^{-1}(z)}{|\sigma(z)|}\left(\frac{1}{h^{-1}_{+}(z)-s}+\frac{1}{h^{-1}_{+}(z)-\frac{\Delta^2}{s}}\right)\cA_{t}(z-\ri\epsilon, h^{-1}_{+}(z-\ri\epsilon))\,.
\end{aligned}
\end{equation}
Taking into account the $\ri\epsilon$'s through the change of variable is of utmost importance as we are interested in the physical discontinuities of the amplitude across its branch cuts, which depend on the sign of the $\ri\epsilon$ terms. The points along segment $\mathtt{C}$ are given, for infinitesimal positive $\epsilon$, by the set $\{\mu=(\Delta+\epsilon)\re^{\ri\psi}\,|\,\pi<\psi<0\}$ which is the image of the set $\{z-\ri\epsilon \,|\,4m_\phi^2<z<4m_\chi^2\}$ under $\mu=h^{-1}_{+}(z)$. In other words, under the change of integration variable, the contour $\mathtt{C}$ becomes a straight line lying an infinitesimal distance below the real $z$ axis, between $4m_\phi^2$ and $4m_\chi^2$. Following the same steps for the segment $\mathtt{F}$ gives,
\begin{equation}
    I_{\mathtt{F}}=-\frac{1}{4\pi}  \int_{4 m_\phi^2}^{4 m_\chi^2} \rd z \frac{h^{-1}_{-}(z)}{|\sigma(z)|}\left(\frac{1}{h_{-}^{-1}(z)-s}+\frac{1}{h_{-}^{-1}(z)-\frac{\Delta^2}{s}}\right)\cA_t(z+\ri \epsilon,h^{-1}_{-}(z+\ri\epsilon))\,.
\end{equation}
Using the identity (arguments of the inverse functions omitted):
\begin{equation}
    h^{-1}_{+}\left(\frac{1}{h_{+}^{-1}-s}+\frac{1}{h_{+}^{-1}-\frac{\Delta^2}{s}}\right)+h_{-}^{-1}\left(\frac{1}{h_{-}^{-1}-s}+\frac{1}{h_{-}^{-1}-\frac{\Delta^2}{s}}\right)=2\,,
\end{equation}
we can take the sum of both $\mathtt{C}$ and $\mathtt{F}$ integrals to obtain
\begin{equation}
\begin{aligned} \label{I_CcupF}
    I_{\mathtt{C}\cup\mathtt{F}}&=-\frac{1}{2\pi}\int^{4m_\chi^2}_{4m_\phi^2}\rd z\frac{1}{|\sigma(z)|}\cA_{t}(z-\ri\epsilon, h^{-1}_{+}(z-\ri\epsilon))\\
    &+\frac{1 }{2\pi}  \int_{4 m_\phi^2}^{4 m_\chi^2} \rd z\, \frac{h^{-1}_{-}(z)}{\sigma(z)}\left(\frac{1}{h_{-}^{-1}(z)-s}+\frac{1}{h_{-}^{-1}(z)-\frac{\Delta^2}{s}}\right)\operatorname{Im}\cA_t(z+\ri \epsilon,h^{-1}_{-}(z+\ri\epsilon))\,,
\end{aligned}
\end{equation}
where we note that the integral on the first line is a constant and so will drop out of any derivatives of the backwards limit $\phi\chi$ amplitude. Repeating the above steps for the remaining two circular segments $\mathtt{G}$ and $\mathtt{L}$ we obtain a similar answer,
\begin{equation}
    I_{\mathtt{G}\cup\mathtt{L}}=\text{constant}+\frac{-1}{2 \pi} \int_{4 m_\phi^2}^{4 m_\chi^2} \rd z \,\frac{h_{+}^{-1}(z)}{\sigma(z)}\left(\frac{1}{h_{+}^{-1}(z)-s}+\frac{1}{h_{+}^{-1}(z)-\frac{\Delta^2}{s}}\right) \operatorname{Im} \cA_t(z+\mathrm{i} \epsilon, h^{-1}_{+}(z+\ri\epsilon))\,.
\end{equation}
Finally taking the sum of these four segments $\mathtt{C,F,G,L}$ we obtain the remarkably simple result for the integral enveloping the circular branch cut:
\begin{equation}
    I_{\text{circular}} = c + \frac{1}{\pi}\int_{4 m_\phi^2}^{4 m_\chi^2}\rd \mu\,\frac{1}{\mu-h(s)}\operatorname{Im}\cA_{t}(\mu+\ri\epsilon;1)\,,
\end{equation}
where $c$ is some constant. We have used the property that for the process $\phi\phi\xr\chi\chi$ the scattering amplitude is equal at angles $\theta$ and $\theta+\pi$ since the out-going particles are indistinguishable. \\

%%%%%%%%%%%%%%%%%%%%%%%%%%%%%%%%%%%%%%%%%%%%%%%%%%%%%%%%5

\paragraph{Right hand cut (segments $\mathtt{A,B}$):} Here no change of variable or crossing is required and the integrals can be combined to give the discontinuity or imaginary part of the backwards limit $\phi\chi$ amplitude,
\begin{equation}
    I_{\text{RH cut}}=\frac{1}{2\pi}\int_{(m_\phi+m_\chi)^2}^{\infty}\rd\mu\,\left(\frac{1}{\mu-s}+\frac{1}{\mu-\frac{\Delta^2}{s}}\right)\operatorname{Im} B(\mu+\ri\epsilon)\,.
\end{equation}
%%%%%%%%%%%%%%%%%%%%%%%%%%%%%%%%%%%%%%%%%%%%%%%%%%%%%%%%5

\paragraph{Left hand cut (segments $\mathtt{D,E}$):} Changing variables using $\mu=h_{-}^{-1}(z)$ which implies $\rd\mu=h^{-1}_-(z)/\sigma(z)$ gives the integral along $\mathtt{D}$ as,
\begin{equation}
\begin{aligned}
    I_{\mathtt{D}}&= \frac{1}{4\pi\ri}\int_{-\infty}^{-\Delta}\rd\mu\,\left(\frac{1}{\mu-s}+\frac{1}{\mu-\frac{\Delta^2}{s}}\right)B(\mu+\ri\epsilon)\\
    &= \frac{1}{4\pi\ri}\int_{\infty}^{4m_\chi^2}\rd z\,\frac{h^{-1}_-(z)}{\sigma(z)}\left(\frac{1}{h^{-1}_-(z)-s}+\frac{1}{h^{-1}_-(z)-\frac{\Delta^2}{s}}\right)B(h^{-1}_-(z)+\ri\epsilon)\,.
\end{aligned}
\end{equation}
Along the integral the amplitude in the integrand is being evaluated outside of its physical region, but as usual crossing can be used to relate it to a physical region process. Taking care of the $\ri \epsilon$'s we have,
\begin{equation}
\begin{aligned}
    B(h^{-1}_-(z)+\ri\epsilon)&=\cA_{\phi\chi}(h^{-1}_-(z)+\ri\epsilon,h(h^{-1}_-(z)+\ri\epsilon))\\
    &=\cA_{\phi\chi}(h^{-1}_-(z-\ri\epsilon),z-\ri\epsilon)\\
    &=\cA_t(z-\ri\epsilon;-1)\,.
\end{aligned}
\end{equation}
where in going from the first line to the second we have used the fact that $h^{-1}_-(z)$ is negative over the region of integration\footnote{In particular,
$$
    h^{-1}(z-\ri\epsilon)=h^{-1}(z)+\ri\epsilon(-1)\frac{h^{-1}_-(z)}{\sigma(z)}+\cO(\epsilon^2)=h^{-1}(z)+\ri\epsilon+\cO(\epsilon^2)\,.
$$}
and we have absorbed any positive factors into $\epsilon$. Therefore, the segment $\mathtt{D}$ is
\begin{equation}
        I_{\mathtt{D}}= -\frac{1}{4\pi\ri}\int^{\infty}_{4m_\chi^2}\rd z\,\frac{h^{-1}_-(z)}{\sigma(z)}\left(\frac{1}{h^{-1}_-(z)-s}+\frac{1}{h^{-1}_-(z)-\frac{\Delta^2}{s}}\right)\cA_t(z-\ri\epsilon;-1)\,.
\end{equation}
The same procedure holds for segment $\mathtt{E}$ and results in an identical expression but with the sign of $\ri\epsilon$ reversed and an overall minus sign due to the orientation of the integral, leading to the total of both segments:
\begin{equation}
    I_{\text{LH cut}} = \frac{1}{2\pi}\int^{\infty}_{4m_\chi^2}\rd \mu\,\frac{h^{-1}_-(\mu)}{\sigma(\mu)}\left(\frac{1}{h^{-1}_-(\mu)-s}+\frac{1}{h^{-1}_-(\mu)-\frac{\Delta^2}{s}}\right)\operatorname{Im}\cA_t(\mu+\ri\epsilon;-1)\,.
\end{equation}

%%%%%%%%%%%%%%%%%%%%%%%%%%%%%%%%%%%%%%%%%%%%%%%%%%%%%%%%5

\paragraph{Inner LH cut (segments $\mathtt{H,K}$):}

Using the exact same procedure as above but instead taking $\mu= h^{-1}_+(z)$ results in the integral,
\begin{equation}
    I_{\text{inner LH}}=\frac{1}{2 \pi} \int_{4 m_\chi^2}^{\infty}\rd \mu\,\frac{-h_{+}^{-1}(\mu)}{\sigma(\mu)}\left(\frac{1}{h_{+}^{-1}(\mu)-s}+\frac{1}{h_{+}^{-1}(\mu)-\frac{\Delta^2}{s}}\right) \operatorname{Im} \cA_t(\mu+\ri \epsilon, \theta=0)
\end{equation}

%%%%%%%%%%%%%%%%%%%%%%%%%%%%%%%%%%%%%%%%%%%%%%%%%%%%%%%%5
\paragraph{Inner RH cut (segments $\mathtt{I,J}$):}
Motivated by the crossing symmetry of $B(s)$ the change of variable $\mu=\Delta^2/z$, results in:
\begin{equation}
    I_{\text{inner RH}}= -\frac{1}{2 \pi} \int_{\left(m_\phi+m_\chi\right)^2}^{\infty} \rd z\, \left(\frac{1}{\frac{\Delta^2}{z}-s}+\frac{1}{\frac{\Delta^2}{z}-\frac{\Delta^2}{s}}\right) \frac{\Delta^2}{z^2}\operatorname{Im} B(z+\ri\epsilon)\,,
\end{equation}
which can be simplified using,
\begin{equation}\label{eq:simpl}
    \left(\frac{1}{\frac{\Delta^2}{z}-s}+\frac{1}{\frac{\Delta^2}{z}-\frac{\Delta^2}{s}}\right) \frac{\Delta^2}{z^2}+\frac{1}{z-s}+\frac{1}{z-\frac{\Delta^2}{s}}=\frac{2}{z}
\end{equation}
to shift some of the integral into a constant,
\begin{equation}
    I_{\text{inner RH}}=\frac{1}{2 \pi} \int_{(m_\phi+m_\chi)^2}^{\infty} \rd \mu\, \left(\frac{1}{\mu-s}+\frac{1}{\mu-\frac{\Delta^2}{s}}\right)\operatorname{Im} B(\mu+\ri \epsilon)+\text { const. }
\end{equation}

%%%%%%%%%%%%%%%%%%%%%%%%%%%%%%%%%%%%%%%%%%%%%%%%%%%%%%%%5

\paragraph{Contours at $0$ and $\infty$:} The contour $\mathcal C_0$ extends to asymptotic infinity in the complex $\mu$ plane giving a contribution,
\begin{equation}
    I_{\infty} = \frac{1}{4\pi\ri}\int_{|\mu|\xr\infty}\rd\mu\,\left(\frac{1}{\mu-s}+\frac{1}{\mu-\frac{\Delta^2}{s}}\right)B(\mu)\,,
\end{equation}
while the contour $\mathcal C_1$ encloses the origin, which due to the crossing property $B(s)=B(\Delta^2/s)$ should strictly be viewed as an asymptotically high energy UV integral in the same way as $I_{\infty}$. By changing variable $\mu=\Delta^2/z$ and using crossing symmetry we can map this contour around the origin to a contour at infinity:
\begin{equation}
    I_0 = \frac{-1}{4\pi\ri}\int_{|z|\xr\infty}\rd z\,\frac{\Delta^2}{z^2}\left(\frac{1}{\frac{\Delta^2}{z}-s}+\frac{1}{\frac{\Delta^2}{z}-\frac{\Delta^2}{s}}\right)B\left(z\right)\,,
\end{equation}
where the contour is exactly the same as in $I_{\infty}$. Again by using the trick in \eqref{eq:simpl} we get,
\begin{equation}
    I_0= \frac{1}{4\pi\ri}\int_{|\mu|\xr\infty}\rd \mu\,\left(\frac{1}{\mu-s}+\frac{1}{\mu-\frac{\Delta^2}{s}}\right)B\left(\mu\right)+\text{const.}\,,
\end{equation}
which combines exactly with the integral at $\infty$ so that
\begin{equation}
    I_0+I_\infty= \frac{1}{2\pi\ri}\int_{|\mu|\xr\infty}\rd \mu\,\left(\frac{1}{\mu-s}+\frac{1}{\mu-\frac{\Delta^2}{s}}\right)B\left(\mu\right)+\text{const.}\,,
\end{equation}\\

Recalling from the main text that the four segments enveloping the circular branch cut gave the integral,
\begin{equation}
    I_{\text{circular}} = \text{const.} + \frac{1}{\pi}\int_{4 m_\phi^2}^{4 m_\chi^2}\rd \mu\,\frac{1}{\mu-h(s)}\operatorname{Im}\cA_{t}(\mu+\ri\epsilon;1)\,,
\end{equation}
the total of all segments is,
\begin{equation}
    \begin{aligned}
        B(s) &= \frac{1}{\pi}\int_{4 m_\phi^2}^{4 m_\chi^2}\rd \mu\,\frac{1}{\mu-h(s)}\operatorname{Im}\cA_{t}(\mu+\ri\epsilon;1)\\
        &+\frac{1}{2\pi}\int_{(m_\phi+m_\chi)^2}^{\infty}\rd\mu\,\left(\frac{1}{\mu-s}+\frac{1}{\mu-\frac{\Delta^2}{s}}\right)\operatorname{Im} B(\mu+\ri\epsilon)\\
        &+\frac{1}{2 \pi} \int_{(m_\phi+m_\chi)^2}^{\infty} \rd \mu\, \left(\frac{1}{\mu-s}+\frac{1}{\mu-\frac{\Delta^2}{s}}\right)\operatorname{Im} B(\mu+\ri \epsilon)\\
        &+\frac{1}{2\pi}\int^{\infty}_{4m_\chi^2}\rd \mu\,\frac{h^{-1}_-(\mu)}{\sigma(\mu)}\left(\frac{1}{h^{-1}_-(\mu)-s}+\frac{1}{h^{-1}_-(\mu)-\frac{\Delta^2}{s}}\right)\operatorname{Im}\cA_t(\mu+\ri\epsilon;-1)\\
        &-\frac{1}{2 \pi} \int_{4 m_\chi^2}^{\infty}\rd \mu\,\frac{h_{+}^{-1}(\mu)}{\sigma(\mu)}\left(\frac{1}{h_{+}^{-1}(\mu)-s}+\frac{1}{h_{+}^{-1}(\mu)-\frac{\Delta^2}{s}}\right) \operatorname{Im} \cA_t(\mu+\ri \epsilon, \theta=0)\\
        &+\frac{1}{2\pi\ri}\int_{|\mu|\xr\infty}\rd \mu\,\left(\frac{1}{\mu-s}+\frac{1}{\mu-\frac{\Delta^2}{s}}\right)B\left(\mu\right)+\text{const.}\,.
    \end{aligned}
\end{equation}
Making use of the fact that the $t$-channel amplitude is the same for angles $0$ and $\pi$, the fourth and fifth lines combine to give the exact same integrand as the first line, but with limits of integration $4m_\phi^2$ to $\infty$. Hence the final result is,
\begin{equation}
    \begin{aligned}
        B(s) &= \frac{1}{\pi}\int_{4 m_\phi^2}^{\infty}\rd \mu\,\frac{1}{\mu-h(s)}\operatorname{Im}\cA_{t}(\mu;1)+\frac{1}{\pi}\int_{(m_\phi+m_\chi)^2}^{\infty}\rd\mu\,\left(\frac{1}{\mu-s}+\frac{1}{\mu-\frac{\Delta^2}{s}}\right)\operatorname{Im} B(\mu)\\
        &+\frac{1}{2\pi\ri}\int_{|\mu|\xr\infty}\rd \mu\,\left(\frac{1}{\mu-s}+\frac{1}{\mu-\frac{\Delta^2}{s}}\right)B\left(\mu\right)+\text{const.}\,.
    \end{aligned}
\end{equation}

%%%%%

\section{Proof of generalised superposition bound}\label{app:gen_bound_proof}
We begin by writing down an integral expression for the following quantity:
\begin{equation}
    \begin{aligned}
    \cO_2&\equiv\partial_s^2[\cG(s)]+\sfc_2\frac{2 \Delta^2}{\hat{S}^2 H} \partial_s \left[\mathcal{A}_{\phi \chi}(H ; 1)\right]\\
    &=\sfc_1 \partial_s^2\left[\mathcal{A}_{\phi \phi}(s ; 1)\right]+\sfc_3 \partial_s^2\left[\mathcal{A}_{\chi \chi}(s ; 1)\right]+\sfc_t \partial_s^2\left[\mathcal{A}_t(s ; 1)\right] \\
    &+\sfc_2\left\{\partial_s^2\left[\mathcal{A}_{\phi \chi}(H ; 1)\right]+\frac{2 \Delta^2}{\hat{S}^2 H} \partial_s\left[\mathcal{A}_{\phi \chi}(H ; 1)\right]\right\}+\sfc_b \partial_s^2\left[\mathcal{A}_{\phi \chi}(H ;-1)\right] .
\end{aligned}
\end{equation}
It is again instructive to separate the integrals that appear into ‘left-hand' (LH) and ‘right-hand' (RH) cut contributions depending on the $\Delta\xr0$ limit of each integrand. If the integrand has a factor of $1/(\mu-s)^3$ in this limit it is designated RH cut integral as it arises due to a cut in the right-hand side of the complex plane, and vice versa if the integrand has a factor $1/(\mu+s-\Sigma)^3$ it is designated a LH cut integral. Then, $\cO_2=I_{\rm LH}+I_{\rm RH}$ with the right-hand contributions:
\begin{equation}
    \begin{aligned}
        I_{\rm RH}=\sfc_1\frac{2}{\pi} &\int_{\Lambda_1}^{\infty}\rd\mu\,\operatorname{Im} \cA_{\phi\phi}(\mu;1)\,\left[\frac{1}{(\mu-s)^3}\right]\\+
        \sfc_2\frac{2}{\pi} &\int_{\Lambda_2}^{\infty}\rd\mu\,\operatorname{Im} \cA_{\phi\chi}(\mu;1)\,\left[\frac{1}{(\mu-H)^3}\right]\left(\frac{-\Delta^2+(s+2\Delta)H}{\hat{S}^2}\right)\\+
        \sfc_3\frac{2}{\pi} &\int_{\Lambda_1}^{\infty}\rd\mu\,\operatorname{Im} \cA_{\chi\chi}(\mu;1)\,\left[\frac{1}{(\mu-s)^3}\right]\\+
        \sfc_b\frac{2}{\pi} &\int_{\Lambda_2}^{\infty}\rd\mu\,\operatorname{Im}\cA_{\phi\chi}(\mu;-1)\left[\frac{1}{(4m_\phi^2-s-h(\mu))^3}\right]\left(1-\frac{\Delta^2}{\mu^2}\right)\\+
        \sfc_t\frac{2}{\pi} &\int_{\Lambda_1}^{\infty}\rd\mu\,\operatorname{Im} \cA_{t}(\mu;1)\,\left[\frac{1}{(\mu-s)^3}\right]\,,
    \end{aligned}
\end{equation}
where $\Lambda_1\equiv 4m_\phi^2$ and $\Lambda_2\equiv (m_\phi+m_\chi)^2$. We can replace $\Lambda_2$ with $\Lambda_1$ in integrals where $\operatorname{Im}\cA_{\phi\chi}(\mu;\pm1)$ appears because by assumption these imaginary parts disappear below the normal threshold at $(m_\phi+m_\chi)^2$ and above $4m_\phi^2$. We can rearrange this into the form
\begin{equation}
	\begin{aligned}\label{eq:RHI}
		I_{\rm RH}=\frac{2}{\pi} &\int_{\Lambda_1}^{\infty}\rd\mu\,\operatorname{Im} \cA_{\rm S}(\mu)\,\left[\frac{1}{(\mu-s)^3}\right]\\+
		\sfc_2\frac{2}{\pi} &\int_{\Lambda_2}^{\infty}\rd\mu\,\operatorname{Im} \cA_{\phi\chi}(\mu;1)\,\left\{\left[\frac{1}{(\mu-H)^3}\right]\left(\frac{-\Delta^2+(s+2\Delta)H}{\hat{S}^2}\right)-\frac{1}{(\mu-s)^3}\right\}\\+
		\sfc_b\frac{2}{\pi} &\int_{\Lambda_2}^{\infty}\rd\mu\,\operatorname{Im}\cA_{\phi\chi}(\mu;-1)\left\{\left[\frac{1}{(4m_\phi^2-s-h(\mu))^3}\right]\left(1-\frac{\Delta^2}{\mu^2}\right)-\frac{1}{(\mu-s)^3}\right\}\,.
	\end{aligned}
\end{equation}
The term on the first line is positive by \eqref{eq:im_super} for $s<\Lambda_1$. Proving that the latter two lines of the above sum to something positive however requires the additional fact that in the physical region, unitarity combined with the partial wave expansion implies that $\operatorname{Im}\cA_{\phi\chi}(\mu;1)\geq|\operatorname{Im}\cA_{\phi\chi}(\mu;-1)|$. So given that $\sfc_2\geq|\sfc_b|$ and $\operatorname{Im}\cA_{\phi\chi}(\mu;1)\geq|\operatorname{Im}\cA_{\phi\chi}(\mu;-1)|$, the last two lines will sum to something positive provided
\begin{equation}\nonumber
    \left[\frac{1}{(\mu-H)^3}\right]\left(\frac{-\Delta^2+(s+2\Delta)H}{\hat{S}^2}\right)-\frac{1}{(\mu-s)^3}\stackrel{?}{>}\left|\left(1-\frac{\Delta^2}{\mu^2}\right)\left[\frac{1}{(4m_\phi^2-s-h(\mu))^3}\right]-\frac{1}{(\mu-s)^3}\right|
\end{equation}
within the region: $4\Delta<s<4m_\phi^2$, $0<\Delta\leq1$ and $\mu>\Lambda_2$. The expression inside the absolute value is in fact positive in this region and so we can remove the absolute value. The resulting inequality
\begin{equation}\label{eq:remainder_ineq}
    \left[\frac{1}{(\mu-H)^3}\right]\left(\frac{-\Delta^2+(s+2\Delta)H}{\hat{S}^2}\right)>\left(1-\frac{\Delta^2}{\mu^2}\right)\left[\frac{1}{(4m_\phi^2-s-h(\mu))^3}\right]
\end{equation}
can be checked using Mathematica's \texttt{Reduce} function which determines it to be satisfied in the given parameter region. Therefore, we can conclude that the sum of the final two lines in \eqref{eq:RHI} is positive and hence $I_{\rm RH}>0$ \textit{at least} in the region $4\Delta<s<4m_\phi^2$.\\

Moving on to the left-hand cuts we will encounter the same issue as for the $s'(s)=s$ case whereby the integrals do not neatly combine into an integral over $\operatorname{Im}\cG^{\times}(\mu)$. The expression for $I_{\rm LH}$ is,
\begin{equation}
    \begin{aligned} \label{IL_contr}
        I_{\rm LH}=\sfc_1\frac{2}{\pi} &\int_{\Lambda_1}^{\infty}\rd\mu\,\operatorname{Im} \cA_{\phi\phi}(\mu;1)\,\left[\frac{1}{(\mu+s-4m_\phi^2)^3}\right]\\+
        \sfc_2\frac{2}{\pi}&\int_{\Lambda_2}^{\infty}\rd\mu\operatorname{Im}\cA_{\phi\chi}(\mu;1)\left[\frac{1}{(\mu+H-\Sigma)^3}\right]\left(\frac{-\Delta^2(s+2\Delta)+(\hat{S}^2+2\Delta^2)H}{\hat{S}^3}\right)\\+
        \sfc_3\frac{2}{\pi} &\int_{\Lambda_1}^{\infty}\rd\mu\,\operatorname{Im} \cA_{\chi\chi}(\mu;1)\,\left[\frac{1}{(\mu+s-4m_\chi^2)^3}\right]\\
        +\sfc_b \frac{2}{\pi}&\int_{\Lambda_1}^{\infty}\rd\mu\,\operatorname{Im}\cA_t(\mu;1)\left[\frac{1}{(\mu+s-4m_\phi^2)^3}\right]\\
        +\sfc_t \frac{2}{\pi}&\int_{\Lambda_2}^{\infty} \mathrm{d} \mu \operatorname{Im} \mathcal{A}_{\phi \chi}(\mu ;-1)\left[\left(1-\frac{\Delta^2}{\mu^2}\right) \frac{1}{(s-h(\mu))^3}\right]\,.
    \end{aligned}
\end{equation}
By manipulating this expression we arrive at
\begin{equation}\label{eq:gen_lh_cut}
    \begin{aligned}
        \frac{\pi}{2}I_{\rm LH}= &\int_{\Lambda_1}^{\infty}\rd\mu\,\left\{\frac{\operatorname{Im} \cG^{\times}(\mu)}{(\mu+s-4m_\phi^2)^3}+\sfc_3\operatorname{Im} \cA_{\chi\chi}(\mu;1)\,\left[\frac{1}{(\mu+s-4m_\chi^2)^3}-\frac{1}{(\mu+s-4m_\phi^2)^3}\right]\right\}\\
        +&\int_{\Lambda_2}^{\infty}\rd\mu\,\left[\sfc_2K_f(\mu;s,\Delta)\operatorname{Im}\cA_{\phi\chi}(\mu;1)
        +\sfc_tK_b(\mu;s,\Delta)\operatorname{Im} \mathcal{A}_{\phi \chi}(\mu ;-1)\right]
    \end{aligned}
\end{equation}
with
\begin{equation}
\begin{aligned}
    K_f(\mu;s,\Delta)\equiv&\left[\frac{1}{(\mu+H-\Sigma)^3}\right]\left(\frac{-\Delta^2(s+2\Delta)+(\hat{S}^2+2\Delta^2)H}{\hat{S}^3}\right)-\frac{1}{(\mu+s-4m_\phi^2)^3}\,,\\
    K_b(\mu;s,\Delta)\equiv&\left[\left(1-\frac{\Delta^2}{\mu^2}\right) \frac{1}{(s-h(\mu))^3}\right]-\frac{1}{(\mu+s-4m_\phi^2)^3}\,,
\end{aligned}
\end{equation}
where we define the crossed superposition amplitude $\cG^{\times}$ as
\begin{equation} \label{crossedS1}
\cG^{\times}(s) = \cG(s) \big|_{\varphi_A \to - \varphi_A} \quad \propto \langle \psi^{\times} | T | \psi^{\times} \rangle \,.\\
\end{equation}
Taking $\varphi_A \to -\varphi_A$ as in the above defines a new state $|\psi^{\times}\rangle$ (as in \S\ref{sec:eqmass_proof}), and modifies the superposition amplitude only by interchanging the roles of $\sfc_b$ and $\sfc_t$ relative to $\cG(s)$ (see Eq.~(\ref{eq:anglecoeff})). Since this is an elastic matrix element its imaginary part in the physical region is also positive.

The first line of \eqref{eq:gen_lh_cut} is manifestly positive for $s>0$ since $m_\phi<m_\chi$, hence it remains only to show that the second line is positive. Given that $\sfc_2\geq|\sfc_t|$ and $\operatorname{Im}\cA_{\phi\chi}(\mu;1)\geq|\operatorname{Im}\cA_{\phi\chi}(\mu;-1)|$, for this integral to be positive it is sufficient that $K_f>K_b>0$ within the limits of integration. However in general $K_f-K_b$ is negative at low values of $\mu$ and switches sign to become positive as $\mu\xr\infty$,   therefore we cannot conclude that $I_{\rm LH}$ is positive. 

Similarly to \eqref{eq:modification_1} we may construct a positive quantity by adding to $I_{\rm LH}$ a positive multiple ‘$x(s,\Delta)$' of $\del_s^2\cA_{\phi\chi}(s;1)$; the question then becomes, how small can we make this multiple whilst keeping the observable positive. To this end consider the quantity $\widetilde I_{\rm LH}\equiv I_{\rm LH}+x(s,\Delta)\sfc_2\del_s^2\cA_{\phi\chi}(s;1)$, which will have the same dispersion relation as \eqref{eq:gen_lh_cut} however with the replacement $K_f\xr \widetilde K_f$ where
\begin{equation}
    \widetilde K_f\equiv K_f + x(s,\Delta)\left(\frac{1}{(\mu-s)^3}+\frac{1}{(\mu+s-\Sigma)^3}\right)\,.
\end{equation}
Now working in units where $m_\phi=1$ (so $\Sigma=4+2\Delta$) we set $x(s,\Delta)$ to be the smallest value that satisfies the following inequality for all $\mu>\Lambda_2=2+\Delta+2\sqrt{1+\Delta}$:
    \begin{equation}
        x(s,\Delta)>\frac{K_b-K_f}{\frac{1}{(\mu-s)^3}+\frac{1}{(\mu+s-\Sigma)^3}}\equiv R(\mu;s,\Delta,\Sigma)\,.
    \end{equation}
The above inequality implies $\widetilde I_{\rm LH}>0$ which in turn implies the superposition positivity bound:
\begin{equation}
\cO_2(s;\theta_{A,B},\varphi_{A,B})+\max_{\mu>(m_\phi+m_\chi)^2}\left\{R(\mu;s,\Delta,\Sigma)\right\}\sfc_2\,\del_s^2\cA_{\phi\chi}(s;1) > 0 \,.
\end{equation}
%%%%%%%

\section{Superposition bounds with IR integrals}
\label{App:Alt_Bound1}

As mentioned in \S\ref{sec:furthergen}, one can leverage superposition amplitudes like $\mathcal{G}$ in the main text in a different way, expressed as bounds involving $\partial_s^2\mathcal{G}(s)$ as well as some non-trivial IR integrals. We provide an example of such a bound here, which turns out to be
\begin{equation}\label{Gsbound_final_IRint}
\boxed{
     \partial_s^2 \cG(s) -  \sfc_2 \mathcal{I}^{\phi\chi}(s) + \bar r_{\max}(s) \cdot \sfc_2 \partial_s^2 \mathcal{A}_{\phi\chi}(s)  > 0  }
\end{equation}
where the IR integral is given by
\begin{eqnarray}
\mathcal{I}^{\phi\chi}(s) & := & \frac{2}{\pi} \int_{-4 \Delta}^{0} \exd\mu \; \frac{ \operatorname{Im} \mathcal{A}_{\phi\chi}\big(\Delta + \tfrac{\mu }{2}+\tfrac{\sqrt{\mu } \sqrt{\mu +4 \Delta }}{2} ; 1 \big)  }{(\mu - s)^3} \label{Iphichi_def}\\
& \ & \quad + \frac{2}{\pi} \int_{(m_\phi + m_\chi)^2}^{m_\phi^2 + 3 m_{\chi}^2} \exd\mu \; \bigg[ \overline{r}_{\max} \bigg( \frac{1}{(\mu-s)^3} + \frac{1}{(\mu-\Sigma+s)^3}  \bigg)  - \frac{ 1 - \frac{\Delta^2}{\mu^2} }{( - h(\mu) + s )^3 } \bigg]  \operatorname{Im} \mathcal{A}_{\phi\chi}( \mu;1) \notag
\end{eqnarray}
and one also defines 
\begin{equation}
\overline{r}_{\max}(s) \equiv \max_{\mu > m_\phi^2 + 3m_\chi^2} \overline{r}(\mu; s,\Delta,\Sigma ) \qquad \qquad \mathrm{with}\ \ \overline{r}(\mu; s, \Delta, \Sigma) \equiv \frac{ \frac{ 1 - {\Delta^2}/{\mu^2} }{( - h(\mu) + s )^3} - \frac{ 1 - {\Delta^2}/{(\mu - \Sigma)^2} }{( h(\Sigma - \mu) - [4m_\phi^2 - s] )^3} }{ \frac{1}{(\mu-s)^3} + \frac{1}{(\mu-[\Sigma-s])^3} }  \ .
\end{equation}
First we derive a dispersion relation for $\mathcal{G}(s)$, then using it to derive the above bound, and we end off with applying to the toy EFT.

\subsection{Dispersion relation for \texorpdfstring{$\wtg(s)$}{tilde-G}}

% In this section we provide the details of the proof given in \S\ref{sec:alt_bound_IRint} which make use of the generalized bound $\mathscr{G}(s,s')$ defined in Eq.~(\ref{Gdef}). We provide here the missing steps which give rise to the dispersion relation (\ref{S1bound}) for the amplitude \SJc{Re-write the prev. paragraph}\\

We start by defining a generalised superposition amplitude $\wtg$ as,
\begin{equation}\label{eq:Gtilde_terms}
    \wtg \equiv \mathscr{G}(s,s'(s))\,,\quad\text{where}\quad s'(s)=\frac{s}{2}+\Delta+\frac{1}{2} \sqrt{s} \sqrt{s+4 \Delta} 
\end{equation}
which importantly is defined in terms of $s'(s)$ as opposed to $H(s)$. Ultimately the bound (\ref{Gsbound_final_IRint}) above is expressed in terms of $\mathcal{G}$ (from Eq.~(\ref{eq:ghs_def}) in the main text) because the two amplitudes are the identical where the bound is evaluated. However the distinction becomes important in the derivation that follows since we will perform contour integrations in the complex $s$-plane where $s'(s)$ and $H(s)$ differ, since
\begin{eqnarray} \label{S1def_APP}
\wtg(s) & = &  \sfc_1 \; \mathcal{A}_{\phi\phi}( s ; 1  ) + \sfc_3  \; \mathcal{A}_{\chi\chi}\big( s;1 ) + \sfc_t \;  \mathcal{A}_{t}( s;1) + \sfc_2 \; \mathcal{A}_{\phi\chi}( s'(s) ; 1) + \sfc_b \; \mathcal{A}_{\phi\chi}( s'(s); -1 )  \ .
\end{eqnarray}
%
% \begin{eqnarray}
% \widetilde\cG(s) & \equiv & \mathscr{G}\big(s,h_{+}^{-1}(4m_\phi^2 - s)\big) \\
% & = &  \sfc_1 \; \mathcal{A}_{\phi\phi}( s ; 1   ) + \sfc_3  \; \mathcal{A}_{\chi\chi}\big( s ; 1 ) + \sfc_t \;  \mathcal{A}_{t}( s ; 1 ) \notag \\
% && \hspace{10mm} + \sfc_2 \; \mathcal{A}_{\phi\chi}\big( h_{+}^{-1}(4m_\phi^2 - s); 1 \big) + \sfc_b \; \mathcal{A}_{\phi\chi}\big( h_{+}^{-1}(4m_\phi^2 - s) ; -1 \big) \label{S1def_APP}
% \end{eqnarray}
% %
Our aim in this sub-section is to write this quantity (more precisely its second derivative) as a dispersion relation.
The strategy for deriving Eq.~(\ref{S1bound}) is to sum the individual dispersion relations for $\partial_s^2$ of the sub-amplitudes in Eq.~(\ref{S1def_APP}). The contributions from the right-hand branch cuts naturally combine into an integral over $\operatorname{Im}\cG(\mu)$ with the kernel $(\mu - s)^{-3}$, ensuring that this term is manifestly positive (see the first term of Eq.~(\ref{S1bound})). This initial step follows the same strategy outlined in \S\ref{sec:superproof_1}. The key difference here is that the sub-amplitude dispersion relations are manipulated so that the contributions from the left-hand branch cuts integrate over $\operatorname{Im}\cG^{\times}_{1}(\mu)$ with the kernel $(\mu - 4m_\phi^2 + s)^{-3}$, making the positivity of this term equally transparent (see the second term of Eq.~(\ref{S1bound})). The remaining terms in Eq.~(\ref{S1bound}) arise as residual contributions from the individual dispersion relation, which crucially only involve elastic amplitudes. 

For example, one can begin with Eq.~(\ref{tchan_Adisp}) from the main text,
\begin{equation}
\cA_t(s;1) = \frac{1}{\pi} \int_{4 m_\phi^2}^{\infty}\rd\mu \; \frac{\operatorname{Im} \cA_t(\mu;1)}{\mu-s}+\frac{1}{\pi} \int_{-\infty}^0\rd\mu \; \frac{\operatorname{Im} \cA_t(\mu;1)}{\mu-s} +\frac{1}{2 \pi i} \int_{|\mu|\xr\infty}\rd\mu\;  \frac{\cA_t(\mu;1)}{\mu-s} \ , 
\end{equation}
repeated here with the integral at infinity written explicitly. Applying two $s$-derivatives to this expression, one can drop the integral at infinity, and then change variables $\mu \to 4m_\phi^2 - \mu$ in the second integral over the left-hand branch cut to arrive at the expression
\begin{eqnarray} \label{funnyAt}
\partial_s^2 \cA_t(s; 1) & = & \frac{2}{\pi} \int_{4 m_\phi^2}^{\infty}\rd\mu \; \frac{\operatorname{Im} \cA_t(\mu; 1)}{(\mu-s)^3}+\frac{2}{\pi} \int_{4m_\phi^2}^{\infty} \rd\mu \; \frac{\operatorname{Im} \cA_{\phi\chi}\big( s'(\mu); - 1 \big)}{(\mu - 4m_\phi^2 + s)^3} 
\end{eqnarray}
where we have used the crossing relation $\mathcal{A}_{t}(s,t,u) = \mathcal{A}_{\phi\chi}(t,s,u)$ to relate $\mathcal{A}_{t}$ to the backwards sub-amplitude in the last term.  

One also has the standard dispersion relations $\mathcal{A}_{\phi\phi}$ and $\mathcal{A}_{\chi\chi}$ for the elastic amplitudes, whose expression for $\partial_s^2$ we list for completeness:
\begin{eqnarray} 
\partial_s^2 \mathcal{A}_{\phi\phi}\big( s ; 1 \big) & = & \frac{2}{\pi} \int_{4m_\phi^2}^{\infty} \exd\mu \; \frac{ \operatorname{Im} \mathcal{A}_{\phi\phi}( \mu ; 1) }{(\mu - s)^3} +  \frac{2}{\pi} \int_{4m_\phi^2}^{\infty} \exd\mu \; \frac{ \operatorname{Im} \mathcal{A}_{\phi\phi}( \mu ; 1 ) }{(\mu - 4m_\phi^2 + s )^3} \label{ordinaryphiphi} \\
\partial_s^2 \mathcal{A}_{\chi \chi}( s ; 1 ) & = & \frac{2}{\pi} \int_{4m_\phi^2}^{\infty} \exd\mu \; \frac{ \operatorname{Im} \mathcal{A}_{\chi\chi}( \mu ; 1) }{(\mu - s)^3} +  \frac{2}{\pi} \int_{4m_\phi^2}^{\infty} \exd\mu \; \frac{ \operatorname{Im} \mathcal{A}_{\chi\chi}(  \mu ; 1 ) }{(\mu - 4m_\chi^2 + s)^3} \label{ordinarychichi}
\end{eqnarray}
We now require dispersion relations for the remaining two sub-amplitudes $\mathcal{A}_{\phi\chi}\big( s'(s); \pm 1 \big)$ in Eq.~(\ref{S1def_APP}). One possible approach is to start from the dispersion relations for $\mathcal{A}_{\phi\chi}(s,\pm 1)$ derived earlier, and substitute $s \to s'(s)$ explicitly, and then differentiate, which is what was in done in the main text. 

An alternative route to derive a dispersion relation is to consider $\mathcal{A}_{\phi\chi}(s'(s),\pm 1)$ as a function of $s$ and use Cauchy's integral formula directly in the $s$ plane. As mentioned earlier, in this approach the distinction between $s'(s)$ and $H(s)$ given by,
\begin{equation}
\begin{aligned}
    s'(s)&=\frac{s}{2}+\Delta+\frac{1}{2} \sqrt{s} \sqrt{s+4 \Delta}\\
    H(s)&=\frac{s}{2}+\Delta+\frac{1}{2} \sqrt{s(s+4 \Delta)}\,.
\end{aligned}
\end{equation}
becomes important as they are not equal to one another in the whole complex $s$ plane. For real values of $s>-4\Delta$ they are equal. The two amplitudes under consideration explicitly the forward and backwards limits:
\begin{eqnarray}
 \mathcal{A}_{\phi\chi}\big( s'(s); 1\big) & = & \mathcal{A}_{\phi\chi}\big( \; \Delta + \tfrac{ s }{2}+\tfrac{\sqrt{ s } \sqrt{ s +4 \Delta }}{2}  \; , 0 \; , \; \Sigma - [\Delta + \tfrac{ s }{2}+\tfrac{\sqrt{ s } \sqrt{ s +4 \Delta }}{2}] \; \big) \\
 \mathcal{A}_{\phi\chi}\big( s'(s); -1 \big) & = & \mathcal{A}_{\phi\chi}\big( \; \Delta + \tfrac{ s }{2}+\tfrac{\sqrt{ s } \sqrt{ s +4 \Delta }}{2}  \; , \; 4m_\phi^2 - s \; , \; \Delta + \tfrac{ s }{2} - \tfrac{\sqrt{ s } \sqrt{ s +4 \Delta }}{2} \; \big) \label{backwards_Mandelstams}
\end{eqnarray}

\paragraph{Forwards limit:}The standard branch cuts of the forward limit amplitude lie at $s'(s) > (m_\phi+m_\chi)^2$ and $\Sigma-s'(s) > (m_\phi+m_\chi)^2$ which correspond to $s > 4m_\phi^2$ and $s<-4\Delta$. There is a further discontinuity in this function due to the branch cut of $s'(s)$ itself which lies along $-4\Delta<s<0$. Taking two derivatives with respect to $s$ then gives 
\begin{equation}
\partial_s^2 \cA_{\phi\chi}(s'(s);1) = \frac{2}{\pi} \int_{4 m_\phi^2}^{\infty}\rd\mu \; \frac{\operatorname{Im} \cA_{\phi\chi}(s'(\mu) ;1)}{(\mu-s)^3} + \frac{2}{\pi} \bigg[ \int_{-\infty}^{-4\Delta} + \int_{-4\Delta}^0  \bigg] \rd\mu  \; \frac{\operatorname{Im}\cA_{\phi\chi}(s'(\mu) ;1)}{(\mu-s)^3} \ .
\end{equation}
We isolate the integral from $-4\Delta<\mu<0$ and denote it by $\mathcal{I}_{2}^{\phi\chi}$, and in the remainder the change of variable $\mu \to 4m_\phi^2 - \mu$ yields
\begin{small}
\begin{equation} \label{funnyphichi}
\partial_s^2 \cA_{\phi\chi}\big(s'(s);1\big) = \frac{2}{\pi} \int_{4 m_\phi^2}^{\infty}\rd\mu \; \frac{\operatorname{Im} \cA_{\phi\chi}\big(s'(\mu);1\big)}{(\mu-s)^3} + \mathcal{I}_{2}^{\phi\chi} + \frac{2}{\pi} \int_{4m_\chi^2}^{\infty} \rd\mu  \; \frac{\operatorname{Im}\cA_{\phi\chi}\big(\Sigma -s'(4m_\phi^2-\mu-\ri\epsilon);1\big)}{(\mu - 4m_\phi^2 +s)^3}\ ,
\end{equation}
\end{small}\ignorespaces
Note that $\mathcal{I}_{2}^{\phi\chi}$ can also be expressed naturally as a circular contour integral of radius $\Delta+\epsilon$ centered at the origin:
\begin{equation}
\mathcal{I}_{2}^{\phi\chi}(s) = \frac{2}{\pi} \int_{-4 \Delta}^{0} \exd\mu \; \frac{ \operatorname{Im} \mathcal{A}_{\phi\chi}\big(s'(\mu) ;1  \big)  }{(\mu - s)^3} =\frac{2!}{2\pi i} \oint_{|\mu| = \Delta+\epsilon} \exd \mu\;  \frac{\left(1 - \frac{\Delta^2}{\mu^2} \right)\mathcal{A}_{\phi\chi}(\mu; 1)}{\big( s - 4m_\phi^2 + h(\mu) \big)^3} \ . \label{I2circle}
\end{equation}
If a tree-level amplitude (i.e. a polynomial) is inserted into this integral it will evaluate to a non-zero answer despite the amplitude not having any non-analyticity itself. This is essentially an artefact of inserting a non-analytic function (i.e. $s'(s)$) into an analytic function $\cA_{\rm tree}$.

\paragraph{Backwards limit:} 
We can immediately use crossing to relate this amplitude to $\cA_t$ as follows:
\begin{equation}
    \cA_{\phi\chi}(s'(s);-1)=\cA_t(4m_\phi^2-s;+1)\,.
\end{equation}
Then since $\del_s^2[\cA_t(4m_\phi^2-s;1)]=\del_s^2[\cA_t](4m_\phi^2-s;1)$ by the chain rule we can use \eqref{funnyAt} to directly obtain
\begin{equation}
   \del_s^2[\cA_{\phi\chi}(s'(s);-1)] =\frac{2}{\pi} \int_{4 m_\phi^2}^{\infty}\rd\mu \; \frac{\operatorname{Im} \cA_t(\mu; 1)}{(\mu+s-4m_\phi^2)^3}+\frac{2}{\pi} \int_{4m_\phi^2}^{\infty} \rd\mu \; \frac{\operatorname{Im} \cA_{\phi\chi}\big( s'(\mu); - 1 \big)}{(\mu -s)^3} \,.
\end{equation}

\paragraph{Total expression:}Having obtained integral expressions for each term in \eqref{eq:Gtilde_terms} we can massage their sum into
\begin{eqnarray} 
\partial_s^2 \wtg(s) & = & \frac{2}{\pi} \int_{4m_\phi^2}^{\infty} \exd\mu \; \frac{\operatorname{Im} \wtg(\mu) }{(\mu - s)^3}  +  \frac{2}{\pi} \int_{4m_\phi^2}^{\infty} \exd\mu \; \frac{ \operatorname{Im} \wtg^{\times}(\mu) }{(\mu+s-4m_\phi^2 )^3} \\
& \ & + \frac{2 \sfc_3}{\pi} \int_{4m_\phi^2}^{\infty} \exd\mu \;  \frac{\operatorname{Im} \mathcal{A}_{\chi\chi}( \mu ; 1 )}{(\mu - 4m_\chi^2 +s )^3} - \frac{2 \sfc_3}{\pi} \int_{4m_\phi^2}^{\infty} \exd\mu \; \frac{ \operatorname{Im} \mathcal{A}_{\chi\chi}( \mu ; 1 ) }{(\mu - 4m_\phi^2 + s)^3}    \notag \\
&& + \frac{2 \sfc_2}{\pi} \int_{4m_\chi^2}^{\infty} \exd \mu\; \frac{ \operatorname{Im} \mathcal{A}_{\phi\chi}( \Sigma -s'(4m_\phi^2-\mu) ; 1 ) }{(\mu - 4m_\phi^2 + s)^3} - \frac{2 \sfc_2}{\pi} \int_{4m_\phi^2}^{\infty} \exd \mu\; \frac{ \operatorname{Im} \mathcal{A}_{\phi\chi}( s'(\mu) ; 1 ) }{(\mu - 4m_\phi^2 + s)^3} \notag \\
&& + \sfc_2 \mathcal{I}^{\phi\chi}_2(s) \notag
\end{eqnarray}
where $\wtg^{\times}(s)$ is defined by swapping $\sfc_b$ and $\sfc_t$ in $\wtg(s)$, and itself also corresponds to an elastic $S$-matrix element. The terms involving $\mathcal{A}_{\phi\phi}(\mu;1)$, $\mathcal{A}_{t}(\mu;1)$, and $\mathcal{A}_{\phi\chi}\big(s'(\mu);-1\big)$ align perfectly in the expressions above as they appear in $\cG^{\times}(s)$. The remaining two sub-amplitudes are however not in the precise form required as defined in $\cG^{\times}(s)$. To correct this, one adds and subtracts the appropriate terms --- this is why the subtracted contributions appear in the final terms of lines 2 and 3 above.

Changing variables such that $\mu \to \Sigma - h_{-}^{-1}(\mu)$ in the first and $\mu \to h_{+}^{-1}(4m_\phi^2 - \mu)$ in the second term of line 3 gives
% \begin{eqnarray} 
% \partial_s^2 \wtg(s) & = & \frac{2}{\pi} \int_{4m_\phi^2}^{\infty} \exd\mu \; \frac{\operatorname{Im} \wtg(\mu) }{(\mu - s)^3}  \qquad +  \frac{2}{\pi} \int_{4m_\phi^2}^{\infty} \exd\mu \; \frac{ \operatorname{Im} \wtg^{\times}(\mu) }{(\mu+s-4m_\phi^2 )^3} \\
% & \ & + \frac{2 \sfc_3}{\pi} \int_{4m_\phi^2}^{\infty} \exd\mu \;  \frac{\operatorname{Im} \mathcal{A}_{\chi\chi}( \mu ; 1 )}{(\mu - 4m_\chi^2 +s )^3} - \frac{2 \sfc_3}{\pi} \int_{4m_\phi^2}^{\infty} \exd\mu \; \frac{ \operatorname{Im} \mathcal{A}_{\chi\chi}( \mu ; 1 ) }{(\mu - 4m_\phi^2 + s)^3}    \notag \\
% &&  + \frac{2 \sfc_2}{\pi}   \int_{m_\phi^2 + 3 m_{\chi}^2}^{\infty} \exd\mu \; \frac{\big( 1 - \frac{\Delta^2}{(\mu - \Sigma)^2} \big) \operatorname{Im} \mathcal{A}_{\phi\chi}( \mu ; 1) }{\big( h(\Sigma - \mu) - 4m_\phi^2 + s \big)^3}  - \frac{2 \sfc_2}{\pi}   \int_{(m_\phi+m_\chi)^2}^{\infty} \exd\mu \; \frac{ \big( 1 - \frac{\Delta^2}{\mu^2} \big) \operatorname{Im} \mathcal{A}_{\phi\chi}( \mu ; 1)  }{( - h(\mu) + s )^3}   \notag \\
% && + \sfc_2 \mathcal{I}^{\phi\chi}_2(s) \notag
% \end{eqnarray}
% Now one notices that the lower limit of the two integrals in line 3 do not align --- simply cut off the lower end of the latter integral to define $\mathcal{I}_{1}^{\phi\chi}$ in Eq.~(\ref{I1_phichi}) and one is left with
\begin{eqnarray} 
\partial_s^2 \wtg(s) & = & \frac{2}{\pi} \int_{4m_\phi^2}^{\infty} \exd\mu \; \frac{ \operatorname{Im} \wtg(\mu) }{(\mu - s)^3}  +  \frac{2}{\pi} \int_{4m_\phi^2}^{\infty} \exd\mu \; \frac{ \operatorname{Im} \wtg^{\times}_{1}(\mu) }{(\mu+s-4m_\phi^2 )^3} \label{S1bound} \\
& \ & + \frac{2 \sfc_3}{\pi} \int_{4m_\phi^2}^{\infty} \exd\mu \; \bigg( \frac{1}{(\mu - 4m_\chi^2 + s)^3} - \frac{ 1 }{(\mu - 4m_\phi^2 + s)^3} \bigg) \operatorname{Im} \mathcal{A}_{\chi\chi}(  \mu; 1 )   \notag \\
&& + \frac{2 \sfc_2}{\pi}   \int_{m_\phi^2 + 3 m_{\chi}^2}^{\infty} \exd\mu \;   \bigg( \frac{ 1 - \frac{\Delta^2}{(\mu - \Sigma)^2} }{\big( h(\Sigma - \mu) - 4m_\phi^2 + s \big)^3} - \frac{ 1 - \frac{\Delta^2}{\mu^2} }{( - h(\mu) + s )^3}   \bigg) \operatorname{Im} \mathcal{A}_{\phi\chi}( \mu ; 1) \notag \\
&& -  \sfc_2 \mathcal{I}^{\phi\chi}_{1}(s)  + \sfc_2 \mathcal{I}^{\phi\chi}_2(s) \notag
\end{eqnarray}
where the new IR integral is defined:
\begin{eqnarray}
\mathcal{I}^{\phi\chi}_{1}(s) & \equiv & \frac{2}{\pi}  \int_{(m_\phi + m_\chi)^2}^{m_\phi^2 + 3 m_{\chi}^2} \exd\mu \; \bigg( 1 - \frac{\Delta^2}{\mu^2} \bigg) \frac{ \operatorname{Im} \mathcal{A}_{\phi\chi}( \mu;1) }{( - h(\mu) + s )^3} \label{I1_phichi}
\end{eqnarray}
Notice that in the limit $\Delta \to 0$ the last three lines of Eq.~(\ref{S1bound}) vanish as expected. Note, this integral is positive. 

\subsection{Positivity bound on \texorpdfstring{$\wtg(s)$}{tilde-G}}
\label{sec:alt_bound_IRint}
Since $\mathrm{Im}\wtg$ and $\mathrm{Im}\wtg^{\times}$ are positive within the limits of the integrals in which they appear, the first line of equation \eqref{S1bound} is positive when $0< s < 4m_\phi^2$. Furthermore,
\begin{equation}
  \begin{split}
    &\frac{1}{(\mu+s-4m_\chi^2)^{3}} - \frac{1}{(\mu+s-4m_\phi^2)^{3}} > 0 \\
    &\frac{ 1 - \frac{\Delta^2}{(\mu - \Sigma)^2} }{( h(\Sigma - \mu) - [4m_\phi^2 - s] )^3} - \frac{ 1 - \frac{\Delta^2}{\mu^2} }{( - h(\mu) + s )^3} < 0
  \end{split}
\hspace{20mm}
  \begin{split}
   & \mathrm{when}\ 4 \Delta < s < 4m_\phi^2
  \end{split}
\end{equation}
so that the second line of \eqref{S1bound} is positive and the third negative. Ignoring the two IR integrals for the moment, the only negative term in $\partial_s^2 \wtg(s) $ is coming from an integral over $\operatorname{Im}\cA_{\phi\chi}$, implying that if we add a sufficiently large positive multiple of $\del_s^2\cA_{\phi\chi}(s;1)$ we will obtain something positive. The strongest such positivity bound is given by finding the smallest value of the multiple, denoted $x(s)$, that satisfies
% %
% \begin{equation} \label{addx_phichi}
% x(s) \cdot \sfc_2 \partial_s^2 \mathcal{A}_{\phi\chi} = x(s) \cdot \frac{2\sfc_2 }{\pi} \int_{(m_\phi+m_\chi)^2}^\infty \exd \mu \; \bigg( \frac{1}{(\mu-s)^3} + \frac{1}{(\mu-\Sigma+s)^3}  \bigg) \operatorname{Im} \cA_{\phi\chi}(\mu;1)
% \end{equation}
% %
% to both sides of the relation (\ref{S1bound}) with the quantity $x(s)$ chosen so that
({\it cf.} Eq.~(\ref{firstxs}))
\begin{equation}
x(s) \bigg( \frac{1}{(\mu-s)^3} + \frac{1}{(\mu-[\Sigma-s])^3}  \bigg) +  \frac{ 1 - \frac{\Delta^2}{(\mu - \Sigma)^2} }{( h(\Sigma - \mu) - [4m_\phi^2 - s] )^3} - \frac{ 1 - \frac{\Delta^2}{\mu^2} }{( - h(\mu) + s )^3} > 0 
\end{equation}
In other words, we want to find $\overline{r}_{\max} \leq x(s)$ such that
\begin{equation}
\overline{r}_{\max}(s) \equiv \max_{\mu > m_\phi^2 + 3m_\chi^2} \overline{r}(\mu; s ) \qquad \mathrm{with}\qquad \overline{r}(\mu; s, \Delta, \Sigma) \equiv \frac{ \frac{ 1 - {\Delta^2}/{\mu^2} }{( - h(\mu) + s )^3} - \frac{ 1 - {\Delta^2}/{(\mu - \Sigma)^2} }{( h(\Sigma - \mu) - [4m_\phi^2 - s] )^3} }{ \frac{1}{(\mu-s)^3} + \frac{1}{(\mu-[\Sigma-s])^3} } 
\end{equation}
which is maximized over the range of the integral in the third line of (\ref{S1bound}). A more subtle detail is that the standard dispersion relation for $\mathcal{A}_{\phi\chi}$ integrates over $\mu \geq (m_\phi + m_\chi)^2$, while the third line of Eq.~(\ref{S1bound}) whose negativity we are compensating integrates $\mathcal{A}_{\phi\chi}$ for $\mu > m_\phi^2 + 3m_\chi^2$. It is then necessary to split apart integration bounds defining a third IR integral
\begin{equation} \label{I3_phichi}
\mathcal{I}_{3}^{\phi\chi} \equiv \frac{2}{\pi} \int_{(m_\phi+m_\chi)^2}^{m_\phi + 3m_\chi^2} \exd \mu \; \bigg( \frac{1}{(\mu-s)^3} + \frac{1}{(\mu-\Sigma+s)^3}  \bigg) \operatorname{Im} A_{\phi\chi}(s;1)
\end{equation}
which is positive. A positivity bound can be immediately obtained by simply subtracting all three IR integrals from $\del_s^2\wtg(s)$, leaving only positive integrals. Furthermore, if we take $ 4\Delta < s < 4m_\phi^2$ where there is no difference between $s'(s)$ and $H(s)$ one can replace the superposition amplitude $\wtg$ with $\mathcal{G}$ from Eq.~(\ref{eq:ghs_def}). Defining the IR integral $\mathcal{I}^{\phi\chi}(s) = - \mathcal{I}^{\phi\chi}_1(s) + \mathcal{I}^{\phi\chi}_2(s) + \overline{r}_{\max} \mathcal{I}^{\phi\chi}_{3}(s) $ to get (\ref{Iphichi_def}), one then gets the bound (\ref{Gsbound_final_IRint}) quoted above.

%%%%

\subsection{Another application to toy EFT}
\label{sec:AltBound_EFT} 

When applied to a tree-level amplitude, the only IR integral that is non-zero is $\mathcal{I}^{\phi\chi}_2(s)$. This is because the amplitude is composed with a non-analytic function, hence the integral arises from a ``kinematic cut'' i.e. the cut of $s'(s)$, as opposed to a threshold branch cut. As we show below, the combination of the IR integral and the derivative of the superposition amplitude is simple if one is considering the amplitude up to order $s^2$ in the Mandelstam variables.

Inserting the tree-level amplitude of the theory \eqref{EFTaction} into $\sfc_2 \partial_s^2 \mathcal{A}_{\phi\chi}(H(s);1) \subset \partial_s^2 \cG(s)$ gives
\begin{small}
\begin{equation}
\partial_s^2 \mathcal{A}_{\phi\chi}\big( H(s);1\big) \big|_{\mathrm{EFT}} = 
\lambda_1 +\frac{\lambda_1 \sqrt{ s ( 4 \Delta + s )}}{8} \left(\frac{\Sigma -2 \Delta }{s^2}+\frac{8 \Delta -\Sigma }{2 \Delta  s}+\frac{2 \Delta +\Sigma }{(4 \Delta +s)^2}+\frac{8 \Delta +\Sigma }{2 \Delta  (4 \Delta +s)}\right)
\end{equation}
\end{small}\ignorespaces
whilst the IR integral gives:
\begin{equation}
\mathcal{I}_2^{\phi\chi}(s) |_{\mathrm{EFT}} = 
- \lambda_1 +\frac{\lambda_1 \sqrt{ s ( 4 \Delta + s )}}{8} \left(\frac{\Sigma -2 \Delta }{s^2}+\frac{8 \Delta -\Sigma }{2 \Delta  s}+\frac{2 \Delta +\Sigma }{(4 \Delta +s)^2}+\frac{8 \Delta +\Sigma }{2 \Delta  (4 \Delta +s)}\right) \ .
\end{equation}
The other two IR integrals $\mathcal{I}_{1,3}^{\phi\chi}(s) |_{\mathrm{EFT}} = 0$ for tree-level (polynomial) amplitudes. Curiously one then finds that
\begin{equation}
\big[ \partial_s^2 \mathcal{A}_{\phi\chi}\big(H(s);1\big) - \mathcal{I}^{\phi\chi}(s) \big] |_{\mathrm{EFT}} \ = \ 2 \lambda_1
\end{equation}
which is completely independent of $s$ and $\Delta$. Whilst the cancellation between these two terms seems remarkable we emphasise that this is only due to us truncating EFT amplitude to order $s^2$ and the general result for the above quantity for a polynomial $\cA_{\phi\chi}(s;1)$ is,
\begin{equation}\begin{aligned}
    &\big[ \partial_s^2 \mathcal{A}_{\phi\chi}\big(H(s);1\big) - \mathcal{I}^{\phi\chi}(s) \big] |_{\mathrm{EFT}}  = \\&\frac{1}{2S^3}\left[4\Delta^2(\cA'_- - \cA'_+)+(\cA''_{+}-\cA''_{-})(s^3+6s^2\Delta+8s\Delta^2)+(\cA''_+ +\cA''_-)S(2\Delta^2 +4\Delta s +s^2 )\right]
    \end{aligned}
\end{equation}
where we have defined $S\equiv \sqrt{s(s+4\Delta)}$ and $\cA'_\pm$ is the first derivative of $\cA_{\phi\chi}(s)$ evaluated at $h^{-1}_{\pm}(4m_\phi^2-s)$ and $\cA''_\pm$ is the second derivative similarly evaluated.

Returning to the quadratic EFT amplitudes we end up with the positivity bound:
\begin{equation}
    \begin{aligned} 
        (8\lambda_\phi) \sfc_1+(8\lambda_\chi) \sfc_3+(2\lambda_1)\sfc_2\left(1+\overline{r}_{\max} \right)>-(\sfc_t+\sfc_b)\left(\lambda_1+2\lambda_2\right)\,,
    \end{aligned}
\end{equation}
which is precisely of the same form as Eq.~(\ref{eq:eft_superbound2}) but with $\mathfrak{C}(s,\Delta) \to \overline{r}_{\max}$. The value of $\overline{r}_{\max}$ is achieved in the same manner as in \S\ref{sec:EFTprobe1} and \S\ref{sec:gen_bound_1}, where at the largest value of $\Delta/m_\phi^2 = 1$ one finds a value of $\bar r_{\max} \simeq 0.160$ (where we set $s=4m_\phi^2$). This gives a very slightly stronger bound when applied to the EFT at tree level than (\ref{eq:eft_superbound2}), and results in couplings that look almost identical to those depicted in Figure \ref{Fig:Compare2}.

%%%%%%%%%%%%%%%%%%%%%%%%%%%%%%%
\section{Toy EFT details}
\label{App:ToyEFT}

This appendix collects some useful results to do with the toy EFT given first in Eq.~(\ref{EFTaction}), included here with all operators up to dimension-8:
\begin{eqnarray}
\mathcal{L}_{\mathrm{EFT}} & \simeq & - \frac{1}{2} (\partial \phi)^2 - \frac{1}{2} m_\phi^2 \phi^2 - \frac{1}{2} (\partial \chi)^2 - \frac{1}{2} m_\chi^2 \chi^2 \quad + a_1 \phi^4 + a_2 \phi^2 \chi^2 + a_3 \chi^4  + b  \phi \chi (\partial \phi \cdot \partial \chi ) \qquad \quad \label{EFT_APPENDIX} \\
&\ & \qquad \qquad + \lambda_\phi (\partial \phi )^4 + \lambda_1 ( \partial \phi \cdot \partial \chi )^2 + \lambda_2 (\partial \phi)^2 (\partial \chi)^2 + \lambda_\chi (\partial \chi )^4  \notag
\end{eqnarray}
As assumed in \S\ref{sec:setup}, the action is invariant under $\mathbb{Z}_2\times\mathbb{Z}_2$ and so generically this action includes three dimension-4 operators and a single dimension-6 operator. Note however these lower-dimension operators do not contribute to the bounds considered so we ignore them for ease of presentation in the main text. From the above EFT we can derive the tree level amplitudes coming from the $\lambda_j$ interactions where (only considering contributions from dimension-8 operators):
\begin{eqnarray}
\mathcal{A}_{\phi\phi}(s,t,u) & = & 2 \lambda_\phi \; \big[ ( 2m_\phi^2 - s )^2 + ( 2m_\phi^2 - t )^2 + ( 2m_\phi^2 - u )^2 \big] \label{Aphiphi_EFT} \\
\mathcal{A}_{\chi\chi}(s,t,u) & = & 2 \lambda_\chi  \; \big[ ( 2m_\chi^2 - s )^2 + ( 2m_\chi^2 - t )^2 + ( 2m_\chi^2 - u )^2 \big] \label{Achichi_EFT}  \\
\mathcal{A}_{\phi\chi}(s,t,u) & = & \tfrac{\lambda_{1}}{2} \big[  (m_\phi^2 +m_\chi^2-s)^2 +  (m_\phi^2 +m_\chi^2-u)^2 \big] + \lambda_{2} (2m_\phi^2 - t)(2m_\chi^2 - t) \label{Aphichi_EFT}
\end{eqnarray}
With $\mathcal{A}_{u}$ and $\mathcal{A}_{t}$ also trivially obtained from (\ref{Aphichi_EFT}) using crossing relations. 

\subsection{Partial UV completion}
\label{App:partialUV}

To give some intuition about what kind of effective couplings to expect, it is amusing to derive the above EFT from a partial UV completion. A simple theory of this sort is
\begin{small}
\begin{eqnarray}
\mathcal{L} & \simeq &- \frac{1}{2} (\partial \phi)^2 - \frac{1}{2} m_\phi^2 \phi^2 - \frac{1}{2} (\partial \chi)^2 - \frac{1}{2} m_\chi^2 \chi^2 \label{partialUV} \\
& \ &   - \frac{1}{2} (\partial H_1 )^2 - \frac{1}{2} m_\phi^2 H_1^2 - \frac{1}{2} (\partial H_2 )^2 - \frac{1}{2} m_\chi^2 H_2^2  - \frac{1}{2} (\partial H_3 )^2 - \frac{1}{2} M_3^2 H_3^2 - \frac{1}{4} F_{\mu\nu} F^{\mu \nu} - \frac{1}{2} M_{4}^2 A_\mu A^\mu \notag \\
& \ & +  ( \alpha_1 m_\phi \phi^2 + \alpha_2 m_\phi \chi^2 ) H_1 \ + \beta m_\chi \phi \chi H_2 \ + \gamma M_3 \phi^2 H_3 \  + \rho A_{\mu} \partial^\mu \phi \chi  \ . \notag
\end{eqnarray}
\end{small}\ignorespaces
which generalizes a UV completion introduced in \cite{Ye:2024rzr}. This action incorporates three heavy scalars $H_{1}, H_2, H_3$ and a heavy vector $A_{\mu}$ (with corresponding field strength $F_{\mu\nu} = \partial_\mu A_{\nu} - \partial_\nu A_{\mu}$), all of which are integrated out at tree level. The vector field is included here, since integrating out heavy scalars only gives rise to positive values of $\lambda_2$. Note that all couplings $\alpha_1$, $\alpha_2$, $\beta$, $\gamma$ and $\rho$ are dimensionless in the conventions used.

Integrating out the four heavy fields $H_{1}$, $H_{2}$, $H_{3}$ and $A_{\mu}$ at tree-level is then a straightforward exercise. Ignoring the six-point and higher vertices generated in this process, one finds that Eq. (\ref{partialUV}) produces an EFT of the form of Eq.~(\ref{EFT_APPENDIX}) with couplings $\lambda_j$ of the form:
\begin{equation} 
\lambda_\phi = \frac{2\alpha_1^2}{m_\phi^4} + \frac{2\gamma^2}{M_3^4} \ , \qquad \lambda_\chi = \frac{2\alpha_2^2}{m_\phi^4} \ , \qquad \lambda_1 = \frac{2\beta^2}{M_{2}^4} + \frac{\rho^2}{ 2 M_{4}^4 } \ , \qquad \lambda_2 = \frac{4 \alpha_1 \alpha_2}{m_\phi^4} - \frac{\rho^2}{ 2 M_{4}^4 } \label{couplinglambda2UV}
\end{equation}
We also include the lower  dimension operators for completeness:
\begin{eqnarray}
a_1 & = & \alpha_1^2 \bigg(  \frac{1}{2} + \frac{2m_\phi^2}{3 m_\phi^2} + \frac{2m_\phi^4}{3m_\phi^4} \bigg) + \gamma^2 \bigg(  \frac{1}{2} + \frac{2m_\phi^2}{3 M_3^2} + \frac{2 m_\phi^4}{3M_3^4} \bigg)   \\
a_2 & = & \alpha_1 \alpha_2 \bigg( 1 - \frac{4m_\phi^2 m_\chi^2}{M_{1}^4} \bigg) + \beta^2 \bigg(  \frac{1}{2} + \frac{m_\phi^2 + m_\chi^2}{2m_\chi^2} + \frac{(m_\phi^2 + m_\chi^2)^2}{2m_\chi^4} \bigg) - \frac{\rho^2 m_\phi^2}{2M_{4}^2} \\
a_3 & = & \alpha_2^2 \bigg(  \frac{1}{2} + \frac{2m_\chi^2}{3 m_\phi^2} + \frac{2m_\chi^4}{3m_\phi^4} \bigg) \\
b & = & - \frac{ 4 \alpha_1 \alpha_2}{m_\phi^2} \bigg( 1 + \frac{2 m_\phi^2 + 2 m_\chi^2}{M_{1}^4} \bigg)  + \frac{\beta^2}{m_\chi^2} \bigg( 1 + \frac{2m_\phi^2 + 2m_\chi^2}{m_\chi^2} \bigg) - \frac{\rho^2}{M_4^2}
\end{eqnarray}
It is clear that $c$, $d$, and $\lambda_1$ must be nonnegative since they are sums of squares of UV couplings. Additionally, one can check that $(\lambda_1 + 2 \lambda_2)^2 < (\lambda_1 + 4 \sqrt{\lambda_\phi \lambda_\chi})^2$, which leads to the last constraint on $\lambda_2$ in Eq.~(\ref{tightestsec3}). This shows that the partial UV completion fully covers the parameter space described by Eq.~(\ref{tightestsec3}). Interestingly this partial UV completion has couplings $\lambda_j$ which have no dependence on the mass differences of the light particles.

\subsection{Extremizing the superposition bound}
\label{App:EqMass}

We here consider the derived superposition bounds applied to the toy EFT, and extract the strongest bounds. The most general bound which encapsulates all the bounds (\ref{eqMassEFT-b}), (\ref{eq:eft_superbound}), and (\ref{eq:eft_superbound2}) is of the form 
%
%\begin{small}
\begin{eqnarray} \label{appboundstart} 
 && 8 \cos^2 \theta_A  \cos^2 \theta_B\; \lambda_\phi  (1+r) + \big[ 1 - \cos2\theta_{A} \cos 2 \theta_B \big] \; \lambda_{1} (1+\mathfrak{C}) + 8 \sin^2  \theta_A  \sin^2 \theta_B \; \lambda_\chi \\
 && \hspace{45mm} > \ - \tfrac{ 1}{2} \big[ p \cos ( \varphi_A - \varphi_B)  + \cos ( \varphi_A + \varphi_B)  \big] \sin 2 \theta_A  \sin 2 \theta_B\; ( \lambda_1 + 2 \lambda_2 ) \notag
\end{eqnarray}
%\end{small}\ignorespaces
%
where we have made use of the explicit formulae Eq.~(\ref{eq:anglecoeff}) for the angular coefficients $\mathtt{c}_j$. We treat the constants $\mathfrak{C}, r, p > 0$ as fixed here, and extremize over all the angles $\theta_{A}$, $\theta_{B}$, $\varphi_{A}$, $\varphi_{B}$ to extract the strongest possible constraints on the couplings $\lambda_j>0$. 

This bound trivially shows that $\lambda_1,\lambda_{\phi},\lambda_{\chi}>0$ when $(\theta_A,\theta_B)\in\{(0,0),(\frac{\pi}{2},\frac{\pi}{2}),(0,\frac{\pi}{2})\}$ as also given in Eq.~(\ref{eq:stdbound}) (which one finds using standard forward positivity bounds). All that remains is to determine a bound on $\lambda_2$ in terms of the above parameters. To this end, note that the only dependence on $\varphi_{A}$ and $\varphi_{B}$ is on the RHS of (\ref{appboundstart}): varying over all angles $\varphi_{A}$ and $\varphi_{B}$ it turns out that $- (p+1) < p \cos ( \varphi_A - \varphi_B)  + \cos ( \varphi_A + \varphi_B)  < p+1$. This means that the tightest bound is achieved when
\begin{eqnarray}
 && 8 \cos^2 \theta_A  \cos^2 \theta_B\; \lambda_\phi  (1+r) + \big[ 1 - \cos2\theta_{A} \cos 2 \theta_B \big] \; \lambda_{1} (1+\mathfrak{C}) + 8 \sin^2  \theta_A  \sin^2 \theta_B \; \lambda_\chi \\
 && \hspace{45mm} > \ \tfrac{p+1}{2}  \big| \sin 2 \theta_A  \sin 2 \theta_B\; ( \lambda_1 + 2 \lambda_2 ) \big| \ . \notag
\end{eqnarray}
Squaring both sides and isolating all $\theta_{A}$- and $\theta_{B}$-dependence on the RHS, the tightest bound occurs when the RHS is minimized, giving
%
%\begin{small}
%\begin{equation}
% ( \lambda_1 + 2 \lambda_2 )^2 < \frac{4\lambda^2_{1} (1+\mathfrak{C})^2}{(p+1)^2}\underset{\theta_{A},\theta_{B}}{\min}  \frac{ \big[ 8 \frac{\lambda_\phi  (1+r)}{\lambda_{1} (1+\mathfrak{C})} \cos^2 \theta_A  \cos^2 \theta_B\;  + \big[ 1 - \cos2\theta_{A} \cos 2 \theta_B \big] \;  + 8 \frac{\lambda_\chi}{\lambda_{1} (1+\mathfrak{C})} \sin^2  \theta_A  \sin^2 \theta_B \;  \big]^2 }{\sin^2( 2 \theta_A)  \sin^2( 2 \theta_B) }  
%\end{equation}
%\end{small}\ignorespaces
%
%
\begin{equation}
 ( \lambda_1 + 2 \lambda_2 )^2 < \frac{4\lambda^2_{1} (1+\mathfrak{C})^2}{(p+1)^2} \; M\bigg( \frac{\lambda_\phi  (1+r)}{\lambda_{1} (1+\mathfrak{C})}, \frac{\lambda_\chi}{\lambda_{1} (1+\mathfrak{C})} \bigg) 
\end{equation}
with the definition (for fixed $x>0$ and $y>0$)
\begin{eqnarray}
M(x, y) & \equiv & \underset{\theta_{A},\theta_{B}}{\min} \frac{ \left[ 8 x \cos^2 \theta_A  \cos^2 \theta_B\; + 1 - \cos2\theta_{A} \cos 2 \theta_B + 8 y \sin^2  \theta_A  \sin^2 \theta_B  \right]^2 }{ \sin^2 2 \theta_A  \sin^2 2 \theta_B } \\
& = & \underset{-1 \leq A,B \leq 1 \ }{\min} \frac{\left[4 x A B + A + B -2 A B +4 y (1-A) (1-B) \right]^2}{4 A \left(1-A\right) B \left(1-B\right)} \ .
\end{eqnarray}
Minimizing the above function is an exercise in elementary calculus: within the range $-1 \leq A,B \leq 1$, the unique minimum occurs at $A_{\star} = B_{\star} = 1 - (1+\sqrt{y/x})^{-1}$, yielding $M(x,y) = (1 + 4\sqrt{xy})^2$. The corresponding tightest bound is therefore
\begin{equation}
 ( \lambda_1 + 2 \lambda_2 )^2 < \frac{4}{(p+1)^2} \Big( \lambda_1 (1 + \mathfrak{C}) + 4 \sqrt{ (1 + r) \lambda_{\phi} \lambda_\chi } \; \Big)^2 \ .
\end{equation}
This can be expressed more straightforwardly as a bound on $\lambda_2$ such that
\begin{equation} \label{general_boundlambda2}
- \left( \frac{1+\mathfrak{C}}{1+p}+\frac{1}{2}\right) \lambda _1 -\frac{4 \sqrt{(1+r) \lambda _{\chi } \lambda _{\phi }}}{1+p} < \lambda_2 <  \left(\frac{1+\mathfrak{C}}{1+p}-\frac{1}{2}\right) \lambda_1+\frac{4 \sqrt{(1+r) \lambda _{\chi } \lambda _{\phi }}}{1+p} \ .
\end{equation}
To return the equal mass formula (\ref{eqMassEFT-b}) from the main text one takes $(r,p,\mathfrak{C}) \to ( 0,1,0)$, giving the result (\ref{tightestsec3}). Similarly, taking $p \to (1 - \Delta^2/s^2)^2$ and $r \to r_{\max}$ and $\mathfrak{C} \to 0$ in  gives the result (\ref{boundEFTresult_1}). Taking $r \to 0$ and $p \to 1$ gives the result (\ref{boundEFTresult_2}).

Positivity bounds making use of non-linear statements of unitarity will further constrain the allowed region of parameter space, particularly in multi-field EFTs with three or more low-energy modes \cite{Li:2021lpe}. In this work, we simply illustrate how the standard two-state superposition bounds get affected by a difference in mass between the states, paving the way to further generalizing more generic bounds that make use of non-linear unitarity.

%\GKc{comment from introduction:

%Paper that get close to what we have been doing: In Shiu et.~al.~\cite{Andriolo:2020lul}, their Eq.~(3.5) and also Appendix B appears to be doing our EFT calculation --- they seem to recover the same bounds as we do. They seem to use a method from \cite{Andriolo:2018lvp} to ``derive a family of bounds'' which seems to be a superposition of helicity states there. The way they derive their bound is kind of interesting and does not have angles in it the way we have been doing. This calculation \cite{Andriolo:2020lul} is framed as a dilaton and axion scattering, and their action (2.14) is I think the same as ours once they specialize to flat background (not sure what they do with ``$f$''). I think important to cite and compare to them, but we can spin this as we proving the equal mass bound, and also being careful about different masses. 

%}

\subsection{Causality application to toy EFT}
\label{App:causality}

To complement the main discussion involving superposition positivity to the EFT, it is also interesting to explore how a different approach imposes constraints on the couplings from the EFT in Eq.~(\ref{EFTaction}). As discussed in \cite{Adams:2006sv}, while the trivial vacua of EFTs may be perfectly consistent, fluctuations around non-trivial backgrounds can propagate at speeds determined by the signs of higher-dimensional operators, with certain sign choices leading to superluminal modes that conflict with causality in any local quantum field theory. Demanding that these modes travel subluminally then places direct constraints on the EFT couplings of any theory.

To this end, we set the masses $m_\phi = m_\chi = 0$ and perturb the EFT action Eq.~(\ref{EFTaction}) about a very simple Lorentz-breaking background, with fluctuations $\widehat{\phi}$ and $\widehat\chi$ in the EFT such that
\begin{equation} \label{background_phichi}
\phi = g \cos \Theta \cdot t + \widehat{\phi} \qquad \qquad \mathrm{and} \qquad \qquad \chi = g \sin \Theta  \cdot t + \widehat{\chi} 
\end{equation}
where we assume a convenient parametrization of the background with $g >0$ and $\Theta \in \mathbb{R}$. We follow a procedure similar to \cite{Andriolo:2018lvp}, and insert Eq.~\eqref{background_phichi} into the EFT from Eq.~\eqref{EFTaction}. To quadratic order in the fluctuations the action is
\begin{eqnarray}
\mathcal{L} & \simeq & \frac{1}{2} \left[ \begin{matrix} \partial_t \widehat{\phi} & \partial_t \widehat{\chi} \end{matrix} \right] \mathbb{W} \left[ \begin{matrix} \partial_t \widehat{\phi} \\ \partial_t \widehat{\chi} \end{matrix} \right] -  \frac{1}{2} \delta^{ij} \left[ \begin{matrix} \partial_i \widehat{\phi} & \partial_i \widehat{\chi} \end{matrix} \right] \mathbb{K} \left[ \begin{matrix} \partial_j \widehat{\phi} \\ \partial_j \widehat{\chi} \end{matrix} \right] + (\mathrm{cubic\ and\ higher}) \ , \label{L_WK}
\end{eqnarray}
which is organized as a quadratic form with matrices
\begin{equation} 
\mathbb{W} := \left[ \begin{matrix} 1 + 12 \lambda_\phi g^2 \cos^2 \Theta  + 2 ( \lambda_1 + \lambda_2 ) g^2 \sin^2 \Theta & 4 ( \lambda_1 + \lambda_2 ) g^2 \sin \Theta \cos \Theta \\ 4 ( \lambda_1 + \lambda_2 ) g^2 \sin \Theta \cos \Theta & 1 + 2 ( \lambda_1 +  \lambda_2 ) g^2 \cos^2 \Theta  + 12 \lambda_\chi g^2 \sin^2 \Theta  \end{matrix} \right]
\end{equation}
and
\begin{equation} 
\mathbb{K} := \left[ \begin{matrix} 1 + 4 \lambda_\phi g^2 \cos^2 \Theta + 2 \lambda_2 g^2 \sin^2 \Theta  & 2 \lambda_1 g^2 \sin \Theta \cos \Theta \\ 2 \lambda_1 g^2 \sin \Theta \cos \Theta & 1 + 2 \lambda_2 g^2 \cos^2 \Theta + 4 \lambda_\chi g^2 \sin^2 \Theta \end{matrix} \right]  \ .
\end{equation}
Now consider plane-wave solutions for the fluctuations, $\widehat{\phi}(t,\mathbf{x}) \sim u_{\mathbf{k}}(t) e^{i \mathbf{k} \cdot \mathbf{x}}$ and $\widehat{\chi}(t,\mathbf{x}) \sim w_{\mathbf{k}}(t) e^{i \mathbf{k} \cdot \mathbf{x}}$ which solve the equations of motion derived from the quadratic action (\ref{L_WK}). Substituting these into the equations of motion yields
\begin{equation} \label{uwddot}
 \left[ \begin{matrix} \ddot{u}_{\mathbf{k}}(t) \\ \ddot{w}_{\mathbf{k}}(t) \end{matrix} \right] \ = \  - \mathbb{W}^{-1} \mathbb{K} \; |\mathbf{k}|^2 \left[ \begin{matrix} u_{\mathbf{k}}(t) \\ w_{\mathbf{k}}(t) \end{matrix} \right]
\end{equation}
where the matrix $\mathbb{W}$ is invertible when the couplings are treated perturbatively. The general solutions to the equation (\ref{uwddot}) are of course linear combinations of waves $e^{ \pm i \omega t}$ where $\omega$ are the two distinct normal modes of the system in Eq.~(\ref{uwddot}) satisfying\footnote{The relation between $\lambda$ and $\omega$ expressed in Eq.~(\ref{omega_lambda}) follows by rewriting the $2\times2$ second‐order system as a $4\times4$ first‐order one for $\mathbf{x}(t) = \big(u_{\mathbf{k}},w_{\mathbf{k}}, \dot{u}_{\mathbf{k}} ,\dot{w}_{\mathbf{k}} \big)$ such that
\begin{equation}
\dot{\mathbf{x}}(t)=\mathbb{F}\mathbf{x}(t) \qquad \mathrm{with} \qquad 
\mathbb{F}=\begin{bmatrix}0_2 & \mathbb{I}_2\\ -\mathbb{W}^{-1}\mathbb{K}|\mathbf{k}|^2 & 0_2\end{bmatrix} \ .
\end{equation}
The solution $\mathbf{x}(t)$ is a superposition of the plane-wave modes $e^{\pm i\omega t}$ from Eq.~(\ref{uwddot}), where $\bar{\lambda} = \pm i \omega$ are the eigenvalues of the $4 \times 4$ matrix $\mathbb{F}$. Simple matrix manipulation shows that $\det( \mathbb{F} - \bar{\lambda} \mathbb{I}_{4} ) = \det( - \mathbb{W}^{-1}\mathbb{K} |\mathbf{k}|^2 - \bar{\lambda}^2  \mathbb{I}_{2} ) = 0$. This identifies $\bar{\lambda}^2 = - \omega^2 =  - |\mathbf{k}|^2 \lambda$ giving rise to Eq.~(\ref{omega_lambda}).}
\begin{equation} \label{omega_lambda}
\omega^2 = \lambda |\mathbf{k}|^2 \qquad \qquad \mathrm{where} \ \det\left( \mathbb{W}^{-1} \mathbb{K} - \lambda \mathbb{I}_2 \right) =0 \ .
\end{equation}
Requiring subluminality then amounts to simply enforcing that the eigenvalues $\lambda$ of $\mathbb{W}^{-1} \mathbb{K}$ are less than the speed of light such that ${ \omega^2 }/{ |\mathbf{k}|^2 } =  \lambda  \leq  1 $.

From here, the eigenvalues $\lambda$ of the matrix $\mathbb{W}^{-1}\mathbb{K}$ are computing perturbatively to find
\begin{equation}
\lambda \simeq 1 - g^2 a \pm g^2 \sqrt{ a^2 - 4 b \; }
\end{equation}
where 
\begin{eqnarray}
a & := & \lambda_1  + 4 \lambda_\phi \cos^2(\Theta) + 4 \lambda_\chi \sin^2(\Theta) \label{adef} \\
b & := & \left( \lambda_1 \sin^2 \Theta + 4 \lambda_\phi \cos^2\Theta  \right) \left( \lambda_1 \cos^2 \Theta + 4 \lambda_\chi \sin^2 \Theta \right) - (\lambda_1 + 2 \lambda_2)^2 \cos^2\Theta \sin^2 \Theta \qquad \label{bdef}
\end{eqnarray} 
Subliminality of wave propagation $\lambda \leq 1$ then means to impose the bound $a \pm \sqrt{ a^2 - 4 b \; } \geq 0$, which reduces to either (i) $a=b=0$ or (ii) $a>0$ and $0 \leq  b \leq \frac{1}{4} a^2$. In the former case (i) we find all the couplings are zero. In the latter more interesting case (ii) the condition $a>0$ can be simplified to $(\lambda_1 + 4 \lambda_\phi) \cos^2\Theta + (\lambda_1 + 4 \lambda_\chi) \sin^2 \Theta > 0 $, which when varied over $\Theta$ gives rise to $\lambda_{\phi}, \lambda_{\chi}, \lambda_{\phi} \geq 0$ which we use below. The remaining condition $0 \leq  b \leq \frac{1}{4} a^2$ of (ii) reduces to (assuming $\Theta \neq \frac{n\pi}{2}$)
\begin{equation} \label{THETA_lambda2constraint}
 (\lambda_1 + 2 \lambda_2)^2 \ \leq \ \left( \lambda_1 \tan^2 \Theta + 4 \lambda_\phi \right) \left( \lambda_1 \cot^2 \Theta + 4 \lambda_\chi \right) \ .
\end{equation} 
The tightest bound is achieved by minimizing the function on the RHS in $\Theta$, which a trivial calculation shows occurs at the point where $\tan^{4}\Theta_{\star} = \lambda_{\phi}/\lambda_{\chi}$. This means that the tightest bound induced by (\ref{THETA_lambda2constraint}) is $(\lambda_1 + 2 \lambda_2)^2 \leq ( \lambda_1 + 4 \sqrt{\lambda_\phi \lambda_\chi} )^2$, which reduces to precisely the equal mass superposition bound Eq.~(\ref{tightestsec3}) from the main text.

%%%%%%%%%%%%%%%%
\bibliographystyle{JHEP}
\bibliography{biblio}
%%%%%%%%%%%%%%%%

%%%%%%%%%%%%%%%%%%%%%

\end{document}